\documentclass{jfm}
\usepackage{graphicx}
\usepackage{epstopdf, epsfig}

\usepackage{amssymb}
\usepackage[normalem]{ulem}     
\usepackage{chngcntr}
\usepackage[colorlinks=true,linkcolor=blue,citecolor=blue]{hyperref}
\usepackage[capitalise,nameinlink,noabbrev]{cleveref}
\crefname{equation}{}{}
\crefname{figure}{figure}{figures}
\crefname{table}{table}{tables}
\let\ref\cref
\let\eqref\cref
\let\autoref\cref
\usepackage{wrapfig}

\shorttitle{EVF velocity estimate and parameter}
\shortauthor{Swapnil Soni and Avishek Ranjan}

\title{Theoretical estimate and characteristics of electro-vortex flows in cylindrical electrodes}

\author{Swapnil Soni\aff{1}
 \and Avishek Ranjan\aff{1}
  \corresp{\email{avishekr@iitb.ac.in}}}

\affiliation{\aff{1}Department of Mechanical Engineering, Indian Institute of Technology Bombay,
Mumbai 400076, Maharashtra, India}

\begin{document}

\maketitle

\begin{abstract}
Electro-vortex flows (EVF) arise in conducting fluids due to diverging current lines and the non-conservative Lorentz force. They are typically characterized by the $S$ parameter, defined as $S=\mu _0 I^2/4\pi^2 \rho \nu^2$, where it is known that $\Rey \sim \sqrt{S}$ for large currents. However, the strength of the EVF in a confined cylindrical domain with a co-axially placed current collector (CC) depends also on the ratio of the CC radius to the cylinder radius, $K=r_0/R$, in addition to the current magnitude, $I$, fluid density, $\rho$ and kinematic viscosity, $\nu$. For high $\Rey$, using the vorticity transport equation, we derive a new theoretical estimate of the r.m.s. EVF velocity and find that $u \propto I (1-K)/\sqrt{K}$. We validate our estimate with numerical simulations using our custom-built code in \textsc{OpenFOAM} for $K\in[0.1,0.7]$ and $I\in[30,555]$\SI{}{\A}. In addition, for the same range, we compare our numerical results with the estimates of maximum EVF velocity available in the literature. We also discuss the EVF characteristics for varying $K$ using the vorticity dynamics. Finally, we also propose a \emph{modified} EVF parameter ($S_M \propto S (1-K)^2/{K}$) based on our velocity estimate that includes $K$. Our results suggest that the scaling relationship should actually be $\Rey \sim \sqrt{S_M}$.
\end{abstract}

\vspace{-1.5cm}
\section{Introduction}  \label{sec:intro}
Electro-vortex flow (EVF) is a current-driven MHD flow that arises in electrically conducting fluids -- such as liquid metals -- when the current is injected through (or withdrawn from) a concentrated current collector (CC) whose dimensions are smaller than those of the liquid metal domain. In such a configuration, the interaction of current with its own magnetic field produces a Lorentz force that is non-conservative \citep{shercliff1970fluid,bojarevics1988electrically,davidson2017MHD}. These flows are important in several applications such as liquid metal batteries (LMBs) \citep{weber2018electro}, arc-welding \citep{woods1971motion}, electroslag remelting \citep{kharicha2018review}, magnetoplasmadynamic thrusters \citep{xisto2015numerical}, etc. In particular, EVF is expected to occur even at a small current in LMBs, including in the Na-Zn LMB, proposed as an alternative for grid-scale energy storage \citep[see table IV in][]{duczek2024fluid}. Moreover, if the flow is too vigorous, it may lead to rupturing of the negative electrode-electrolyte interface in LMBs \citep{herreman2019numerical}, and an a priori estimate of the EVF velocity can be useful to the battery designers. 

It is important to note that EVF is not an instability. It cannot arise when the current, however large it may be, is homogeneous or uni-directional since in that scenario, the Lorentz force is completely balanced by the (irrotational) pressure gradient. The EVF strength is usually characterized by the EVF parameter, $S=\mu _0 I^2/4\pi^2 \rho \nu^2$, where $\mu_0$ is the magnetic permeability of free space, $I$ is the applied current, $\rho$ is the density, and $\nu$ is the kinematic viscosity. However, this definition of $S$ does not incorporate the diverging nature of current lines, crucial for the existence of EVF. This necessitates specifying the ratio of the current collector radius to the cylindrical domain radius (defined as radius ratio, and denoted by $K=r_0/R$), \emph{separately} to completely describe the EVF. However, the exact relationship of $S$, and therefore, the EVF velocity, $|\boldsymbol{u}|$, on the parameter $K$ is not well known. The EVF velocity magnitude is known to decrease with an increase in $K$ \citep{bojarevics1988electrically}. For $K=1$, no flow is expected as the current lines are uni-directional, and for $K \rightarrow 0$, high-speed flow is expected due to the maximum divergence of the current lines. Despite this, most of the EVF studies in the literature are for a fixed $K$. In this paper, we analytically obtain the dependence of EVF velocity for a cylindrical domain on $K$ and validate the same using numerical simulations.

In their analytical investigation of the flow due to the injection of current at a point ($K=0$), \citet{shercliff1970fluid} obtained an exact similarity solution of the momentum equation for a semi-infinite, inviscid domain. Note that in this case, the work done by the Lorentz force becomes infinite near the current injection point, and there is no energy dissipation due to the absence of sidewalls. Earlier, \citet{lundquist1969hydromagnetic} had investigated a similar problem, and later \citet{sozou1971fluid} included the effect of viscosity. For low $\Rey $ (of $\mathcal{O} (1)$) a balance of the Lorentz and viscous forces leads to $u \propto I^2$ and, therefore, $\Rey \sim S$, where the $\Rey$ is based on the domain radius, $R$. On the other hand, a balance of inertia and the Lorentz force, applicable at larger $\Rey$, leads to an estimate $\Rey \sim \sqrt{S}$ with $u \propto I$ \citep{bojarevics1988electrically, herreman2019numerical}. However, once again, these scalings are incomplete, as they do not include the variation in $K$ applicable for realistic domains.

Since the current collector and the domain have a finite dimension, EVF in confined geometries is of immense interest for the practical applications listed above \citep{moffatt1978some,bojarevics1988electrically}. Using numerical simulation results, \citet{vlasyuk1987effects} had proposed relating $\Rey_{max}$ and $S$ using the following empirical correlations for the maximum EVF speed in a cylindrical liquid metal domain.
\begin{equation}
  Re_{max} = \left\{
    \begin{array}{ll}
      \dfrac{S}{\sqrt{10^{1+5K}}} & \text{for} \: S < 10^{3}, \\[12pt]
      \sqrt{S} \sqrt[3]{10^{3-5K}}         & \text{for} \: S > 10^{5},
    \end{array} \right.
    \label{Vlasyuk_estimates}
\end{equation}
valid for $0.2 \leq K \leq 0.8$. These correlations have been used in the recent scaling laws by \citet{herreman2019numerical} and \citet{mohammad2025current}. However, the physical basis behind these remains unclear. Moreover, as we shall see later, they over-predict the maximum EVF speed in our simulations. \citet{chudnovskii1989evaluating} used the integral constraint on a closed streamline
\begin{equation}
 \oint \boldsymbol{F}_{L} \cdot \mathrm{d} \boldsymbol{l} = -\nu \oint \nabla ^2 \boldsymbol{u} \cdot \mathrm{d} \boldsymbol{l},
 \label{integral}
\end{equation}
where $\boldsymbol{F}_{L} = \boldsymbol{J} \times \boldsymbol{B}/\rho$ is the Lorentz force per unit mass, $\boldsymbol{J}$ is the current density, and $\boldsymbol{B}$ is the magnetic field, to obtain: \mbox{$u \sim \mu _0 I^2 (1-K^2)/2\pi ^2 \rho \nu R K^2$} (details of this derivation are included in \S\labelcref{chu_est}). The relation \labelcref{integral} is essentially a balance between the work done by the Lorentz force on a fluid particle moving on a closed streamline and the energy dissipated by the viscous force near the boundaries. This estimate does have a physical basis, and it also predicts $u \sim 0$ for $K=1$, but it is crucial to note that this requires $\Rey$ to be low and, therefore, it is valid only at small current magnitudes. 

For the turbulent regime $\Rey$, \citet{davidson2017MHD} proposed replacing the laminar shear stress with Reynolds stress in order to correctly determine the (time-averaged) velocity estimate for turbulent flows. This leads to $u \sim \sqrt{\mu/\rho} I/4\pi R$ (details are in \S \labelcref{chu_est}). However, this is valid only for hemispherical domains and requires a priori estimates of velocity fluctuations and boundary layer thickness. Moreover, this estimate requires the assumption that $K \approx 1$ (this may only work for a hemispherical domain, where the current still diverges, albeit lesser than for small $K$). However, with this assumption, there is no EVF at all in a cylindrical domain since this implies perfectly straight current lines. Furthermore, a correct velocity estimate should predict a change in $|\boldsymbol{u}|$ with $K$. 

\begin{figure}
\centering
\begin{tabular}{cc}
\includegraphics[width=6cm]{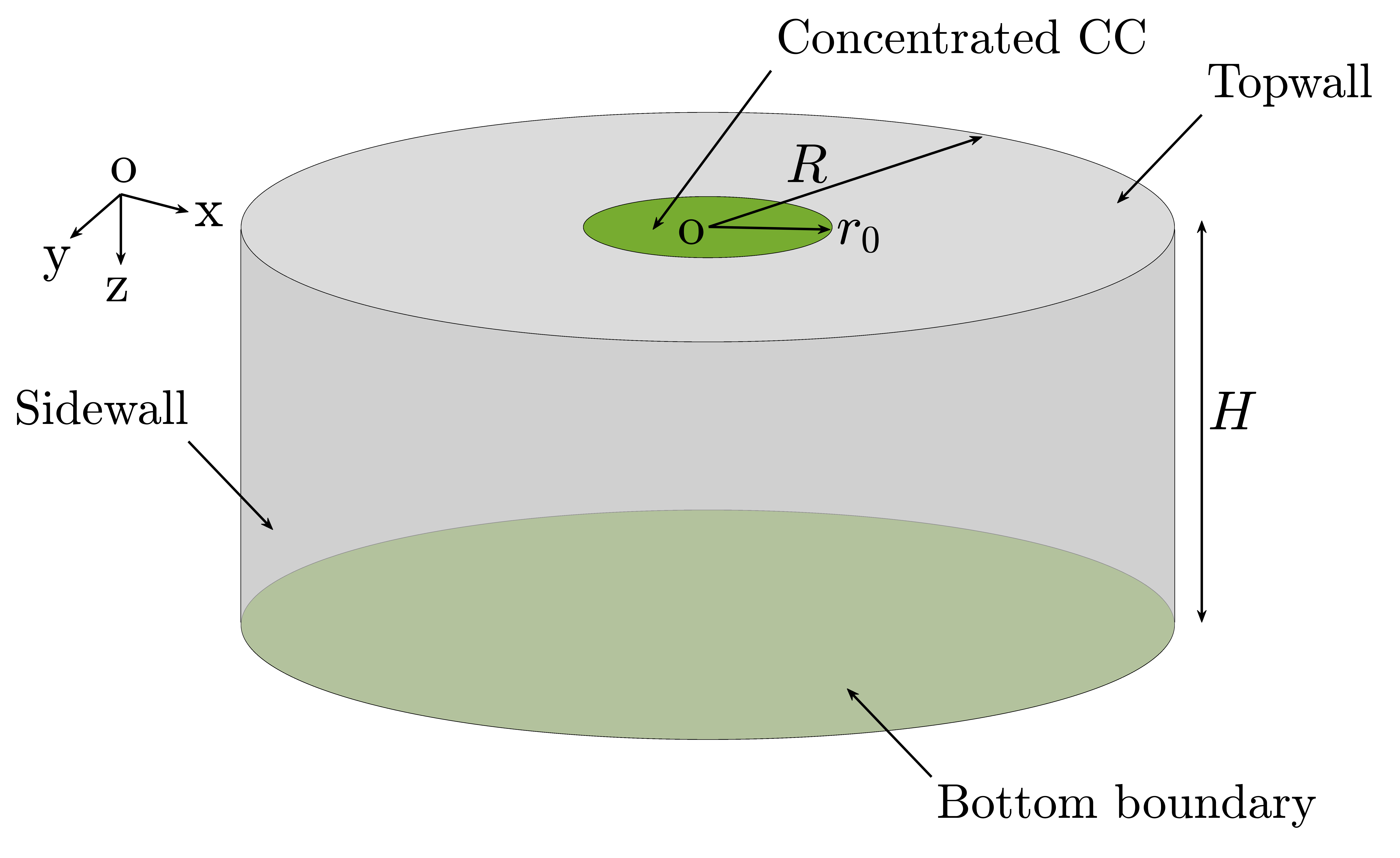} &
\includegraphics[width=5cm]{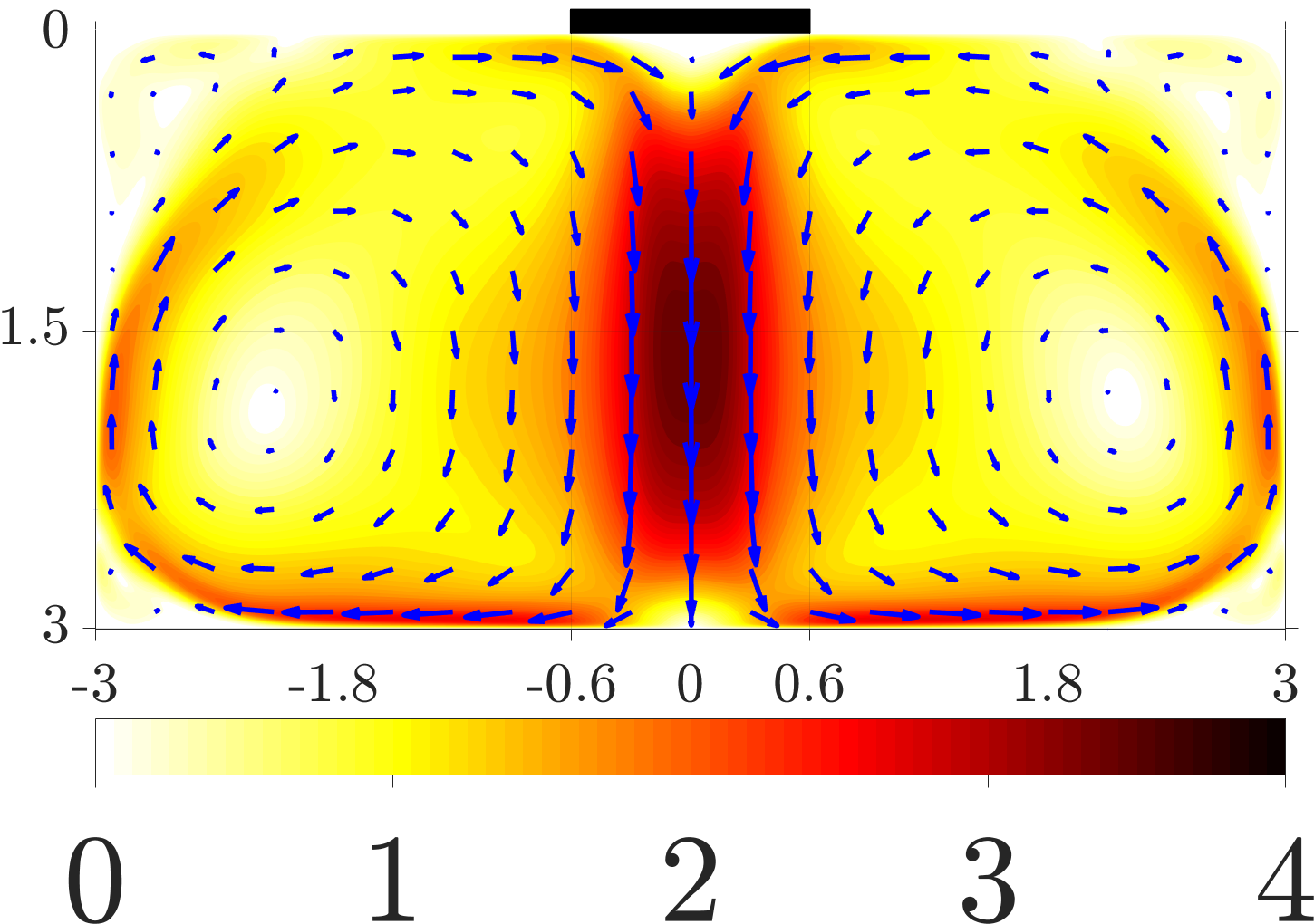} \\
\text{(a)} & \text{(b)}
\end{tabular}
\caption{(a) Schematic of our cylindrical computational domain for EVF simulations. The current enters from the (concentrated) current collector and leaves uniformly from the bottom boundary. Both side and top walls are electrically insulated. (b) Time-averaged EVF velocity magnitude (in cm/s) shown in the xz-plane extracted from our 3D simulations in \textsc{OpenFOAM} for the $K=0.2$ case.}
\label{fig:schematic}
\end{figure}

It is useful to note that in laboratory experiments or practical systems such as LMBs, a (vertical) external magnetic field may always be present. This can create swirl after interacting with the (radial) current, which may be of help in electromagnetic stirring, for instance. Using order of magnitude analysis on the terms in the vorticity equation with the rotational Lorentz force, and defining a forcing length scale related to the CC dimensions, $\delta_f$, \citet{kinnear1994_EM} predicted that at small values of current, the velocity scales as \mbox{$u \sim \delta_f^3 \nabla \times \boldsymbol{F}_{L}/\nu$}. On the other hand, at larger current values the \mbox{$u \sim \delta_f \sqrt{\nabla \times \boldsymbol{F}_{L}}$}. For swirling flows in the presence of an applied current, it is well-known that EVF (due to the current's own magnetic field) is overwhelmed by swirl even for a small external field \citep{davidson1999role,frick2022electro}. Recent attempts have been made to study the effect of $K$ on the swirl \citep{benard2022numerical,mohammad2025current}. In that regard, it is interesting to note that recent experiments of \citet{mohammad2025current} have studied the effect of current collector dimension on the swirl speed, as well as its effect on the transition between EVF and swirl. However, they have used the correlations of Vlasyuk, \labelcref{Vlasyuk_estimates}, that lack physical basis. 

In practical systems, such as LMBs, a multitude of phenomena including swirl and buoyancy are present. An accurate understanding of each phenomenon in isolation is essential before multiphysics phenomena can be analysed. In this article, in \S \labelcref{derivation} we derive a new theoretical estimate of the average EVF velocity for large currents where we include the effect of $K$. As we shall see, our estimate does predict no EVF for $K=1$. Moreover, based on our estimate, we propose a modified EVF parameter $S_M$, as an alternative to $S$, that is applicable for cylindrical domains. Subsequently, in \S \labelcref{chu_est}, we also derive the maximum EVF estimate, valid at high $\Rey$, using the arguments in \citet{chudnovskii1989evaluating,davidson2017MHD}. The predictions for several values of $K$ and $I$ are validated by numerical simulations in \S \labelcref{simulations}. We find that our simulation results compare extremely well with our theoretical estimates. Finally, we conclude in \S \labelcref{sec:conclusion}.

\section{A theoretical estimate of average EVF velocity at high \texorpdfstring{$\Rey$}{ReyNo}}      \label{derivation}
We consider a cylindrical domain, shown in \autoref{fig:schematic}(a), in which the current enters from the co-axially placed concentrated current collector, diverges into the liquid metal domain and leaves uniformly from the bottom boundary. The topwall, outside the CC, and the sidewall are electrically insulated. A typical EVF structure for the radius ratio of $K=0.2$ obtained from our 3D simulation is shown in the xz-plane in \autoref{fig:schematic}(b). Here, we derive a theoretical estimate for the average EVF velocity in isolation with any other phenomenon in the following steps, which start with the calculation of the Lorentz force.

By substituting the applied current density, $\boldsymbol{J}$, from the Amp\`ere's law ($\nabla \times \boldsymbol{B} = \mu_0 \boldsymbol{J}$) into the Lorentz force ($\boldsymbol{F}_{L}$) per unit mass \citep{davidson2017MHD}, we get
\begin{equation}
    \boldsymbol{F}_{L}= \dfrac{\boldsymbol{J} \times \boldsymbol{B}}{\rho} = - \nabla \left( \dfrac{B_{\theta}^2}{2 \rho \mu_0} \right) - \dfrac{B_{\theta}^2}{\rho \mu_0 r} \: \boldsymbol{\hat{e}}_r,      \label{unb_FL}
\end{equation}
where $\boldsymbol{B}$ is the current-induced magnetic field due to the poloidal current density, $\boldsymbol{J}$. The term, ${B_{\theta}^2}/{2 \rho \mu_0}$, inside the ($\cdot$) is the magnetic pressure, which only adds to the fluid pressure and does not influence the fluid motion in the electrode domain. The second term,
\begin{equation}
    \boldsymbol{F}_{L,u} = - \dfrac{B_{\theta}^2}{\rho \mu_0 r} \: \boldsymbol{\hat{e}}_r,     \label{unb_FL2}
\end{equation}
is the non-conservative part of the Lorentz force, and drives EVF in an initially quiescent liquid metal. We call it \emph{unbalanced} Lorentz force.

Let us now take the $\theta-$component of the vorticity transport equation for poloidal flows, which can be obtained by taking the curl of the momentum equation (in a cylindrical coordinate system)
\begin{equation}
    \dfrac{\partial \omega_{\theta}}{\partial t} + \left( \boldsymbol{u} \cdot \nabla \right) \omega_{\theta} - \dfrac{u_r \omega_{\theta}}{r} = \nu \left (\nabla^{2} \omega_{\theta} - \dfrac{\omega_{\theta}}{r^2} \right) + \left[ \nabla \times \boldsymbol{F}_{L,u} \right]_{\theta},
    \label{vorticity_eqn}
\end{equation}
where $\boldsymbol{F}_{L,u}$ is the unbalanced Lorentz force \cref{unb_FL2}, and $\omega_{\theta}$ is the $\theta-$component of the vorticity of the velocity field ($\boldsymbol{\omega} = \nabla \times \boldsymbol{u}$). From the divergence theorem applicable to the curl of $\boldsymbol{F}_{L,u}$, 
\begin{align}
    \int_V \nabla \times \boldsymbol{F}_{L,u} \: \mathrm{d} V        
    & = \int_{\Omega} \boldsymbol{n} \times \left( - \frac{B_{\theta}^2}{\rho \mu_0 r} \right) \boldsymbol{\hat{e}}_r \: \mathrm{d} \Omega \nonumber\\        
    & = \underbrace{\int_{\Omega} - \boldsymbol{\hat{e}}_z \times \left. \left( - \frac{B_{\theta}^2}{\rho \mu_0 r} \right) \right |_{z=0} \boldsymbol{\hat{e}}_r \: \mathrm{d} \Omega}_{\text{Top boundary}} + \underbrace{\int_{\Omega} \boldsymbol{\hat{e}}_z \times \left. \left( - \frac{B_{\theta}^2}{\rho \mu_0 r} \right) \right |_{z=H} \boldsymbol{\hat{e}}_r \: \mathrm{d} \Omega}_{\text{Bottom boundary}} \nonumber\\
    & = 2\pi \left[ \int_0^R \left. \left( \frac{B_{\theta}^2}{\rho \mu_0} \right) \right |_{z=0} \boldsymbol{\hat{e}}_{\theta} \: \mathrm{d} r - \int_0^R \left. \left( \frac{B_{\theta}^2}{\rho \mu_0} \right) \right |_{z=H} \boldsymbol{\hat{e}}_{\theta} \: \mathrm{d} r \right]   \label{curl_FL_eq1}
\end{align}
where $\Omega$ is the bounding surface of the cylindrical liquid domain and $\boldsymbol{n}$ is the unit normal vector to the surface.
Note that i) the origin is defined at the centre of the current collector and $z$ is taken to be positive in the direction of the EVF jet, ii) only the top and bottom boundaries contribute to the integral. The geometry and the coordinate system can be seen in the \cref{fig:schematic}(a). The approximate magnetic field at the top ($z=0$) and at the bottom ($z=H$) boundary can be derived from Amp\`ere’s circuital law \citep{millere1980,bojarevics1988electrically} and reads as
    \refstepcounter{equation}
    $$
        B_{\theta} (r,z=0) = \left\{
        \begin{array}{ll}
        \: \dfrac{\mu_0 I}{2 \pi} \dfrac{r}{r_0^2}, & \text{for} \: r < r_0 \\[12pt]
        \: \dfrac{\mu_0 I}{2 \pi} \dfrac{1}{r},         & \text{for} \: r \geq r_0
    \end{array} \right., \quad\quad\quad
        B_{\theta} (r,z=H) = \dfrac{\mu_0 I}{2 \pi} \dfrac{r}{R^2}.
        \eqno{(\theequation{\mathit{a},\mathit{b}})}
        \label{B0_profiles_eqn} 
    $$
By substituting these magnetic field profiles into \cref{curl_FL_eq1} and simplifying for the $\theta-$component, we get
\begin{eqnarray}        
    \int_V \left[ \nabla \times \boldsymbol{F}_{L,u} \right]_{\theta} \: \mathrm{d} V        
    & = & \dfrac{\mu_0 I^2}{2\pi \rho} \left[ \int_0^{r_0} \left( \dfrac{r^2}{r_0^4} \right) \mathrm{d} r + \int_{r_0}^R \left( \dfrac{1}{r^2} \right) \mathrm{d} r - \int_0^{R} \left( \dfrac{r^2}{R^4} \right) \mathrm{d} r \right] \nonumber\\
    & = & \dfrac{\mu_0 I^2}{2\pi \rho} \left[ \left( \dfrac{4}{3} \right) \left( \dfrac{1}{r_0} - \dfrac{1}{R} \right) \right].
\end{eqnarray}
Or, equivalently, in terms of the radius ratio, $K$,
\begin{equation}
    \int_V \left[ \nabla \times \boldsymbol{F}_{L,u} \right]_{\theta} \: \mathrm{d} V        
    = \dfrac{\mu_0 I^2}{2\pi \rho R} \left[ \left( \dfrac{4}{3} \right) \dfrac{1 - K}{K} \right],   \label{FL_integral}
\end{equation}
where $K=r_0/R$.

At steady state, when the Lorentz force is large, we assume that all the vorticity that is created near the CC is advected into the \emph{bulk region} and also assume that
\begin{equation}
    \left \langle u_r \dfrac{\partial \omega_{\theta}}{\partial r} \right \rangle \sim \left \langle u_z \dfrac{\partial \omega_{\theta}}{\partial z} \right \rangle \sim \left \langle \dfrac{u_r \omega_{\theta}}{r} \right \rangle \sim \dfrac{1}{V} \int_V [ \nabla \times \boldsymbol{F}_{L,u} ]_{\theta} \: \mathrm{d} V,    \label{balance}
\end{equation} 
where $V$ is the domain volume. Here, the $\langle \cdot \rangle$ denotes the time-average. (The time-averaged flow can be assumed axisymmetric in the absence of any external magnetic field since $\boldsymbol{F} _{L,u}$ is also axisymmetric.) Therefore, from \autoref{balance}, taking the balance between
\begin{equation}    
    \left \langle u_z \dfrac{\partial \omega_{\theta}}{\partial z} \right \rangle \sim \dfrac{1}{V} \int_V [ \nabla \times \boldsymbol{F}_{L,u} ]_{\theta} \: \mathrm{d} V.       \label{inertia}
\end{equation}
Replacing the value of the volume-integral from \autoref{FL_integral} in \autoref{inertia}, we have
\begin{equation}
    \left \langle u_z \dfrac{\partial \omega_{\theta}}{\partial z} \right \rangle \sim \dfrac{1}{V} \: \dfrac{\mu_0 I^2}{2\pi \rho R} \left[ \left( \dfrac{4}{3} \right) \dfrac{1 - K}{K} \right].   \label{balance_2}
\end{equation}
For a poloidal flow, $\omega_{\theta} = \left( \partial u_r/ \partial z - \partial u_z/ \partial r \right)$. We take the vorticity magnitude estimate to be $|\omega_{\theta}| \approxeq 2 \: \partial u_z/ \partial r$ for $R=H$ assumed in this study, where $H$ is the domain height (vorticity arises from both radial and vertical velocity gradients in the confined domain). Thus, \autoref{balance_2} becomes
\begin{equation}
    \left \langle 2 \: u_z \dfrac{\partial^2 u_z}{\partial z \partial r} \right \rangle \sim \dfrac{1}{V} \: \dfrac{\mu_0 I^2}{2\pi \rho R} \left[ \left( \dfrac{4}{3} \right) \dfrac{1 - K}{K} \right].   \label{balance_3}
\end{equation}
Now, in order to find a volume-averaged velocity, $u$, we perform a scaling analysis by taking
\begin{equation}
    \Delta r \sim (R - r_0);  \quad
    \Delta z \sim {H};    \quad
     V \sim \pi R^2 H.   \quad
    \label{scaling}
\end{equation}
The scale for $\Delta r$ is inspired by the radial length associated with the diverging current lines, important for the EVF. Another key point to note is that the vorticity magnitude sharply drops with an increase in the CC radius, $r_0$, as observed earlier \citep{vlasyuk1987effects, bojarevics1988electrically}, and this must reflect in the velocity estimate as well. Therefore,
\begin{equation}
    \dfrac{2 u^2}{H(R - r_0)} \sim \dfrac{1}{\pi R^2 H} \: \dfrac{\mu_0 I^2}{2\pi \rho R} \left[ \left( \dfrac{4}{3} \right) \dfrac{1 - K}{K} \right].      
    \label{velocity_scale}
\end{equation}
Or,
\begin{equation}
    u \sim \sqrt{\dfrac{\mu_0 I^2}{4\pi^2 \rho R^2} \left[ \left( \dfrac{4}{3} \right)  \dfrac{(1 - K)^2}{K} \right]} 
    = \sqrt{S \: \dfrac{\nu^2}{R^2} \left[ \left( \dfrac{4}{3} \right) \dfrac{(1 - K)^2}{K} \right]}.       \label{velocity_scale_2}
\end{equation}
Note that this estimate predicts no EVF for $K=1$, and it is valid for large current applications such that $u \propto I$. This estimate does not need any boundary layer thickness, forcing scale or velocity fluctuation scale to be specified, as was the case for the estimates in \citet{kinnear1994_EM,davidson2017MHD}. Alternatively, we can write
 \begin{equation}
    \Rey \sim \sqrt{S \: \left[ \left( \dfrac{4}{3} \right) \dfrac{(1 - K)^2}{K} \right]}.       \label{Re_SM}
\end{equation}  
Here, $\Rey$ is based on the domain radius, $R$. This leads to \mbox{$\Rey \propto \sqrt{S}(1-K)/\sqrt{K}$}; proposing modifications accounting for the radius ratio, $K$, in the earlier relation $\Rey \propto \sqrt{S}$ for large current applications. Thus, our theoretical estimate also leads us to an alternate or \emph{modified} EVF parameter that incorporates the effect of radius ratio, $K$, given by
\begin{equation}
    S_M = \dfrac{\mu_0 I^2}{4\pi^2 \rho \nu^2} \left[ \left( \dfrac{4}{3} \right) \dfrac{(1 - K)^2}{K} \right] = S \left[ \left( \dfrac{4}{3} \right) \dfrac{(1 - K)^2}{K} \right]   \label{mod_evf_param}
\end{equation}

Unlike $S$, the modified EVF parameter, $S_M$, accounts for the effect of converging/diverging current lines near the CC, which is the very basis of EVF. The assumptions we have made to derive our estimate are summarised as follows: i) magnetic field profiles \cref{B0_profiles_eqn} on the top and bottom boundaries assume the domain to be tall, ii) the radial length, $\Delta r$, scales as $(R-r_0)$, and iii) $R=H$. We shall see in \S \ref{simulations} that the r.m.s. of the time-averaged EVF velocity obtained from our numerical simulations compares very well with our theoretical estimate \cref{velocity_scale_2} for various $K$, and $I$ values.

\section{An estimate for the maximum EVF velocity at high \texorpdfstring{$\Rey$}{ReyNo}}        \label{chu_est}
Before deriving an estimate for high $\Rey$, it is useful to first revisit the low $\Rey$ estimate of \citet{chudnovskii1989evaluating}, for steady, laminar flow. On balancing the work done by the Lorentz force with the energy dissipated by the viscous force near the boundaries \labelcref{integral} on a closed streamline, we get
\begin{align}
 \oint \boldsymbol{F_L} \cdot \mathrm{d} \boldsymbol{l} & = \int_R^{r_0} \boldsymbol{F_{L,u}} \cdot \mathrm{d} \boldsymbol{r} + \int_{r_0}^0 \boldsymbol{F_{L,u}} \cdot \mathrm{d} \boldsymbol{r} + \int_0^R \boldsymbol{F_{L,u}} \cdot \mathrm{d} \boldsymbol{r},
 \label{eq1} \\[5pt]
 \oint \boldsymbol{F_L} \cdot \mathrm{d} \boldsymbol{l} & = - \left [\int_R^{r_0} \left ( \dfrac{B_{\theta}^2}{\rho \mu_0 r} \right) \mathrm{d}r + \int_{r_0}^0 \left ( \dfrac{B_{\theta}^2}{\rho \mu_0 r} \right) \mathrm{d}r +\int_0^R \left ( \dfrac{B_{\theta}^2}{\rho \mu_0 r} \right) \mathrm{d}r \right ].
 \label{eq2}
\end{align}
(Recall that that $\boldsymbol{F}_{L,u}$ is radially inward, and it does not contribute to the above line integral along the electrode axis and the sidewall). The flow is accelerated near the CC and then becomes vertical at the axis forming a jet, to satisfy the continuity. Since, typically, the EVF is the strongest on the axis, this integral balance can be associated with a local (maximum) EVF speed. Moreover, using \labelcref{B0_profiles_eqn} and with $K=r_0/R$ we get
\begin{equation}
 \oint \boldsymbol{F_L} \cdot \mathrm{d} \boldsymbol{l} = \dfrac{\mu_0 I^2}{4 \pi^2 \rho R^2} \left [ \dfrac{1 - K^2}{K^2} \right ].
 \label{eq4}
\end{equation}
A scale for the term on the r.h.s. of the \cref{integral} can be obtained as follows. Taking $R=H$, where $H$ is the electrode height, $\nabla \sim 1/R$ and $\int \mathrm{d} \boldsymbol{l} \sim R$, gives
\begin{equation}
\nu \oint \nabla ^2 \boldsymbol{u} \cdot \mathrm{d} \boldsymbol{l} \sim \nu \dfrac{uR}{R^2} = \dfrac{\nu u}{R}.      \label{eq5}
\end{equation}
Substituting the values from \cref{eq4,eq5} into \cref{integral} and ignoring the negative sign (for the scaling analysis), we can get
\begin{equation}
u \sim \dfrac{\mu_0 I^2}{4 \pi^2 \rho \nu R} \left [ \dfrac{1 - K^2}{K^2} \right ].    \label{eq7}
\end{equation} 
Thus, we get the $u \propto I^2$ relationship applicable for the laminar regime. Note that the group of terms outside the $[\cdot]$ give the same estimate as that obtained by \citet[see \S 12.4.3]{davidson2017MHD} for the hemispherical domain. The terms in $[\cdot]$ account for the radius ratio, $K$, for the cylindrical domain. This velocity scale \labelcref{eq7} predicts no flow for $K=1$. Rearranging, we get
\begin{equation}
\Rey \sim \dfrac{\mu_0 I^2}{4 \pi^2 \rho \nu^2} \left [ \dfrac{1 - K^2}{K^2} \right ],    \label{eq8}
\end{equation}
where $\Rey$ here is based on the domain radius, $R$. From this, we can obtain the modified EVF parameter, $S_{M,C}$, of \citet{chudnovskii1989evaluating},
\begin{equation}
S_{M,C} = \dfrac{\mu_0 I^2}{2 \pi^2 \rho \nu^2} \left [ \dfrac{1 - K^2}{K^2} \right ].    \label{eq9}
\end{equation}

This leads to $\Rey \sim S_{M,C}$, applicable for low $\Rey$. To obtain the estimate applicable for high $\Rey$, \citet{chudnovskii1989evaluating} divided the above estimate \labelcref{eq7} by a factor proportional to $\sqrt{S_{M,C}}$, to get $\Rey \sim \sqrt{S_{M,C}}$. This is associated with the maximum velocity estimate
\begin{equation}
u \sim \sqrt{\dfrac{\mu_0 I^2}{4 \pi^2 \rho R^2} \left [ \dfrac{1 - K^2}{K^2} \right ]}.    \label{eq10}
\end{equation}

As noted by \citet{chudnovskii1989evaluating}, this estimate compares well with the empirical results of \citet{vlasyuk1987effects} \labelcref{Vlasyuk_estimates} for $0.2 \leq K \leq 0.8$. Let us attempt to obtain an estimate using a physical basis inspired by the analysis in \citet[\S 12.4.3]{davidson2017MHD} for EVF in hemispherical domains. For the term on the r.h.s. of the \autoref{integral}, as suggested by \citet{davidson2017MHD}, we replace the laminar shear stress by the Reynolds stress for a closed streamline (close to the boundary) in the EVF with high $\Rey$.
\begin{equation}
\nu \oint \nabla ^2 \boldsymbol{u} \cdot \mathrm{d} \boldsymbol{l} = \dfrac{\mu}{\rho} \oint \nabla \cdot (\nabla \boldsymbol{u}) \cdot \mathrm{d} \boldsymbol{l} \sim \dfrac{1}{\rho\delta_w} \tau_w R.
\label{eq12}
\end{equation}
Here $\mu$ is the dynamic viscosity of the liquid metal, and $\tau_w$ \& $\delta_w$ are the wall shear stress and characteristic length scale (here, the boundary layer thickness), respectively, with $\nabla \sim 1/\delta_w$, $\int \mathrm{d} \boldsymbol{l} \sim R$.

Now, we take $\tau_w/\rho \sim (u^{\prime})^2$ to estimate the turbulence level in the metal, where $u^{\prime}$ is the fluctuating velocity component defined as $u^{\prime} = u - \langle u \rangle$. Substituting the scale for $\tau_w/\rho$, and taking $u^{\prime} = \alpha_1 u$, $\delta_w = \alpha_2 R$, where $\alpha_1$ and $\alpha_2$ depend on the flow configuration in the chosen domain. We obtain
\begin{equation}
\dfrac{1}{\rho\delta_w} \tau_w R \sim \dfrac{(u^{\prime})^2}{\delta_w} R = \left( \dfrac{\alpha_1^2}{\alpha_2} \right) u^2.
\label{eq13}
\end{equation}
Substitution of this into \autoref{eq12} yields
\begin{equation}
\nu \oint \nabla ^2 \boldsymbol{u} \cdot \mathrm{d} \boldsymbol{l} \sim \left( \dfrac{\alpha_1^2}{\alpha_2} \right) u^2.
\label{eq14}
\end{equation}
Substitution of this, together with the \autoref{eq4}, into \autoref{integral} culminates
\begin{equation}
\left( \dfrac{\alpha_1^2}{\alpha_2} \right) u^2 \sim \dfrac{\mu_0 I^2}{4 \pi^2 \rho R^2} \left [ \dfrac{1 - K^2}{K^2} \right ]
\Rightarrow
u \sim \sqrt{\left( \dfrac{\alpha_2}{\alpha_1^2} \right) \dfrac{\mu_0 I^2}{4 \pi^2 \rho R^2} \left [ \dfrac{1 - K^2}{K^2} \right ]}.
\label{eq15}
\end{equation}
Alternatively,
\begin{equation}
\Rey \sim \sqrt{\left( \dfrac{\alpha_2}{\alpha_1^2} \right) \dfrac{\mu_0 I^2}{4 \pi^2 \rho \nu^2} \left [ \dfrac{1 - K^2}{K^2} \right ]}.
\label{eq16}
\end{equation}
Interestingly, the values of $\alpha_1 \approx 1/3.5$ and $\alpha_2 \approx 1/10$ used by \citet{davidson2017MHD}, applicable for induction furnace, show that $\left( {\alpha_2}/{\alpha_1^2} \right) \approx 1$. With this, we arrive at the $\Rey \sim \sqrt{S_{M,C}}$ for high $\Rey$. However, as outlined earlier, this requires the knowledge of turbulent intensity and the boundary layer thickness.

\section{EVF characteristics and validation of our theoretical estimate}         \label{simulations}
The EVF is characterized by an inward flow near the current collector culminating in an axial jet and a return flow near the side walls; forming a toroidal vortex and a poloidal flow structure under steady-state conditions within a cylindrical domain with a coaxially positioned current collector. (Please refer to \autoref{fig:schematic}(b) for visualization.) The flow intensity increases with an increase in the strength of the applied current or a decrease in the CC radius (the flow may transition from laminar to turbulent). In contrast, reducing the current strength or increasing the CC radius reduces the flow speed. In this section, we discuss the fundamental characteristics of electro-vortex flow along with the quantitative analysis to validate our theoretical estimate for the average EVF velocity. In addition, we also assess the validity of the estimates of \citet{vlasyuk1987effects} \labelcref{Vlasyuk_estimates} and \citet{chudnovskii1989evaluating} (\S \labelcref{chu_est}) for the prediction of the maximum velocity under high $S$ conditions.

We perform numerical simulations to validate our estimate. Experimental setups -- such as that of \citet{zhilin1986experimental} -- inevitably introduce factors like current feed lines, current leakages, solid-liquid electromagnetic coupling, and external magnetic fields (e.g., the Earth’s magnetic field), all of which can significantly alter the current and magnetic field distributions, thereby affecting the flow. These complications pose considerable challenges to experimentally realizing a \emph{pure} electro-vortex flow -- one that is isolated from other multiphysics phenomena. Furthermore, the results of the 3D simulations helps explain the EVF characteristics which otherwise may not be possible in the experiments due to the opaque nature of the liquid metal.

\subsection{Simulation details}     \label{simulation_details}

Numerical simulations are performed using \textsc{OpenFOAM v6} with our custom solver tailored for simulating current-driven MHD flows \citep[refer][for details]{soni2024evaluating}, which has been validated with the literature. Our geometry, shown in \autoref{fig:schematic}(a), is a cylindrical domain of radius $R$, in which the current enters from a concentrated CC of radius $r_0$, and leaves uniformly from the bottom boundary. All other walls are electrically insulated. The no-slip and no-penetration boundary conditions are assumed for the velocity.

The continuity and momentum equations to simulate incompressible current-driven flows with the Lorentz force per unit mass, $(\boldsymbol{J} \times \boldsymbol{B})/\rho$, under the induction-less approximation are as follows \citep{davidson2017MHD}:
\begin{subeqnarray}
 \nabla \cdot \boldsymbol{u} & = &
    0,		\label{continuity_eqn} \\[3pt]
  \frac{\partial \boldsymbol{u}}{\partial t} + (\boldsymbol{u} \cdot \nabla) \boldsymbol{u} & = &
    - \frac{1}{\rho} \nabla p + \nu \nabla^{2} \boldsymbol{u} + \frac{\boldsymbol{J} \times \boldsymbol{B}}{\rho}.	\label{momentum_eqn}
\end{subeqnarray}

The current density is computed using the Ohm's law and the Laplace equation obtained by assuming the charge conservation ($\nabla \cdot \boldsymbol{J} = 0$): $\boldsymbol{J} = - \sigma \nabla \phi$ and $\nabla^2 \phi = 0$, where $\sigma$ is the electrical conductivity of the liquid metal and $\phi$ is the applied scalar electric potential. Fixed gradient boundary condition on $\phi$ is used at the CC. The magnetic field, $\boldsymbol{B}$ on the walls, is accurately computed using the Biot-Savart law \labelcref{BS_law} in the confined domain,
\begin{equation}
\boldsymbol{B}(\boldsymbol{r}) = \frac{\mu_{0}}{4 \pi} \int \frac{\boldsymbol{J}(\boldsymbol{r}^{\prime}) \times (\boldsymbol{r} - \boldsymbol{r}^{\prime})}{|\boldsymbol{r} - \boldsymbol{r}^{\prime}|^{3}} \mathrm{d} V^{\prime}.		\label{BS_law}
\end{equation}

Thereafter, $\nabla^2 \boldsymbol{B} = 0$ is solved to obtain the interior field to save computation time \citep{weber2018electro}. We ignore any external magnetic field. A fourth-order gradient scheme with cubic interpolation is used to solve the equations. The properties of liquid mercury at room temperature, as mentioned in \citet{weber2018electro}, are employed in the simulations and are assumed constant. More details are in \cite{soni2024evaluating} and \citet{soni2025evf}.

\subsubsection{Computational grid selection}      \label{grid_selection}
For the case with $K=0.2$ and $I=\SI{200}{\A}$, we performed a detailed grid convergence study using three mesh sizes, focusing on the magnitude of instantaneous velocity. These along with validation with the literature is discussed in detail in our earlier work, \citet{soni2024evaluating}. This case is considered as the benchmark, given its extensive treatment in the literature. For cases with lower $K$ and higher current magnitudes, we assessed grid independence of the average velocity. Details of the selected grids for all simulated cases are provided in \autoref{table:cases_K,table:cases_I0}, along with the respective time step sizes, which were chosen to ensure the Courant number remains below 1.

\subsection{Effect of the current collector radius}      \label{K_effect}
\begin{figure}
\centering
\begin{tabular}{ccc} 
\includegraphics[width=4cm]{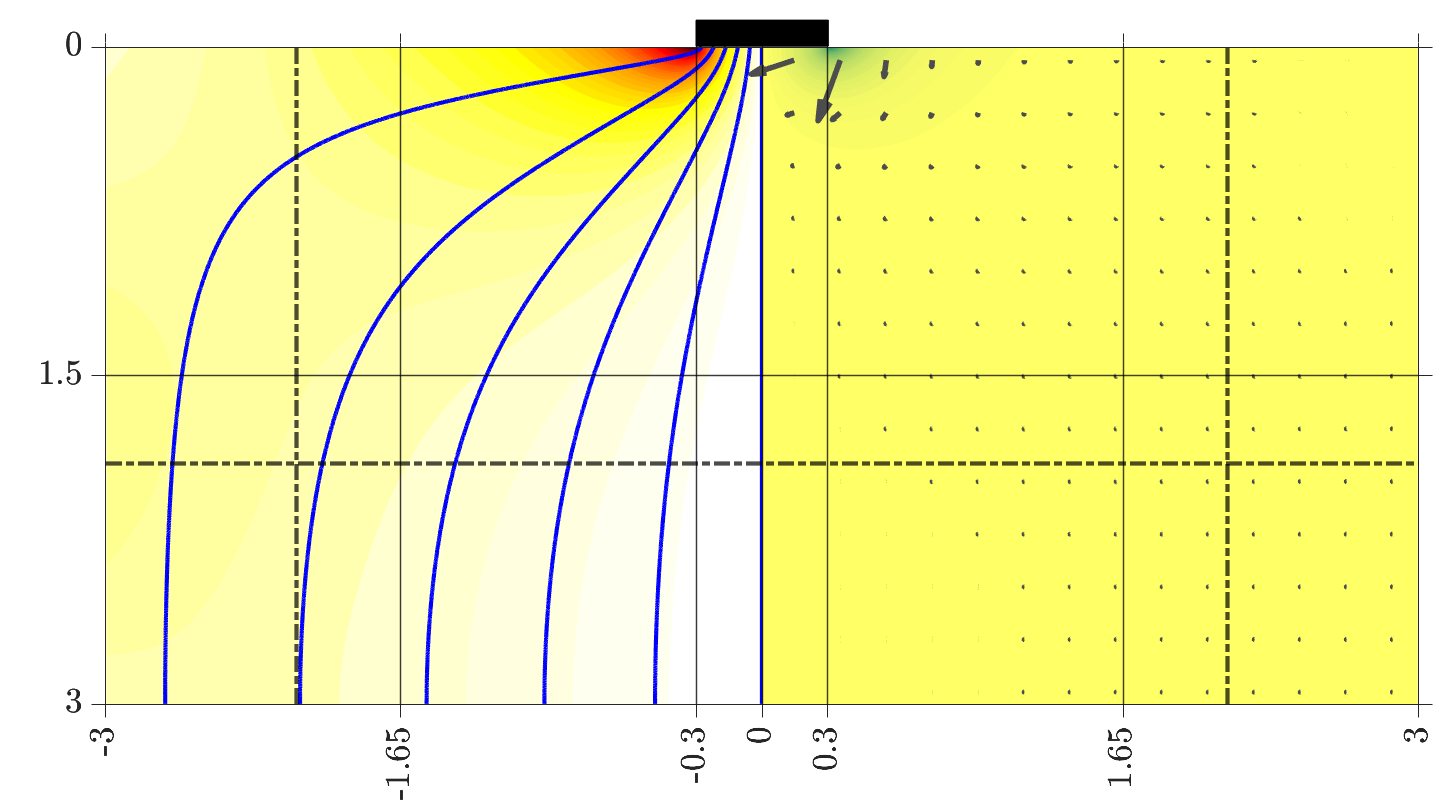} &
\includegraphics[width=4cm]{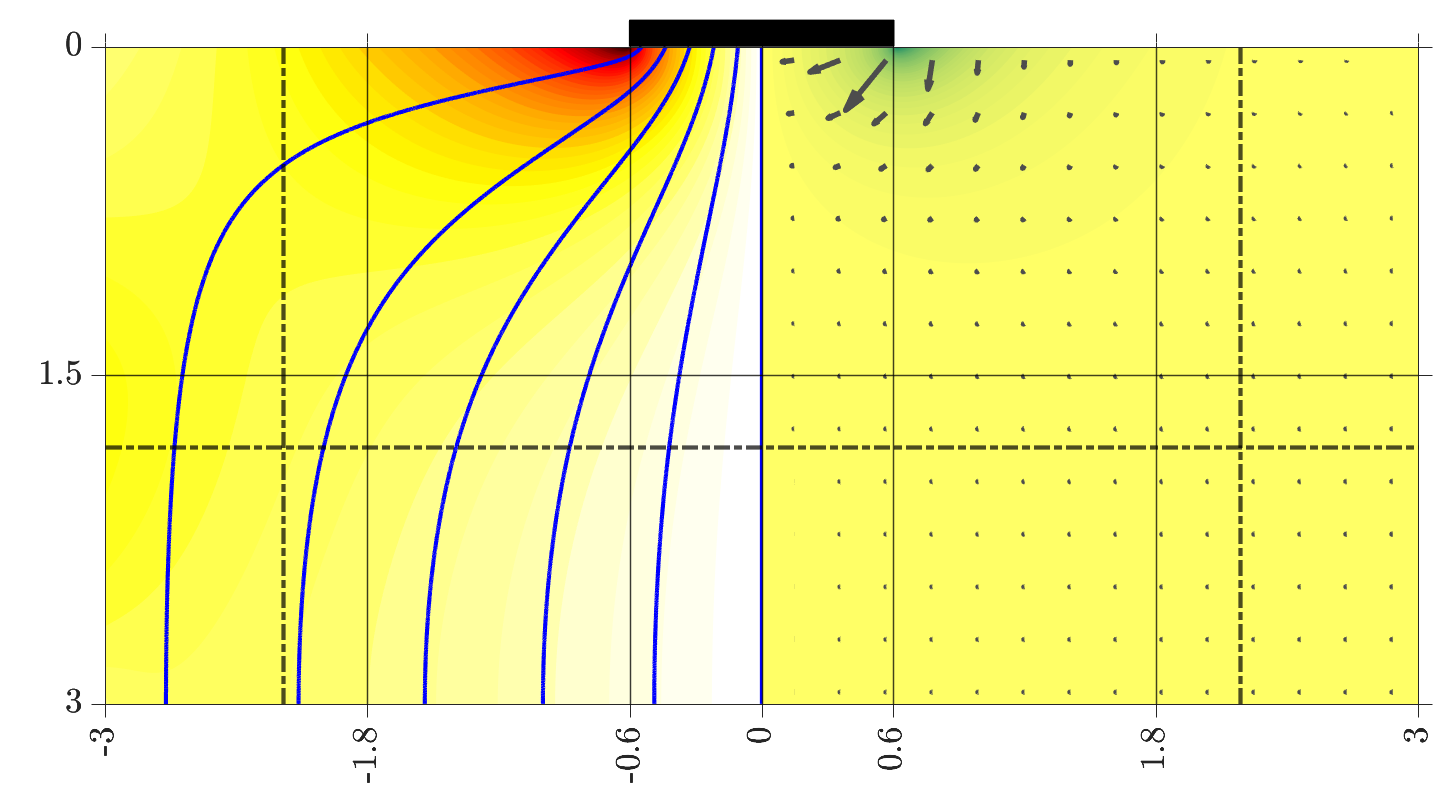} &
\includegraphics[width=4cm]{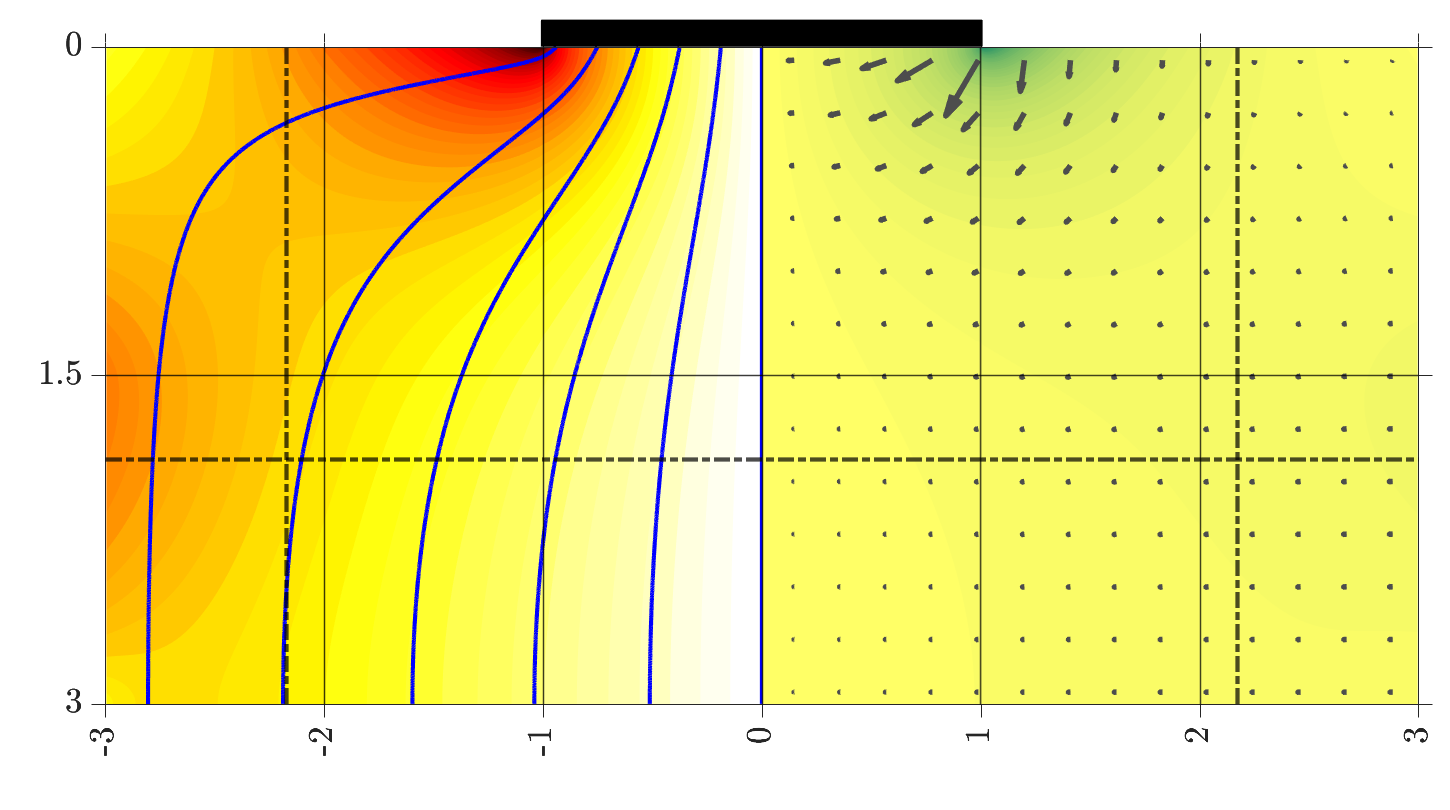} \\
 \text{(a)} $K=0.1$ & \text{(b)} $K=0.2$ & \text{(c)} $K=0.33$ \\ \\
  
\includegraphics[width=4cm]{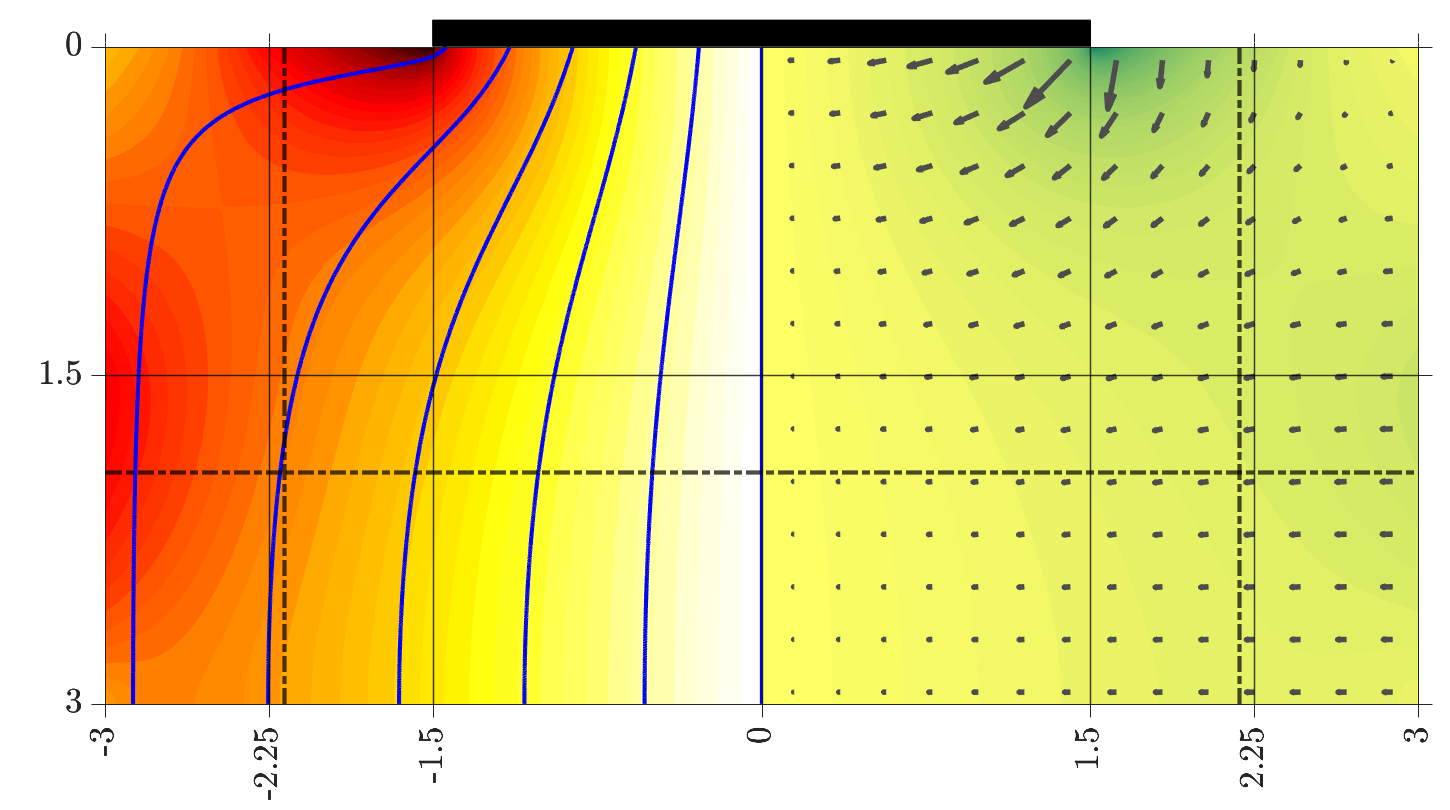} &
\includegraphics[width=4cm]{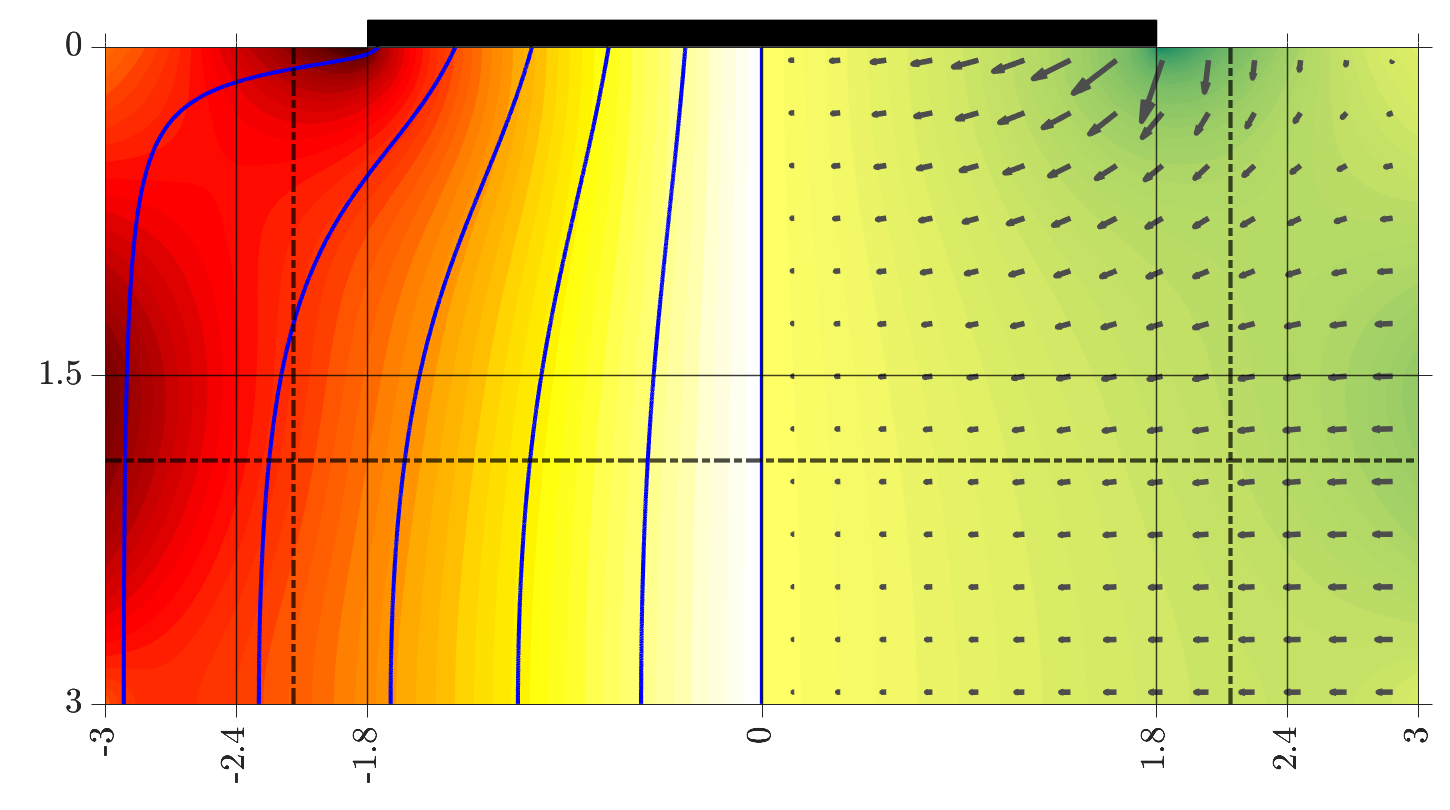} &
\includegraphics[width=4cm]{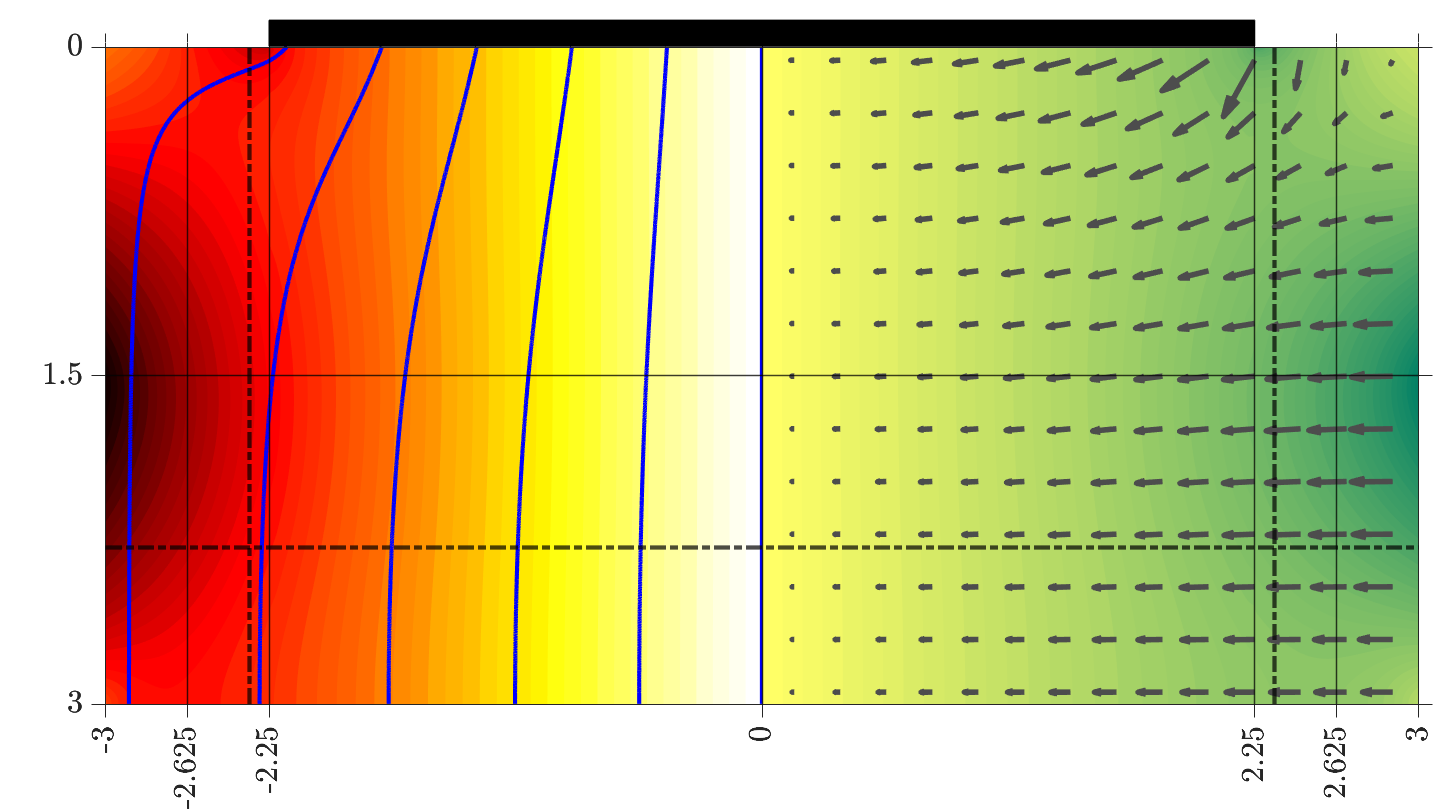} \\
 \text{(d)} $K=0.5$ & \text{(e)} $K=0.6$ & \text{(f)} $K=0.75$ \\ \\
\end{tabular}
\begin{tabular}{cc}
    \centering
\includegraphics[width=5cm]{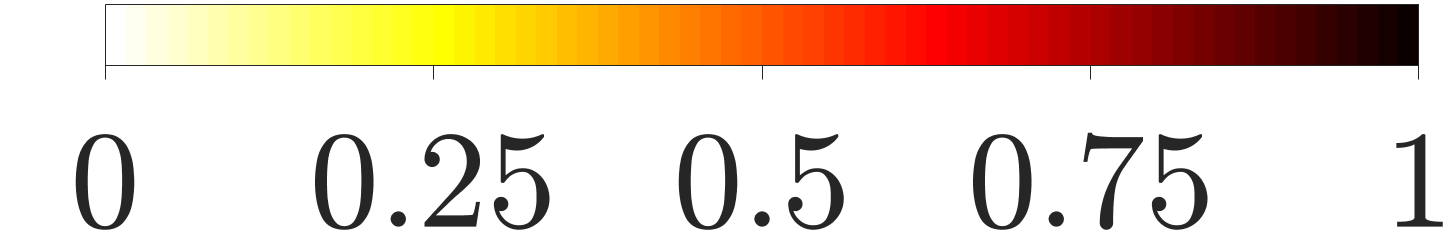} &
\includegraphics[width=5cm]{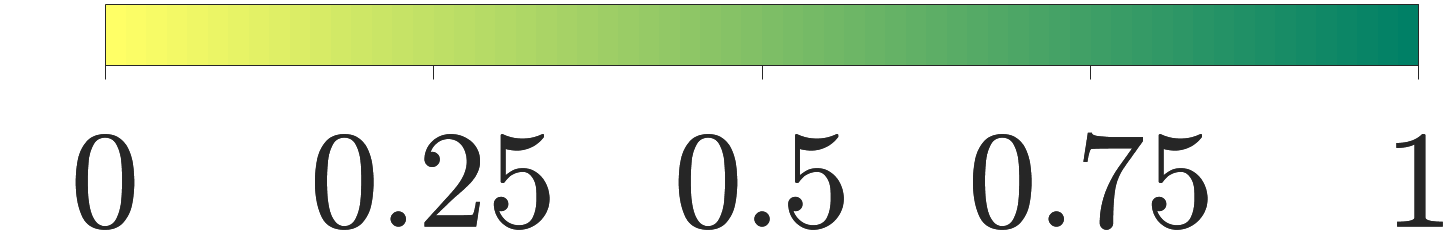} \\[5pt]

\Large{$\frac{|\boldsymbol{B_{0}}|}{|\boldsymbol{B_{0}}|_{max}}$}   &
\Large{$\frac{F_{L,u}}{(F_{L,u})_{max}}$}   \\[10pt]
    \end{tabular}
\caption{Magnetic field overlapped with the current lines (left half), contours of the \emph{unbalanced} Lorentz force ($\boldsymbol{F}_{L,u}$) with arrows showing the direction of the \emph{total} Lorentz force for that particular $K$, $\boldsymbol{F}_{L}$ (right half).}
\label{fig:contour_plots_K_B0}
\end{figure}

For a constant applied current of \SI{200}{\A}, we select various radius ratios (see \mbox{\autoref{table:cases_K}}) within the range $K\in[0.1,\:0.95]$ for the numerical simulation of EVF in a cylindrical domain with $R=H$. The left half in each plot of \cref{fig:contour_plots_K_B0} shows the magnetic field normalized by its maximum (refer \autoref{table:K_max_A1} for the maximum values) and superimposed with the current lines, whereas the right half shows the normalized \emph{unbalanced} Lorentz force, $\boldsymbol{F}_{L,u}$, with arrows indicating the direction of \emph{total} Lorentz force. (These are normalized to identify the regions of relative dominance.) As expected, variations in $K$ lead to changes in the current path, thereby affecting the distributions of magnetic field and Lorentz force within the domain. For lower values of $K$, the magnetic field and unbalanced Lorentz force are concentrated near the CC. However, for $K \geq 0.33$, we observe that the field strength near the sidewall, due to the uniform current within the finite-size domain, becomes comparable to that arising from the diverging current. In fact, for $K\geq0.75$, the field is the largest on the sidewall. It can be observed in all cases with $K < 0.75$ that the $\boldsymbol{F}_{L,u}$ is much more significant in the region $r_0 < r < R$ than in $0 < r < r_0$. This is verified by the ratio, $A/B$, obtained from our numerical simulations (please refer \autoref{tab:FL_int}). Furthermore, we also show that the ratio of the $\nabla\times\boldsymbol{F}_{L}$ in the cylindrical region $0<r<r_0$ to that in the cylindrical shell region $r_0<r<R$, denoted as $C/D$ is $\approx 0.3$. Thus, the predominant contribution to the curl of the Lorentz force comes from the region outside the cylinder circumscribing the CC. This even holds for $K=0.75$.

\begin{table}
  \begin{center}
\def~{\hphantom{3}}
  \begin{tabular}{lcccccccc}
      $K$  &  $I$ (\SI{}{\A}) & $N_d \times N_{\theta} \times N_z$  & $\Delta t$ (\unit{\s})   & $S (\times 10^4)$   &   $S_M (\times 10^4)$  & $\Rey_{theo}$  & $\Rey_{rms}$   &   $\Rey_{max}$
      \\[3pt]
       0.1  &   200 & $285 \times 100 \times 100$ & 0.0005  & ~653   & 7056  &   5712   & 5378 & 17678   \\
       0.15  &   200 & $285 \times 100 \times 100$ & 0.0005  & ~653   & 4196  &   4405   & 4337 & 11882   \\
       0.2  &   200 & $260 \times 120 \times 85~$ & 0.002~  & ~653   & 2787  &   3590   & 3726 & ~8918   \\
       0.25 &   200 & $265 \times 100 \times 100$ & 0.002~  & ~653   & 1960  &   3010   & 3276 & ~7064   \\
       0.33 &   200 & $265 \times 100 \times 100$ & 0.002~  & ~653   & 1185  &   2341   & 2465 & ~4777   \\
       0.4 &   200 & $258 \times 112 \times 100$ & 0.002~  & ~653   & ~784  &   1904   & 1908 & 3614   \\
       0.5  &   200   & $260 \times 120 \times 85~$ & 0.002~ & ~653   & ~436  &   1419   & 1460 & ~2766   \\
       0.6  &   200   & $240 \times 160 \times 85~$ & 0.005~ & ~653   & ~232  &   1037   & ~999 & ~1840   \\
       0.65  &   200   & $240 \times 160 \times 85~$ & 0.005~ & ~653   & ~164  &   ~871   & ~708 & ~1326   \\
       0.7  &   200   & $240 \times 160 \times 85~$ & 0.005~ & ~653   & ~112  &   ~720   & ~462 & ~~886   \\
       0.75 &   200   & $240 \times 160 \times 85~$ & 0.01~~ & ~653   & ~~73  &   ~579   & ~412 & ~~913   \\
       0.95 &   200 & $200 \times 280 \times 85~$ & 0.01~~  & ~653   & ~~~2  &   ~103   & ~744 & ~1504   \\
  \end{tabular}
  \caption{Details on the numerical simulations for various $K$ with the number of cells ($N_d \times N_{\theta} \times N_z$, in $d$, $\theta$, and $z$--directions respectively) and chosen time step to satisfy the \emph{CFL} criteria. The current density, $J_0$, is constant at the bottom boundary and is $\approx$ \SI{7}{\A/\cm^2}. Reynolds number (based on $R$) for these cases is also reported along with the $S$ and $S_M$ values. The $S$, $S_M$, and $\Rey$ values are rounded off to the nearest integer.}
  \label{table:cases_K}
  \end{center}
\end{table}

\begin{table}
  \begin{center}
\def~{\hphantom{3}}
  \begin{tabular}{lcccccccc}
      $K$  &  $|\boldsymbol{B_0}|_{max}$ & $(F_{L,u})_{max}$  & $|\langle \omega_{\theta} \rangle|_{max}$   & $|\langle \boldsymbol{u} \rangle|_{max}$   &   $\langle u_{z} \rangle_{max}$
      \\[3pt]
        &  (G) & (\SI{}{\cm/\s^2})  & (\SI{}{1/\s})   & (\SI{}{\cm/\s})   &   (\SI{}{\cm/\s})
      \\[3pt]
       0.1  &   66.87 & 83.94 & 524.53  & 6.89   & 6.89   \\
       0.15  &   45.79 & 26.84 & 271.53  & 4.58   & 4.58   \\
       0.2  &   34.05 & 11.17 & 160.86  & 3.46   & 3.46   \\
       0.25 &   28.26 & ~6.19 & 110.03  & 2.72   & 2.72   \\
       0.33 &   21.40 & ~2.71 & ~61.87  & 1.79   & 1.79   \\
       0.4 &   17.52 & ~1.50 & ~39.51  & 1.29   & 1.23   \\
       0.5  &   13.72   & ~0.73 & ~26.56 & 1.01   & 0.98   \\
       0.6  &   10.72   & ~0.37 & ~14.35 & 0.65   & 0.63   \\
       0.65  &   ~9.74   & ~0.29 & ~~8.78 & 0.42   & 0.41   \\
       0.7  &   ~9.82   & ~0.22 & ~~6.04 & 0.30   & 0.30   \\
       0.75 &   10.87   & ~0.23 & ~~5.67 & 0.31   & 0.29   \\
       0.95 &   11.29 & ~0.25 & ~11.23  & 0.52   & 0.52   \\
  \end{tabular}
  \caption{The maximum values used in normalizing the quantities. This is for the effect of $K$. The applied current is \SI{200}{\A}, which is kept constant. The values are rounded off to the two significant digits.}
  \label{table:K_max_A1}
  \end{center}
\end{table}

\begin{table}
  \begin{center}
\def~{\hphantom{3}}
  \begin{tabular}{lcccccc}
    $K$  &  $A = \int_{0}^{r_0} \left| \boldsymbol{F}_{L} \right| \mathrm{d}V$ & $B = \int_{r_0}^{R} \left| \boldsymbol{F}_{L} \right| \mathrm{d}V$ &  $C = \int_{0}^{r_0} \left| \nabla \times \boldsymbol{F}_{L} \right| \mathrm{d}V$ & $D = \int_{r_0}^{R} \left| \nabla \times \boldsymbol{F}_{L} \right| \mathrm{d}V$  & $A/B$  & $C/D$ \\[3pt]
       
    0.1  &   \num{7.79e-04}   & \num{6.77e-03}   & 0.24~  &   1.12~   & 0.12   & 0.22   \\
       
    0.2  &   \num{8.11e-04}   & ~~\num{5e-03}   & 0.11~  &   0.47~   & 0.16   & 0.24   \\
       
    0.33  &   \num{9.15e-04}   & \num{3.95e-03}   & 0.062  &   0.22~   & 0.23   & 0.29   \\
       
    0.5  &   \num{1.07e-03}   & \num{2.75e-03}   & 0.035  &   0.092   & 0.39   & 0.38   \\
       
    0.6  &   ~\num{1.2e-03}   & \num{2.24e-03}   & 0.022  &   0.053   & 0.58   & 0.41   \\
       
    0.75  &   \num{1.75e-03}   & \num{1.74e-03}   & \num{9.25e-03}  &   0.039   & 1.01   & 0.24   \\
  \end{tabular}
  \caption{Volume integral of $\boldsymbol{F}_{L}$ and its curl inside and outside the CC region for selected cases. These are obtained from the numerical simulations.}
  \label{tab:FL_int}
  \end{center}
\end{table}

\begin{figure}
\centering
\begin{tabular}{lccc}
\rotatebox{90}{\hspace{0.5cm} ${K=0.1}$} &
\includegraphics[width=4cm]{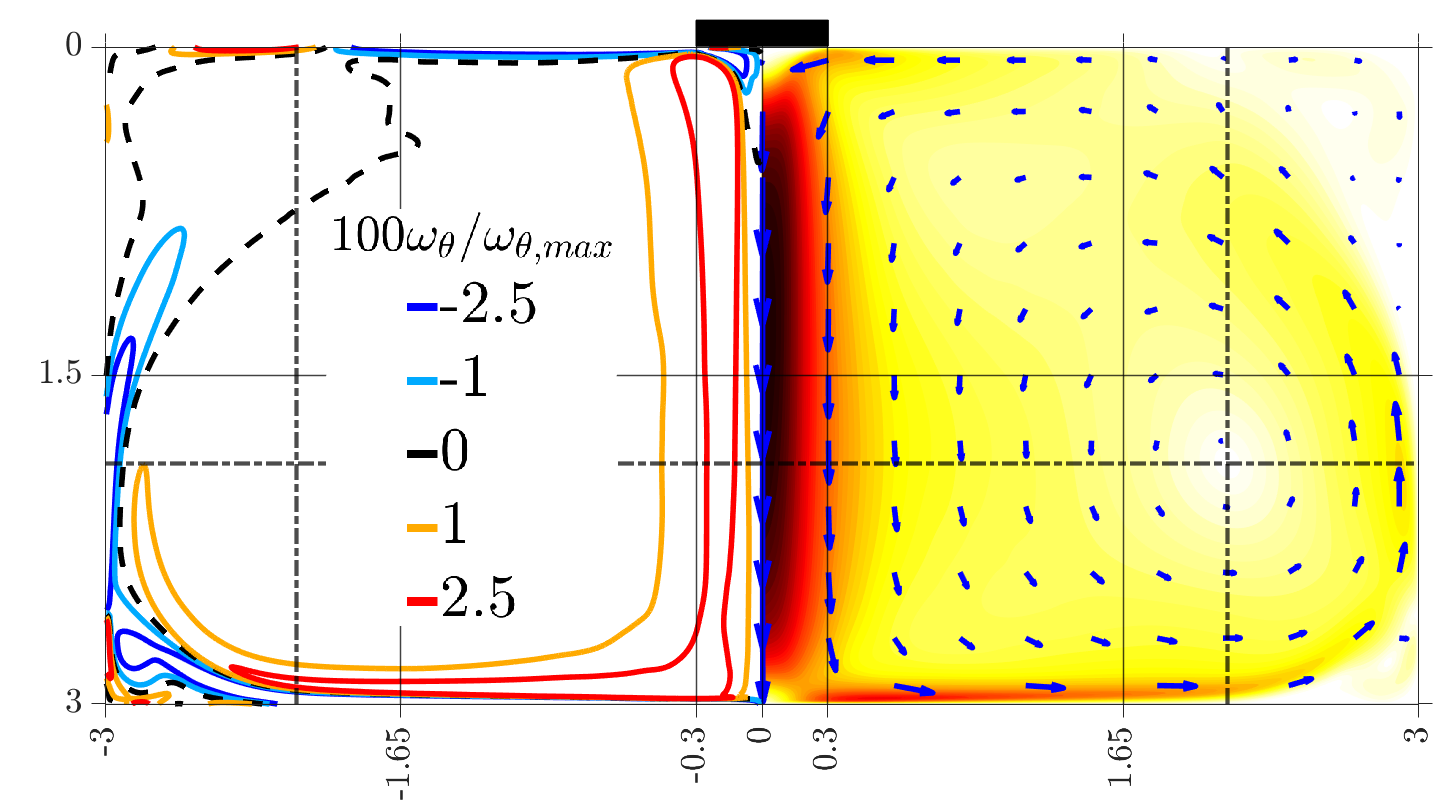} &
\includegraphics[width=4cm]{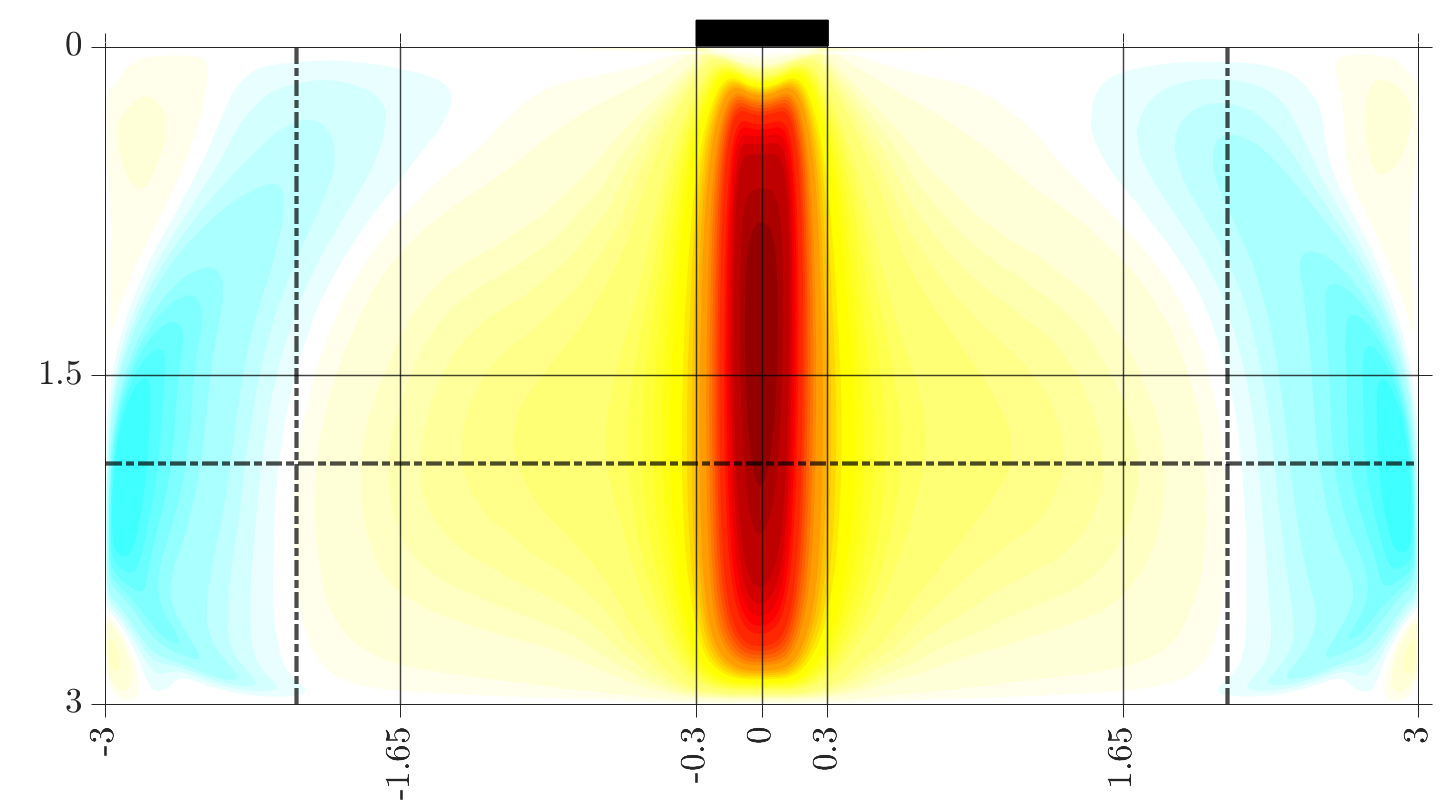} &
\includegraphics[width=4cm]{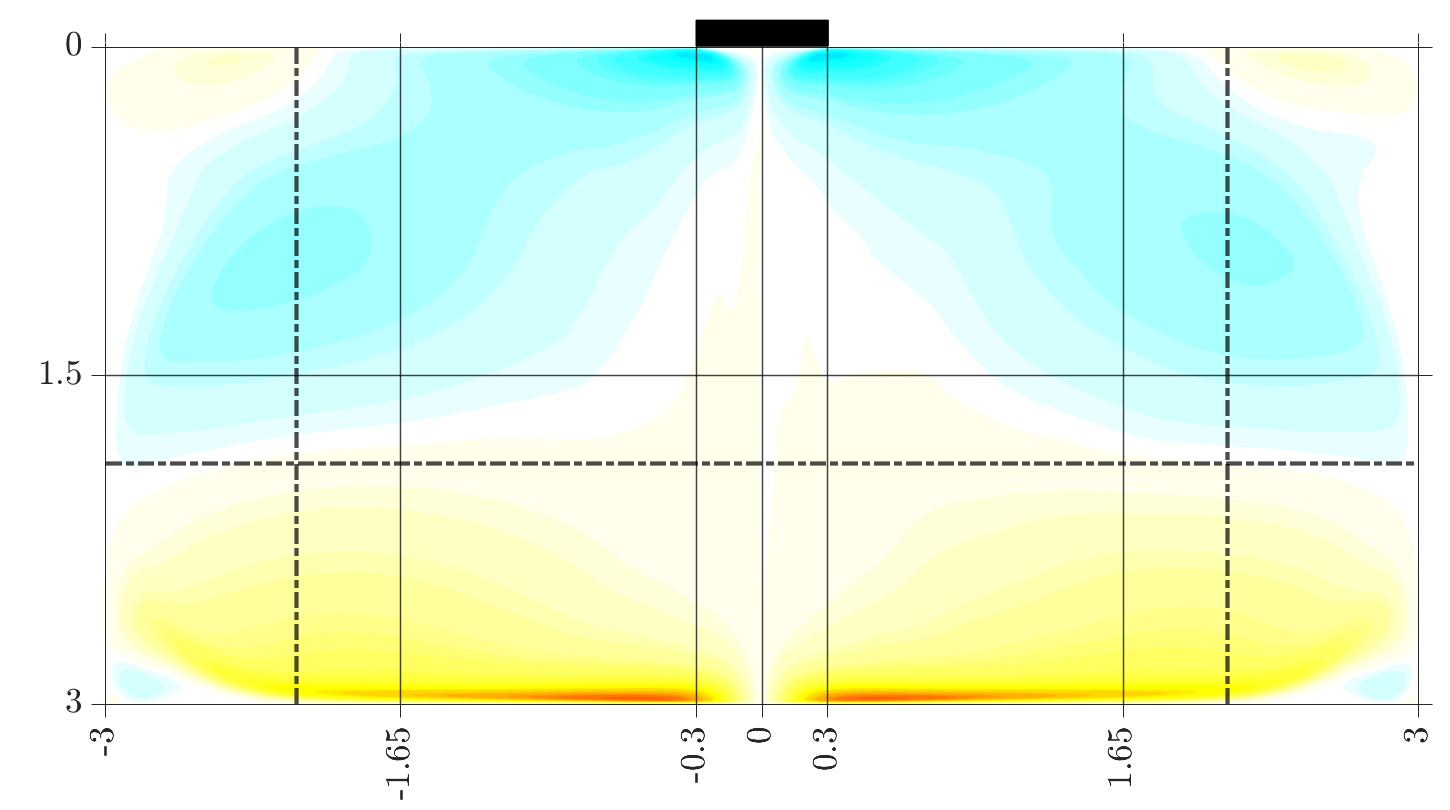} \\
\end{tabular}
\begin{tabular}{cc}
    \centering
\includegraphics[width=3.5cm]{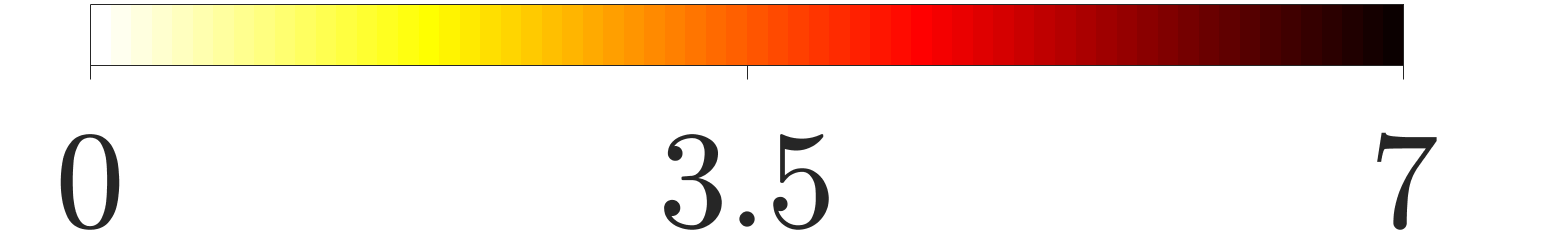} &
\includegraphics[width=3.5cm]{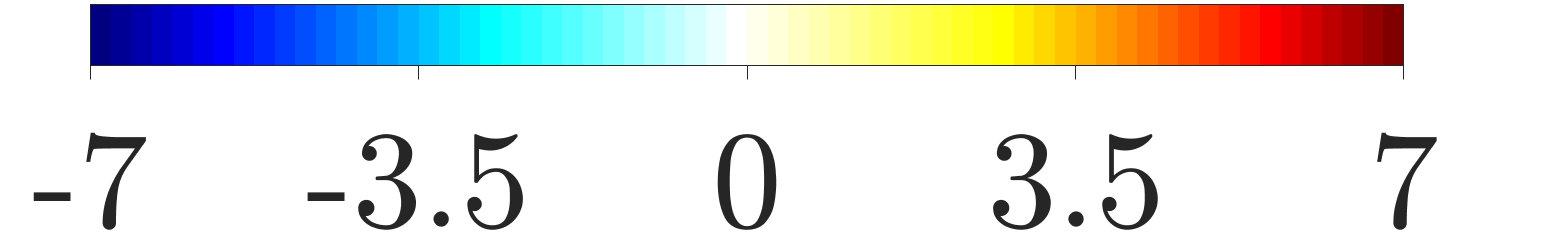} \\
    \end{tabular} \\[2pt]

\begin{tabular}{lccc}
\rotatebox{90}{\hspace{0.5cm} ${K=0.2}$} &
\includegraphics[width=4cm]{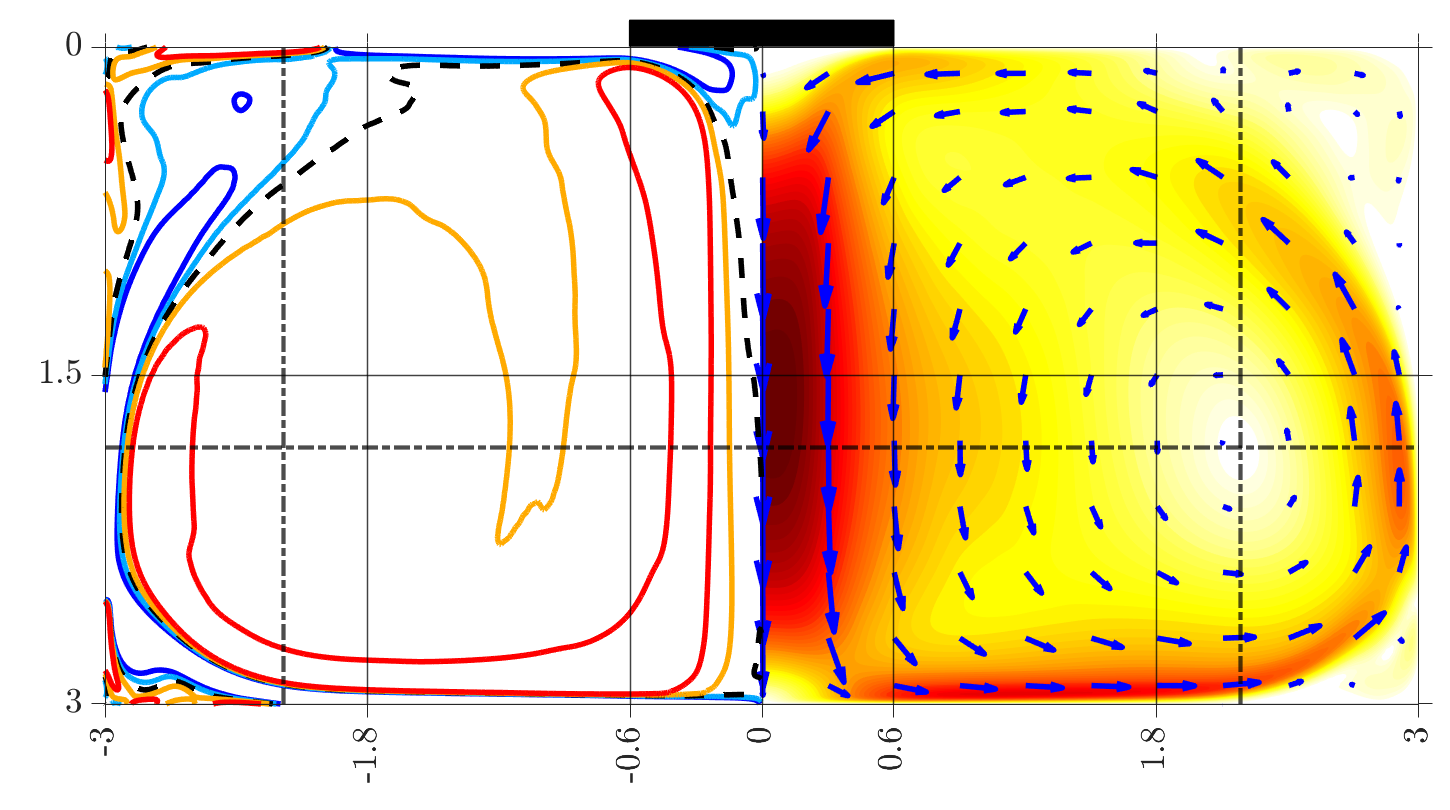} &
\includegraphics[width=4cm]{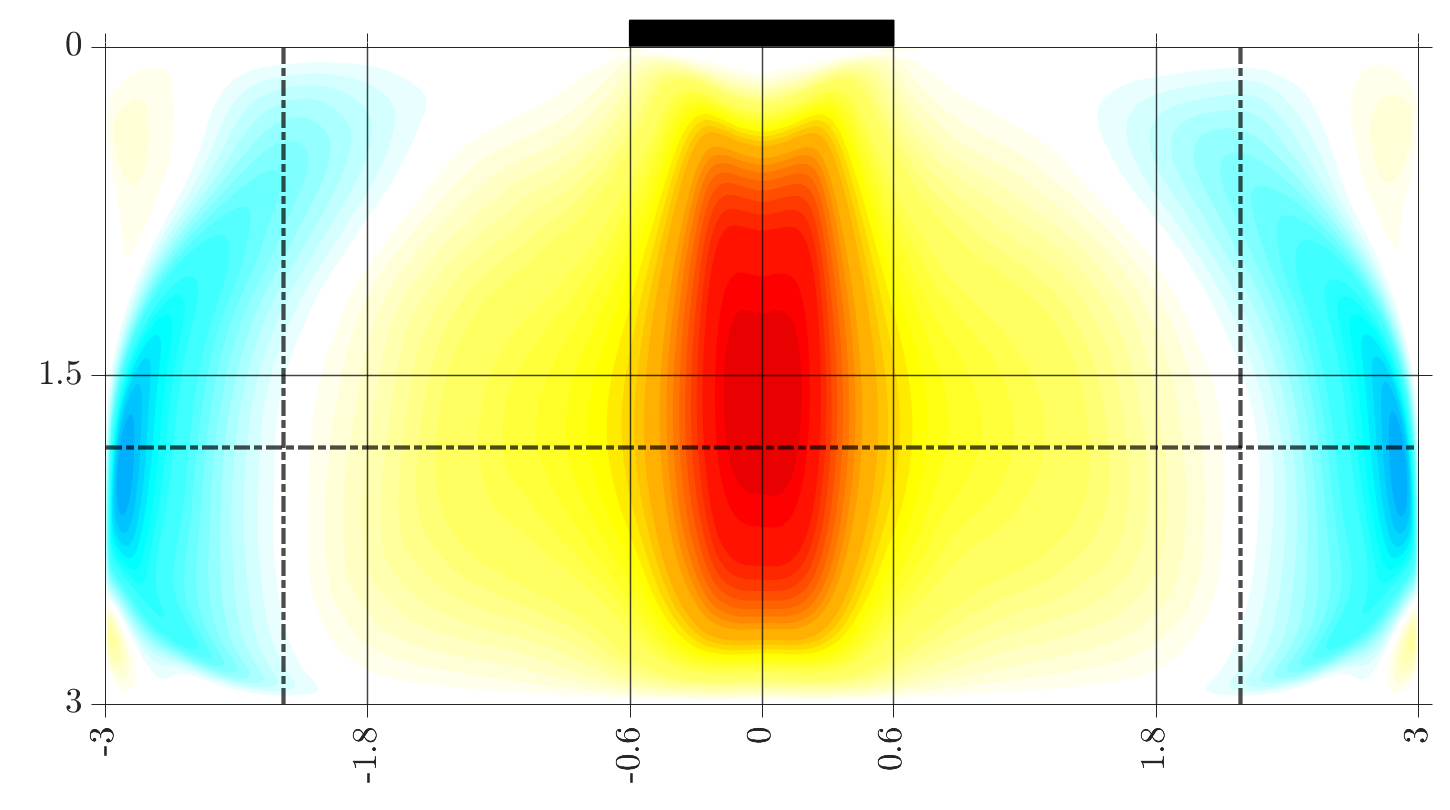} &
\includegraphics[width=4cm]{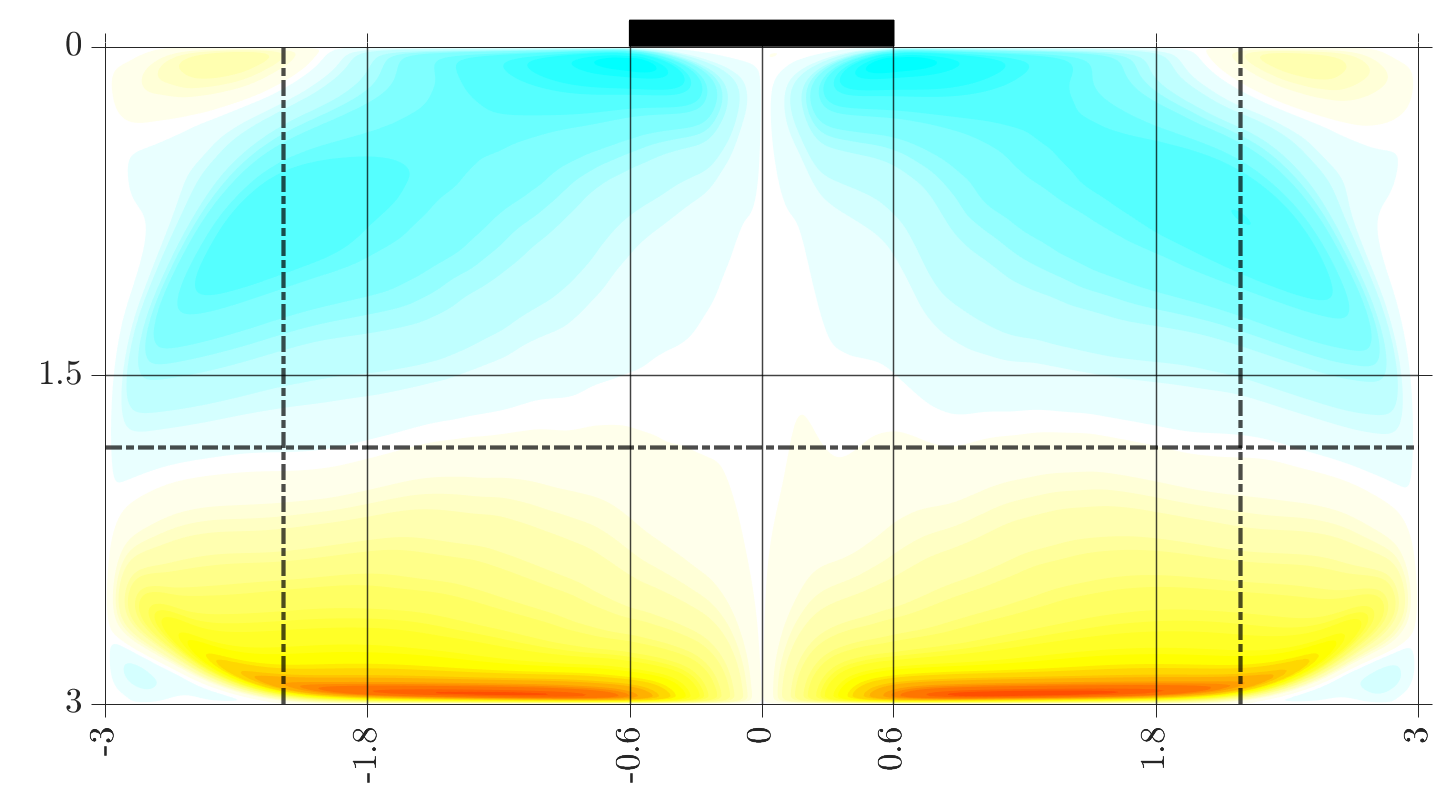} \\
\end{tabular}
\begin{tabular}{cc}
    \centering
\includegraphics[width=3.5cm]{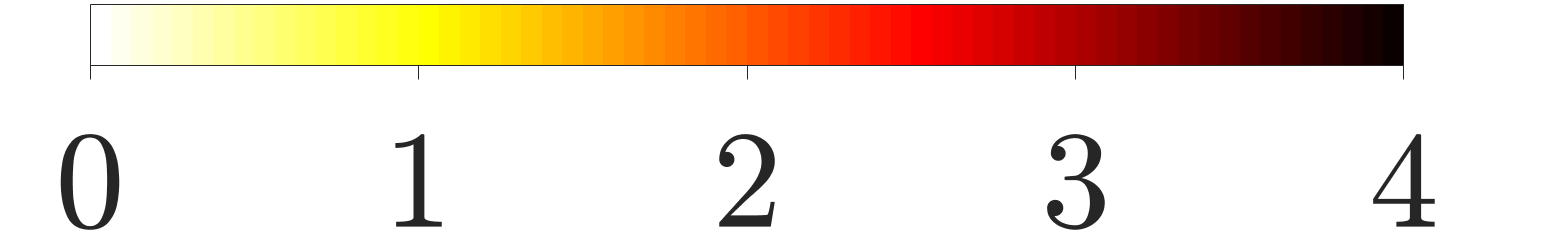} &
\includegraphics[width=3.5cm]{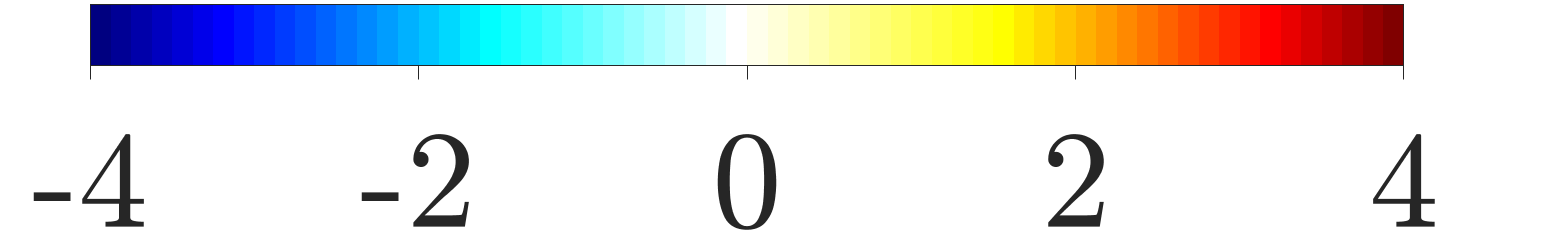} \\
    \end{tabular} \\[2pt]

\begin{tabular}{lccc}
\rotatebox{90}{\hspace{0.4cm} ${K=0.33}$} &
\includegraphics[width=4cm]{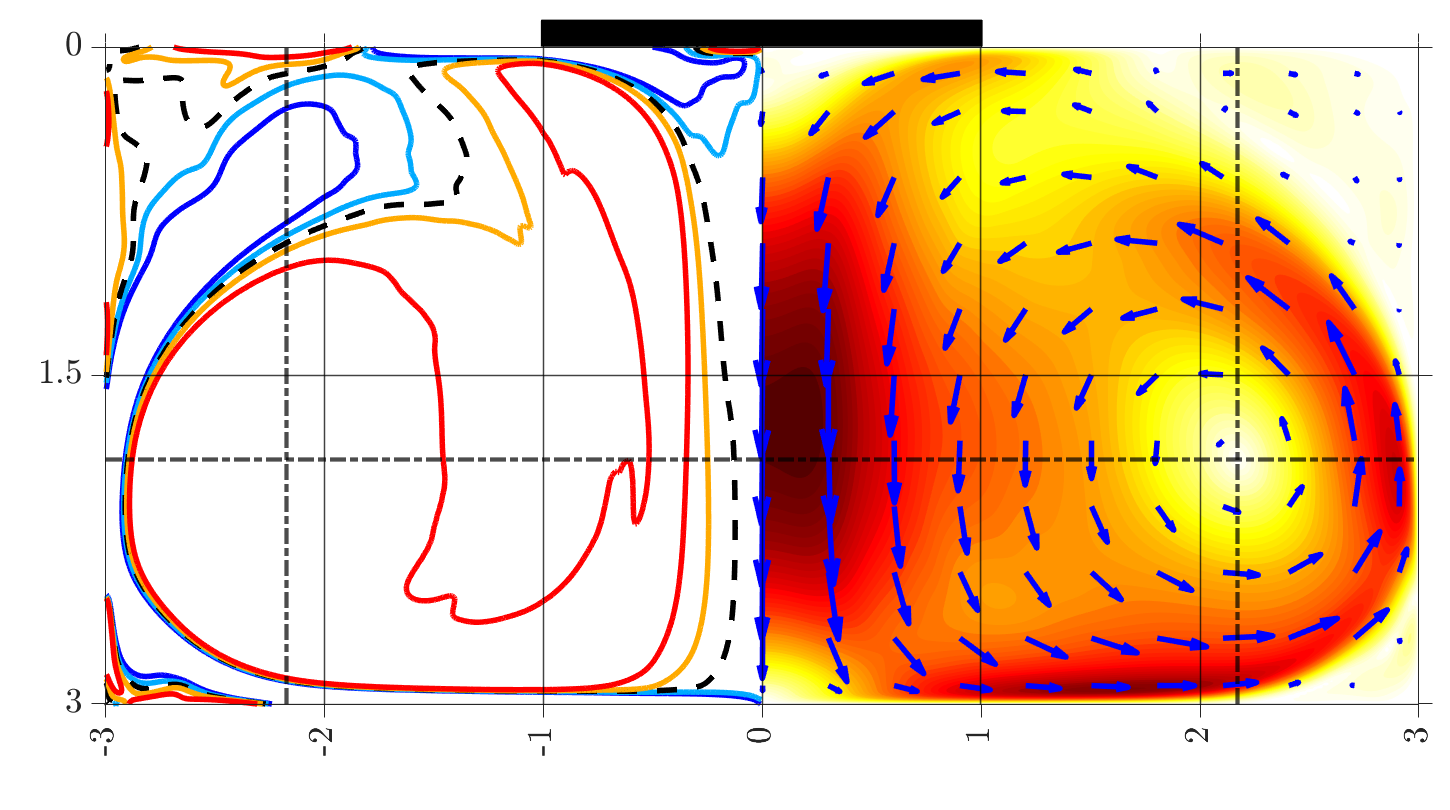} &
\includegraphics[width=4cm]{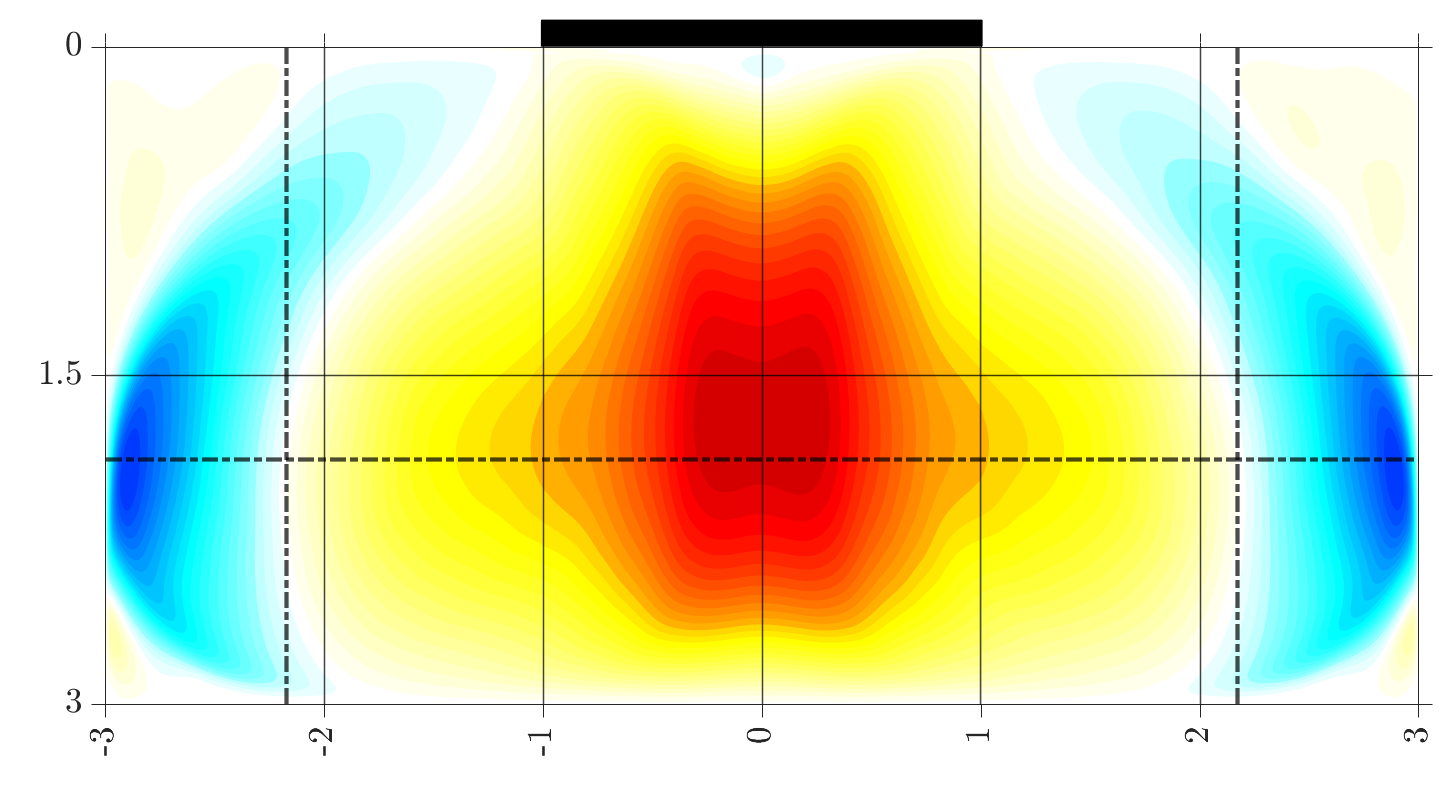} &
\includegraphics[width=4cm]{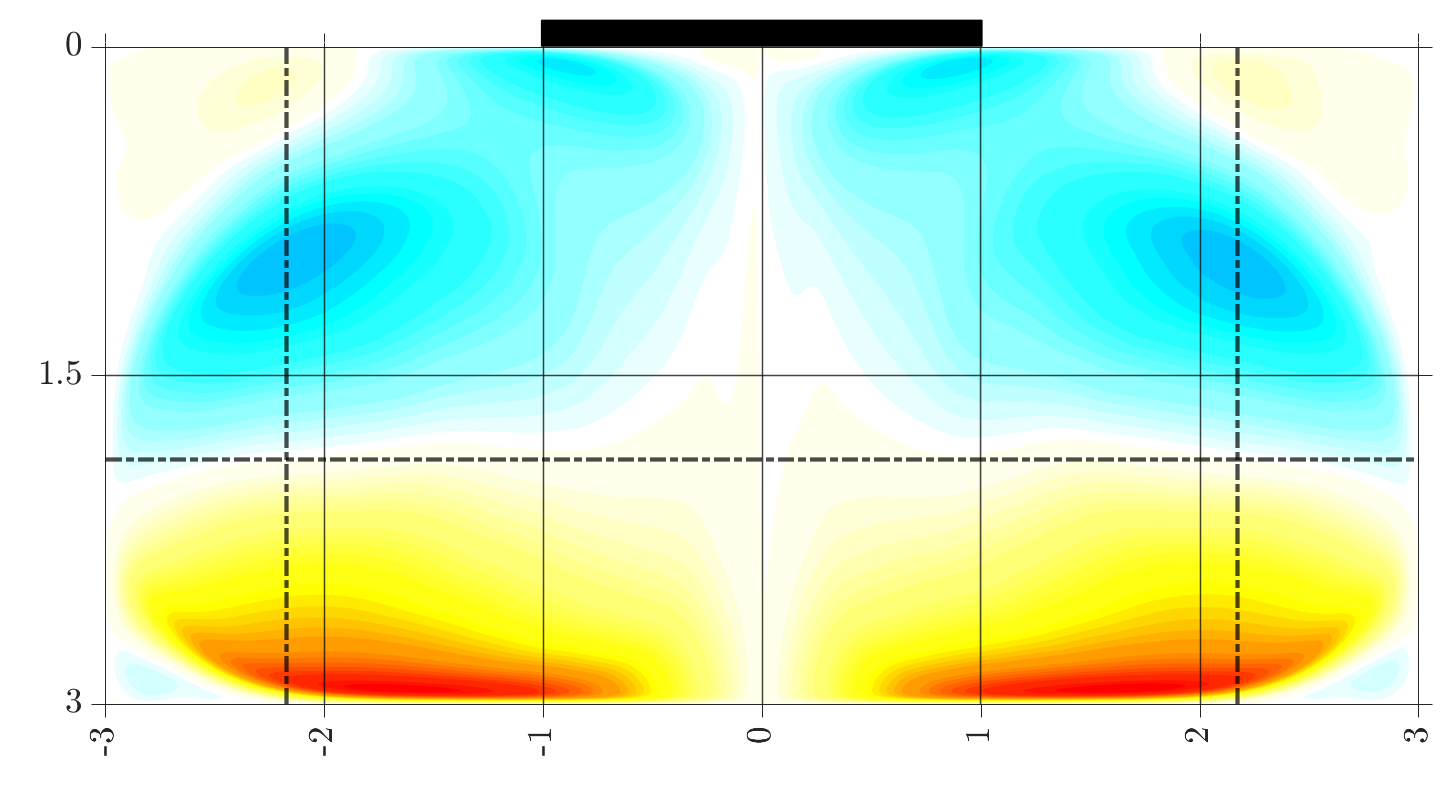} \\
\end{tabular}
\begin{tabular}{cc}
    \centering
\includegraphics[width=3.5cm]{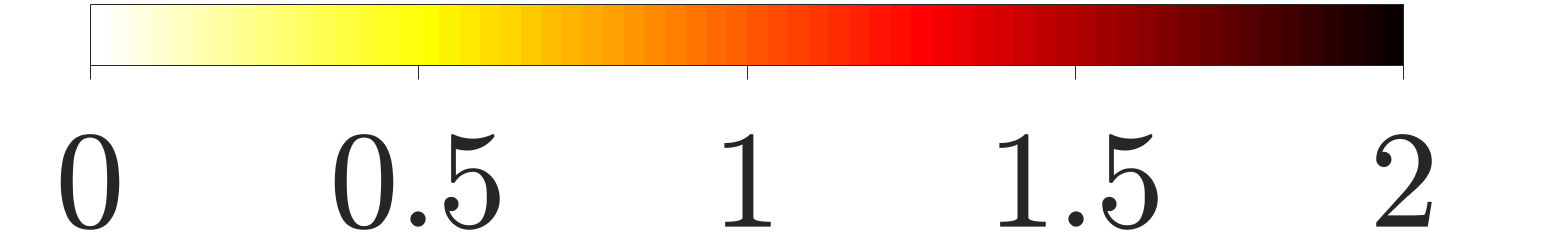} &
\includegraphics[width=3.5cm]{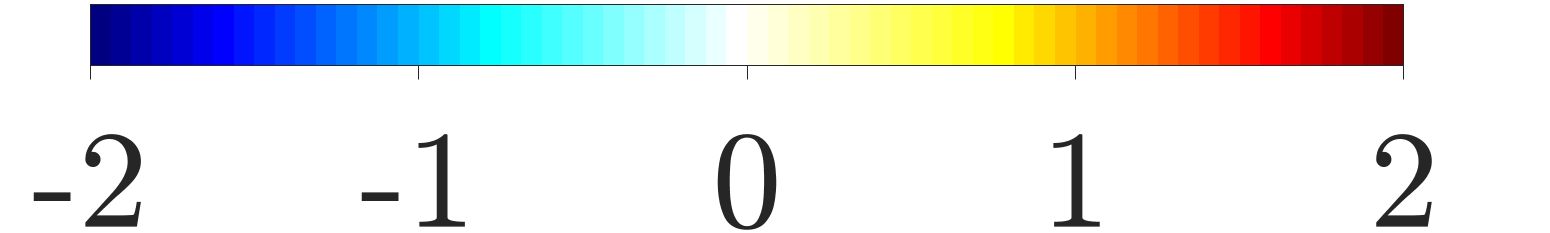} \\
    \end{tabular} \\[2pt]

\begin{tabular}{lccc}
\rotatebox{90}{\hspace{0.5cm} ${K=0.5}$} &
\includegraphics[width=4cm]{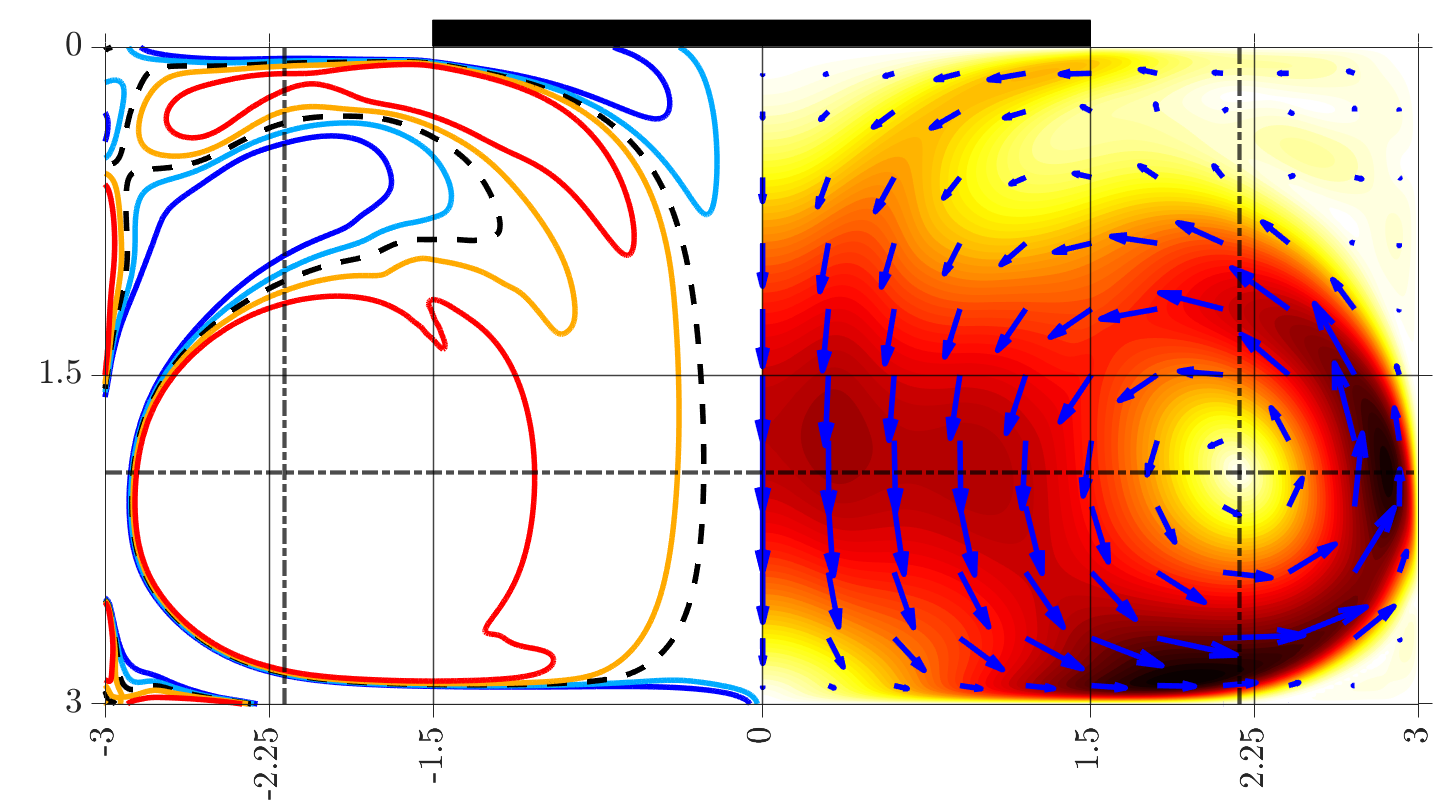} &
\includegraphics[width=4cm]{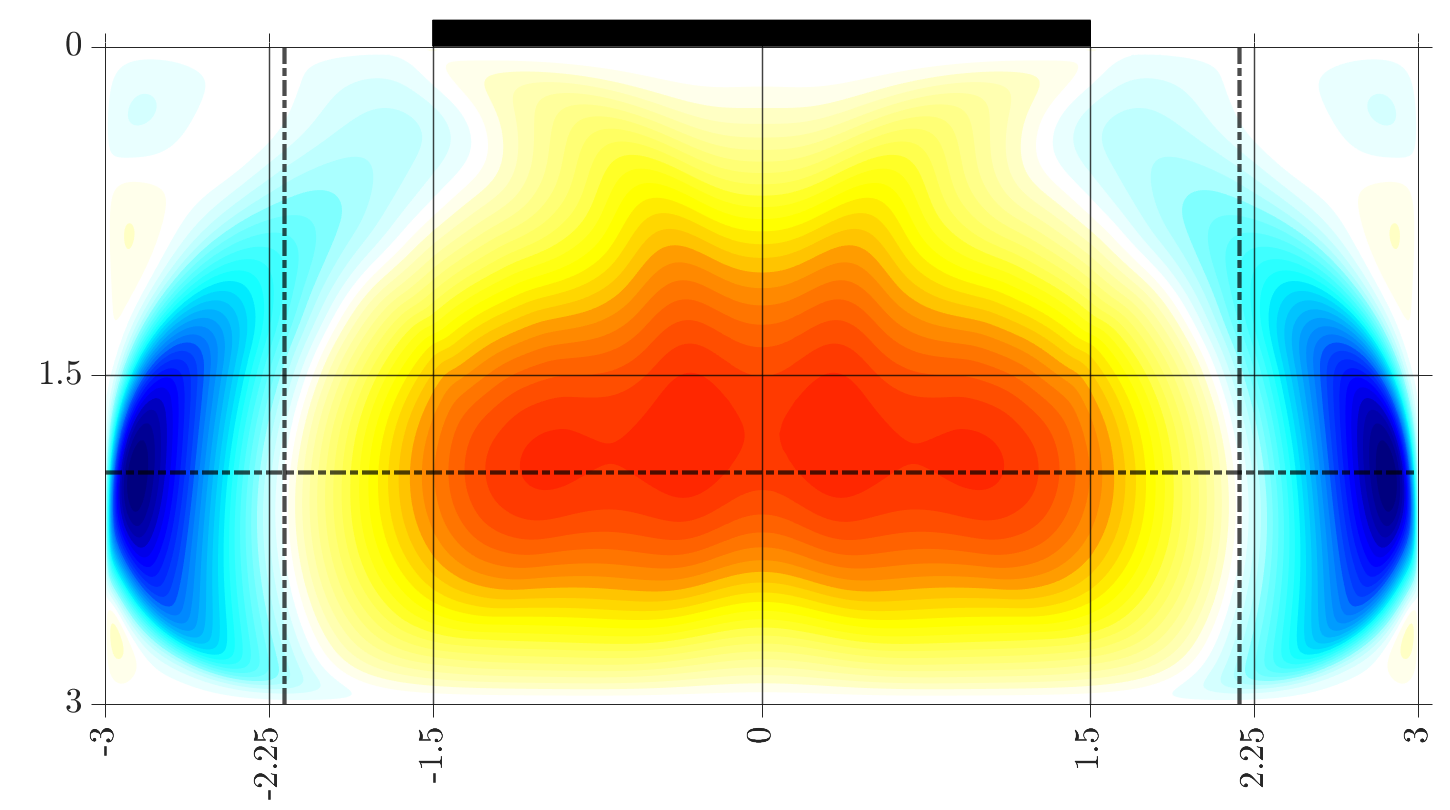} &
\includegraphics[width=4cm]{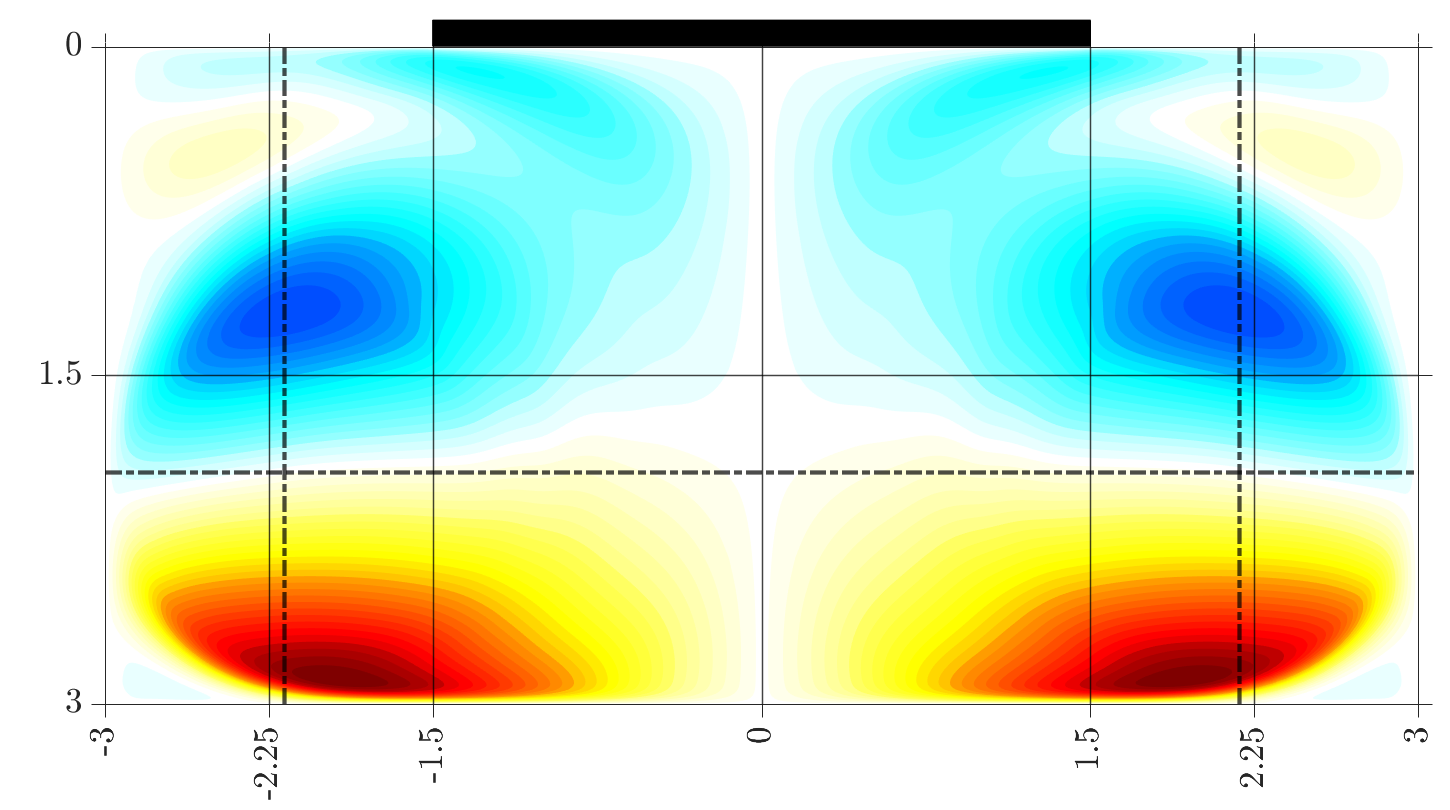} \\
\end{tabular}
\begin{tabular}{cc}
    \centering
\includegraphics[width=3.5cm]{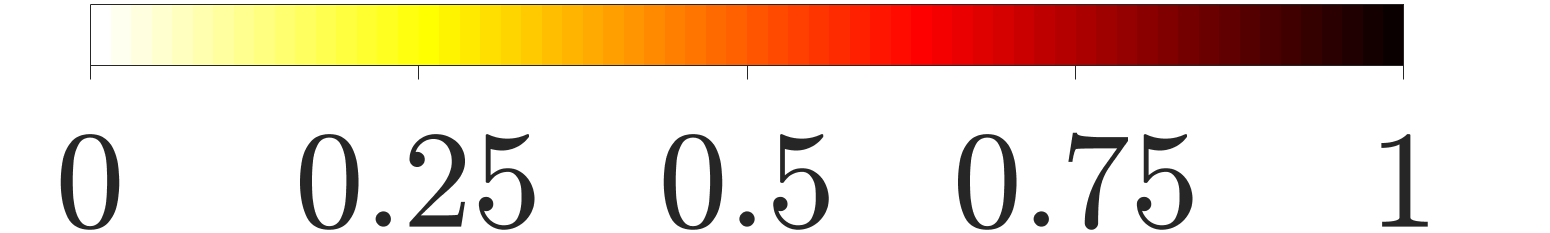} &
\includegraphics[width=3.5cm]{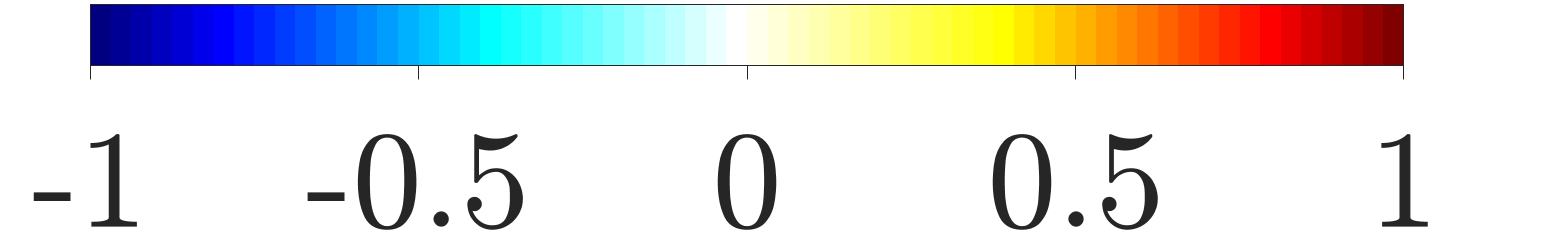} \\
    \end{tabular} \\[2pt]

\begin{tabular}{lccc}
\rotatebox{90}{\hspace{0.5cm} ${K=0.6}$} &
\includegraphics[width=4cm]{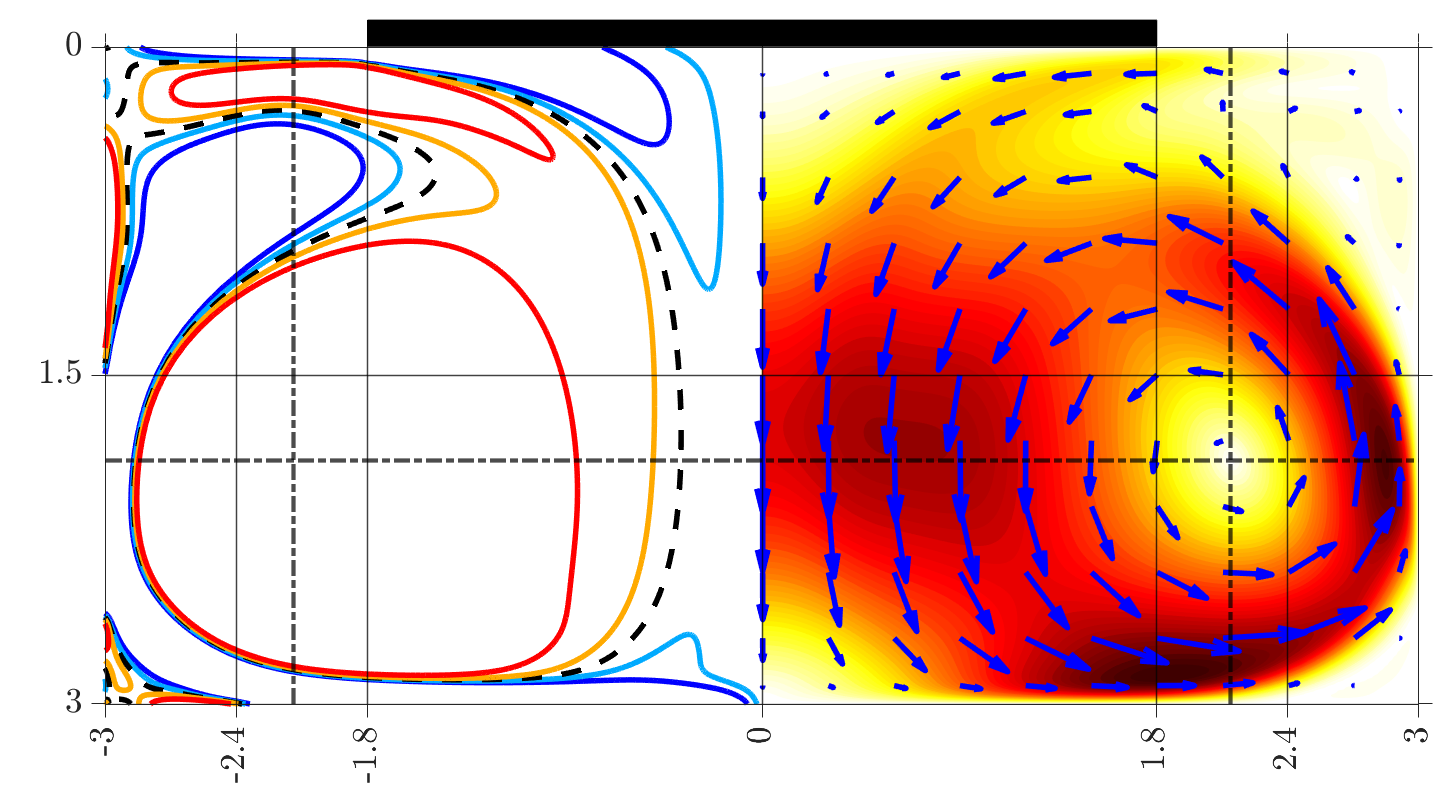} &
\includegraphics[width=4cm]{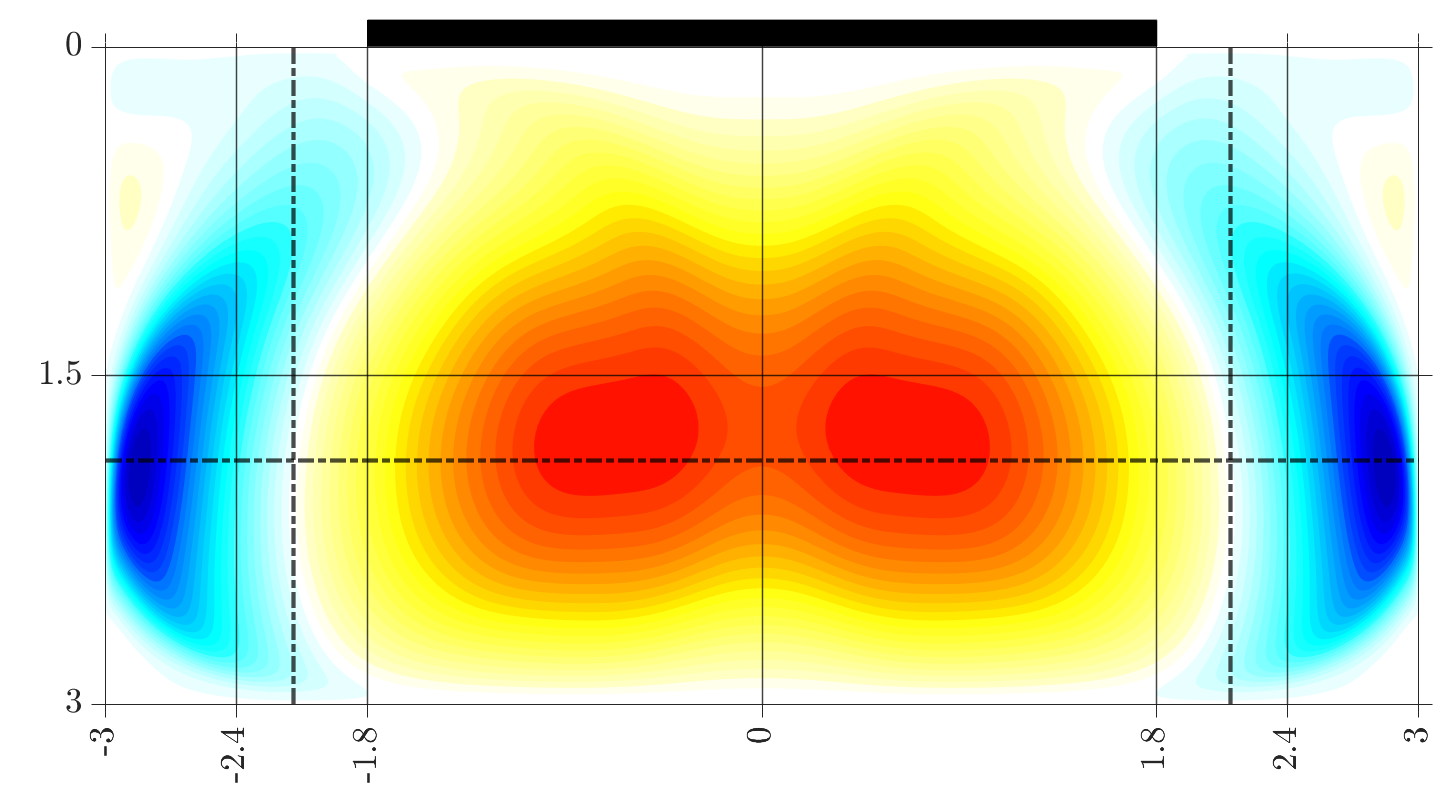} &
\includegraphics[width=4cm]{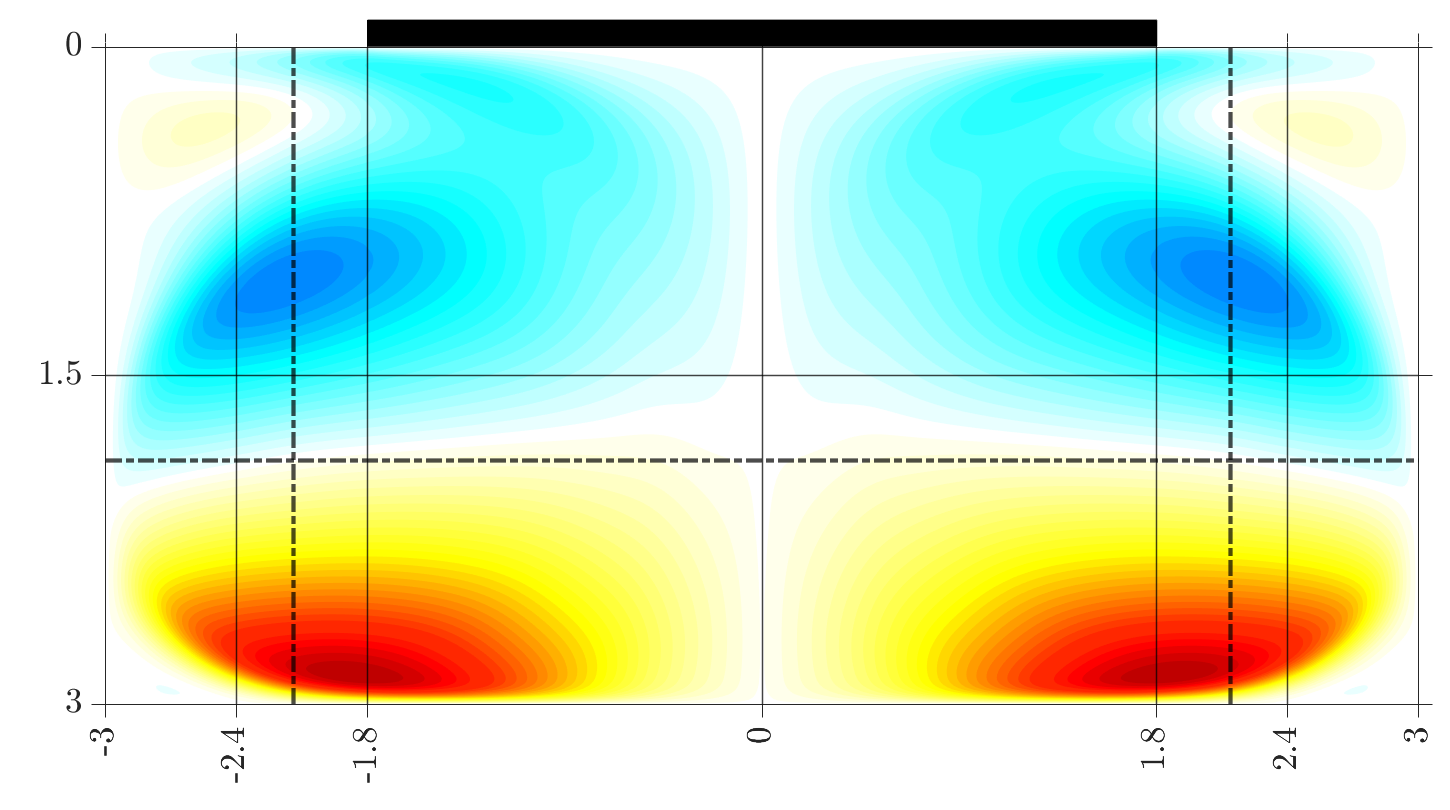} \\
\end{tabular}
\begin{tabular}{cc}
    \centering
\includegraphics[width=3.5cm]{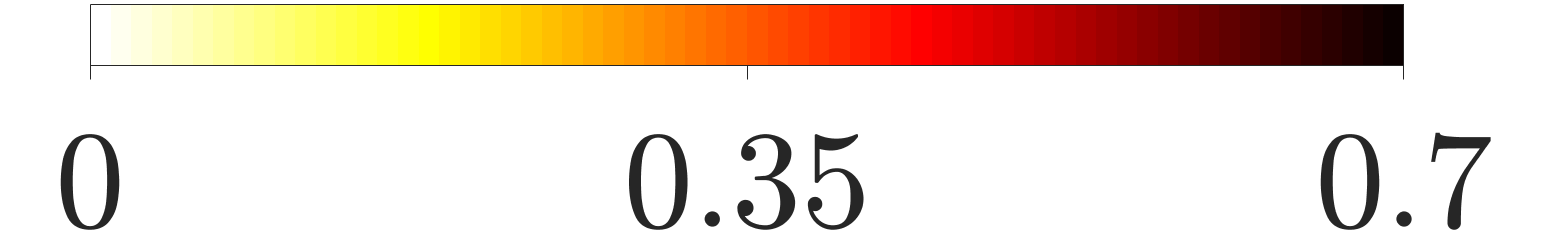} &
\includegraphics[width=3.5cm]{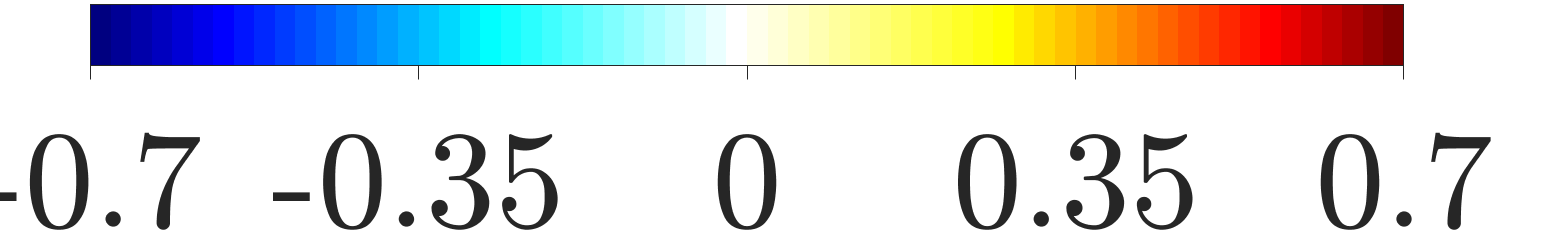} \\
    \end{tabular} \\[2pt]

\begin{tabular}{lccc}
\rotatebox{90}{\hspace{0.4cm} ${K=0.75}$} &
\includegraphics[width=4cm]{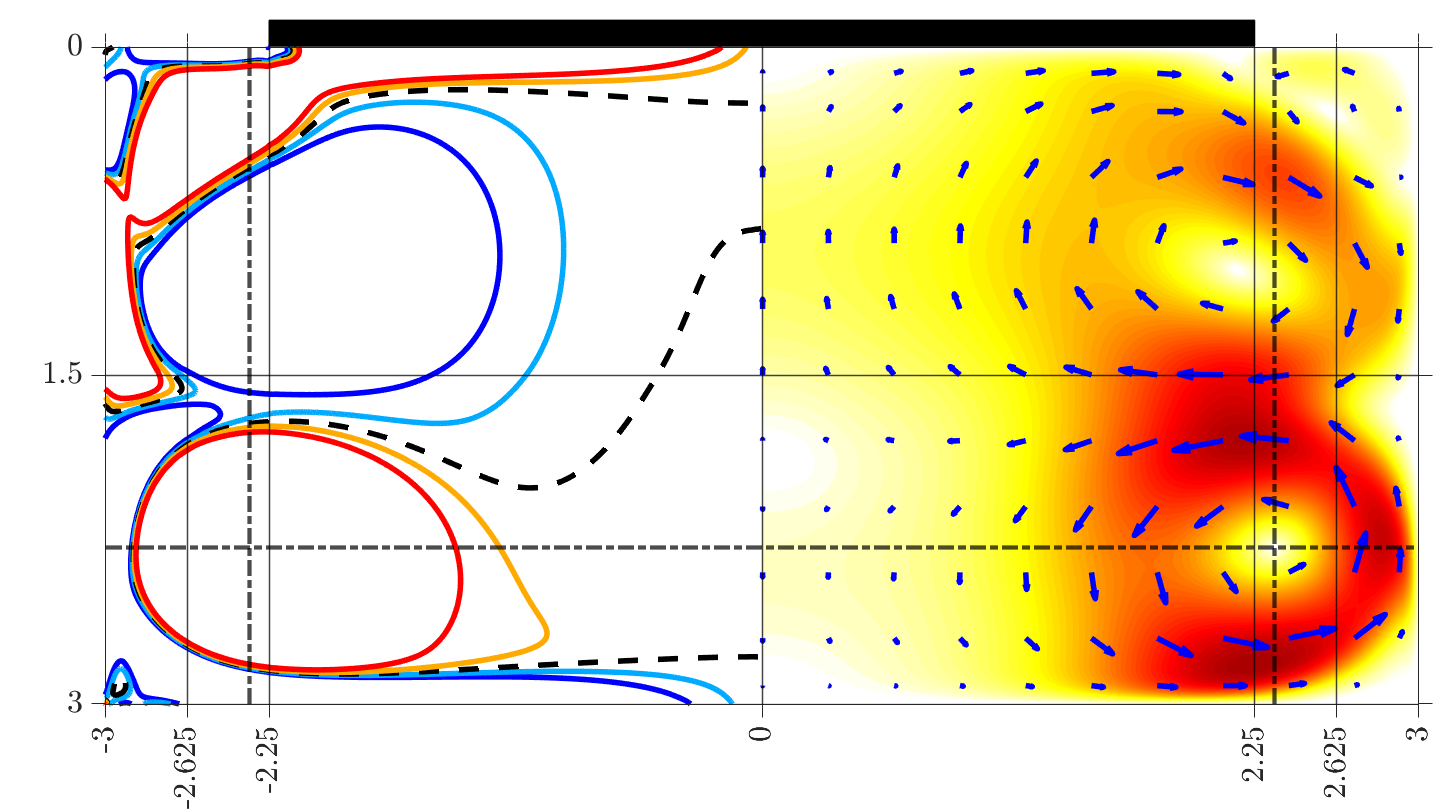} &
\includegraphics[width=4cm]{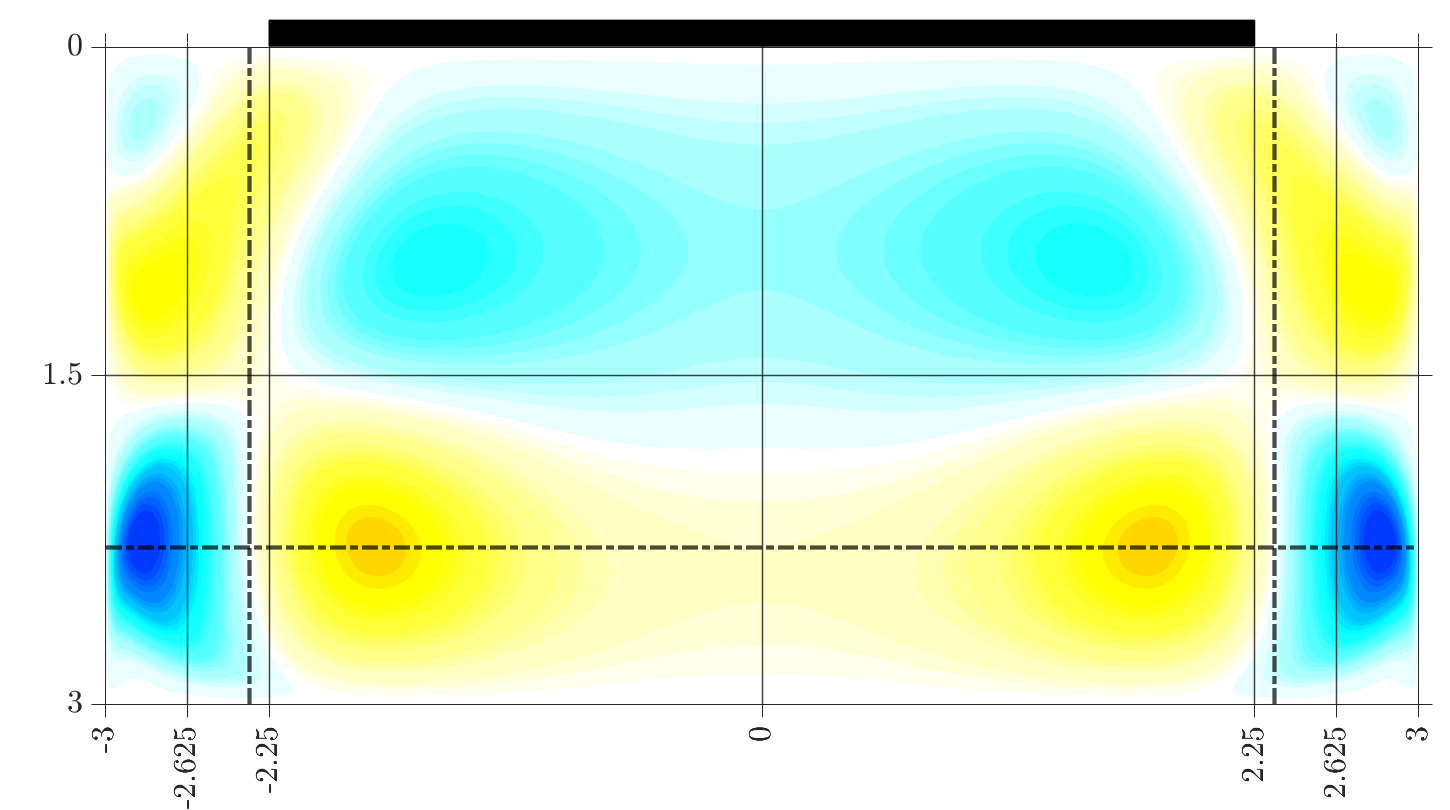} &
\includegraphics[width=4cm]{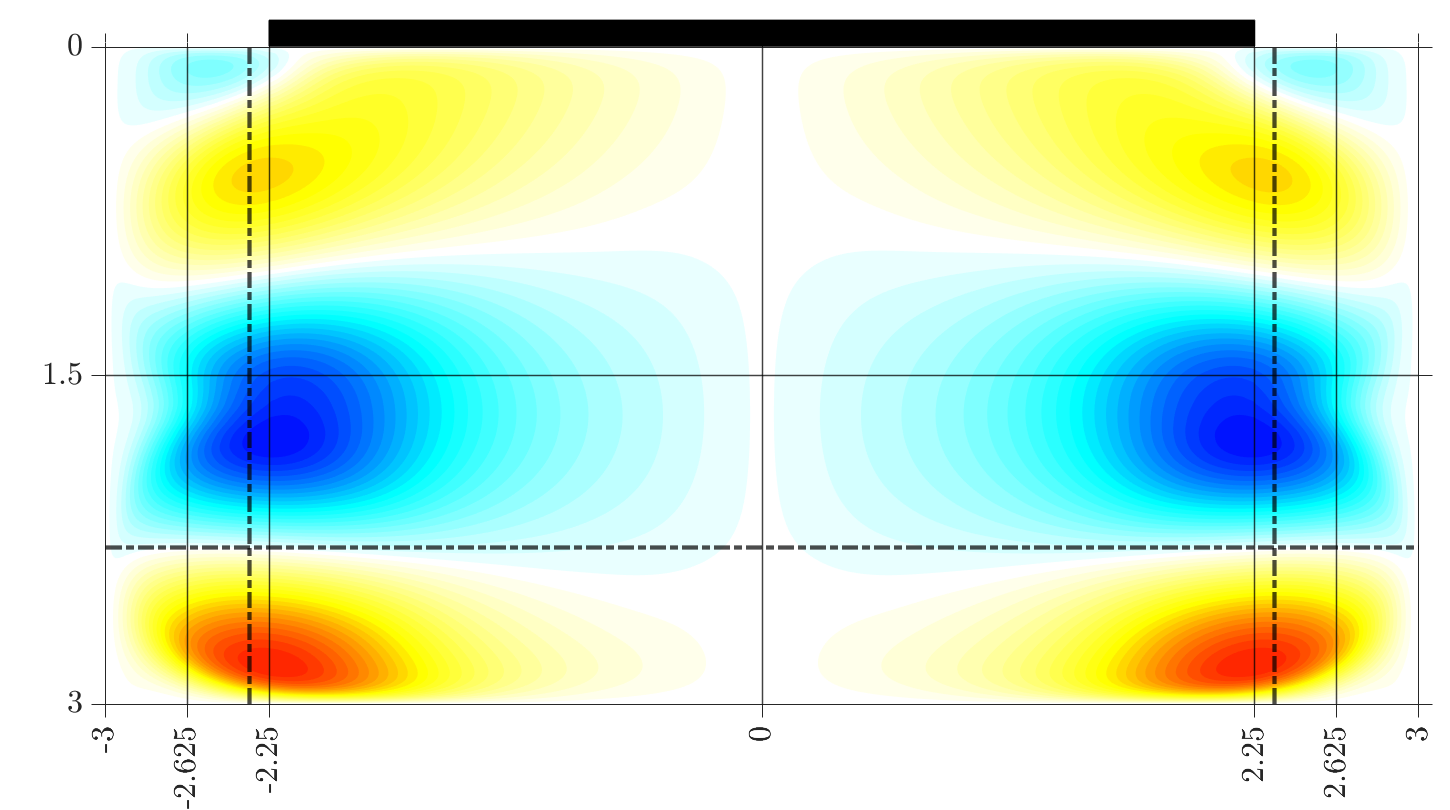} \\
\end{tabular}
\begin{tabular}{cc}
    \centering
\includegraphics[width=3.5cm]{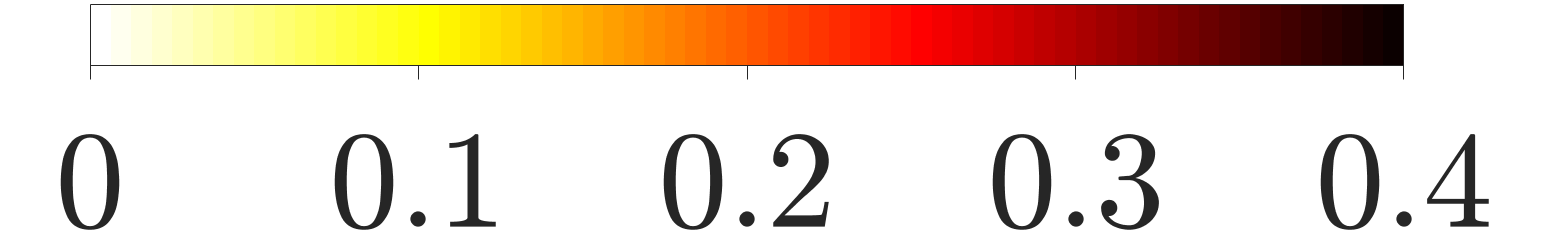} &
\includegraphics[width=3.5cm]{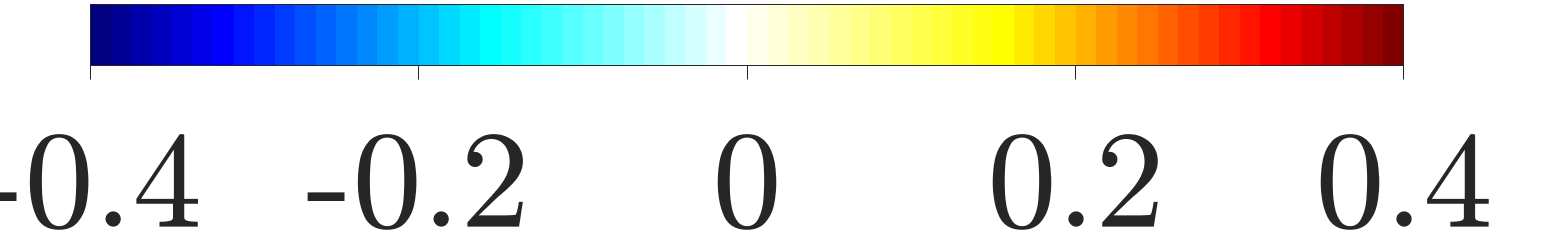} \\
    \end{tabular} \\[2pt]
\caption{(i) Time-averaged vorticity (left) and velocity magnitude (right), (ii) time-averaged axial velocity, and (iii) radial velocity. All the velocities are in \SI{}{\cm/\s}. These contours are shown in the xz-plane extracted from the 3D simulations of EVF for various $K$ values. CC is shown in black. Please note that each case has a different colormap legend.}
\label{fig:contour_plots_K}
\end{figure}

The time-averaged velocity magnitude is plotted in the right half of the \cref{fig:contour_plots_K}(i). Please refer to \autoref{table:K_max_A1} for the maximum values. An intense EVF jet forms for smaller $K$ cases, resulting in a larger axial velocity as compared to that with larger $K$. With an increase in the CC radius, reduction in the current density (for the same total current) leads to a lower jet intensity and also to a lateral broadening of the jet. This results in a relatively higher vertical velocity for the return flow near the sidewall. For $K=0.5$ to $K=0.7$, the radial velocity near the bottom boundary is greater than the jet velocity. Moreover, the jet begins to split as $K$ increases (also see \cref{fig:contour_plots_K}(ii) \& (iii) for the velocity component plots). In the standard EVF, mean flow is characterized by a downward axial jet and an upward return flow near the sidewall resulting in an axisymmetric toroidal vortex of size $\sim R$. This is observed in the cases with $K<0.75$.

To further investigate the change in EVF pattern with $K$, we show the scaled vorticity, $\omega_{\theta}$, in the left half of \cref{fig:contour_plots_K}(i). With an increase in $K$, the negative vorticity generated at the wall near the CC region is advected towards the axis. Thus, the zero-vorticity line shifts away from the axis, signifying a shift in the peak velocity position. This was also observed by \citet{vlasyuk1987effects} in their numerical study. Due to this shift, the positive vorticity gets diffused radially, splitting the (positive) vorticity column into two parts. These changes explain the bifurcation of the axial jet. (The flow and vorticity evolution movie can be found in the supplementary material.) Furthermore, the vorticity remains nearly constant within the (toroidal) vortex region. This is expected according to the Prandtl-Batchelor theorem, which states that for high $\Rey$ flows with closed streamlines, the vorticity should be uniform in the bulk region \citep{batchelor1956steady,davidson2017MHD}. Due to these reasons and the confinement effects, both the terms in $\omega_{\theta}$ have almost equal contributions, i.e., $\partial u_r/ \partial z \approxeq \partial u_z/ \partial r$, which we had assumed in our velocity estimate in \S \ref{derivation}.

\begin{figure}  \centerline{\includegraphics[width=0.6\linewidth]{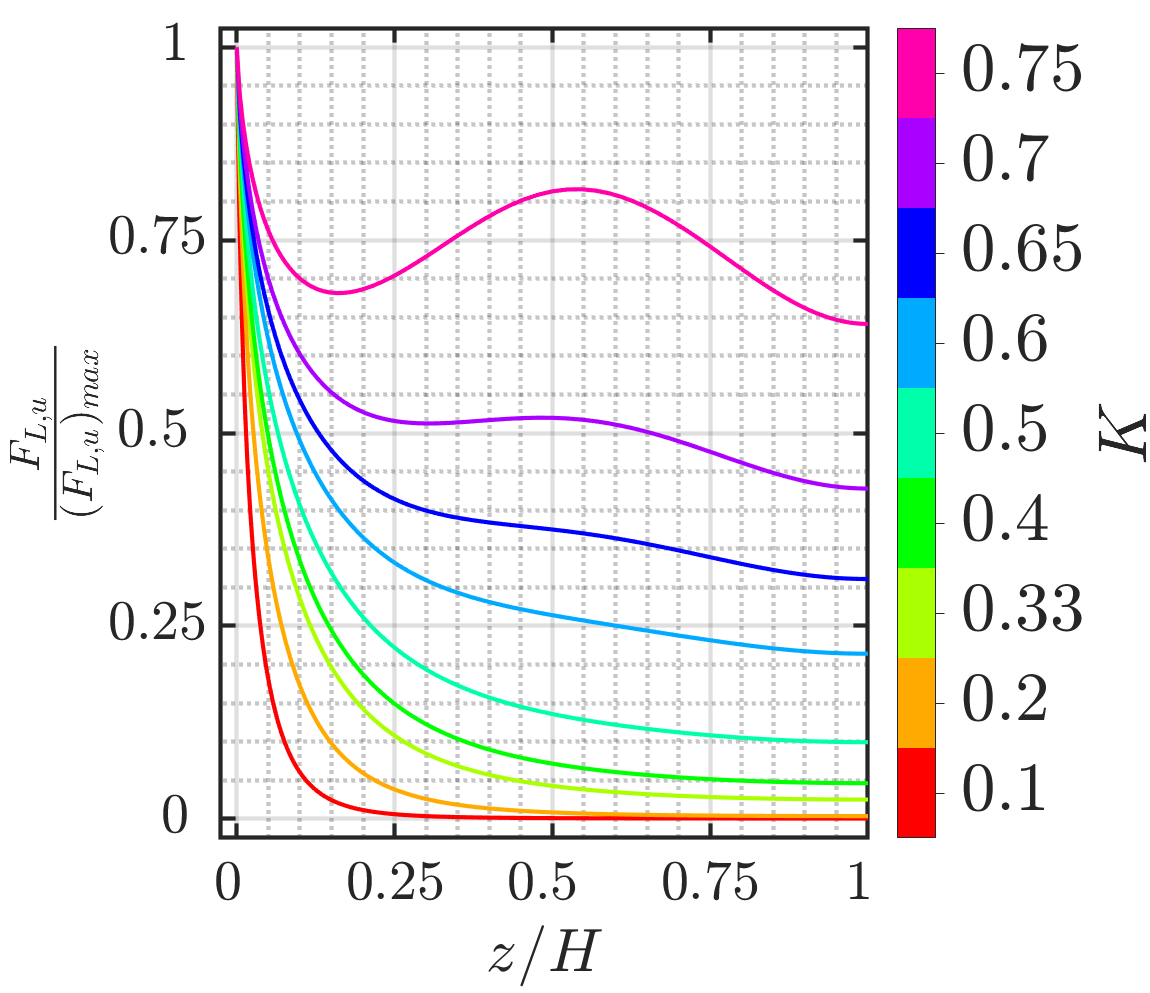}}
  \caption{The $\boldsymbol{F}_{L,u}$ (normalized with its maximum value) along a vertical line, located at the CC corner, for different $K$.}
\label{fig:unb_line}
\end{figure}

\begin{figure}  \centerline{\includegraphics[width=0.48\linewidth]{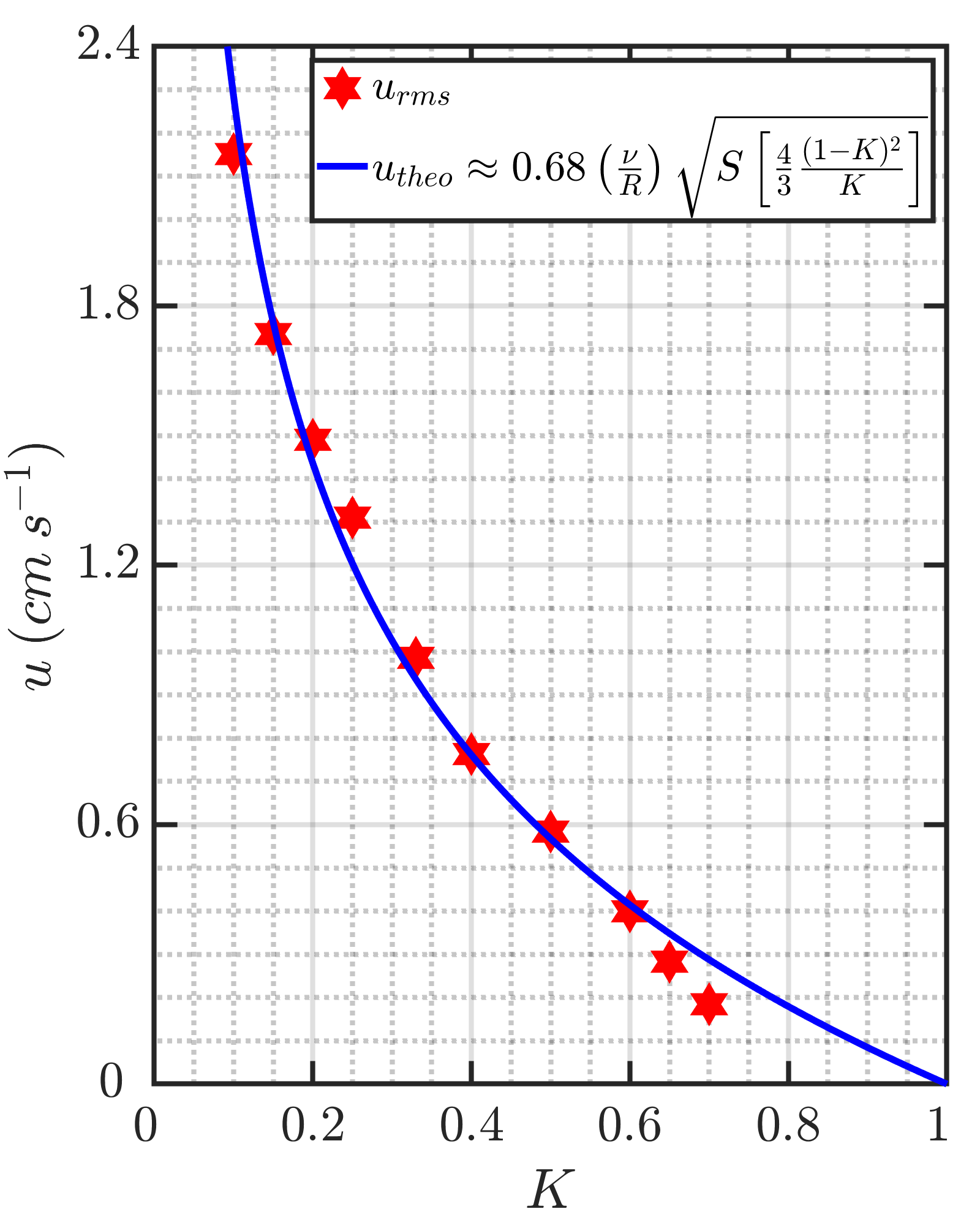} 
\includegraphics[width=0.48\linewidth]{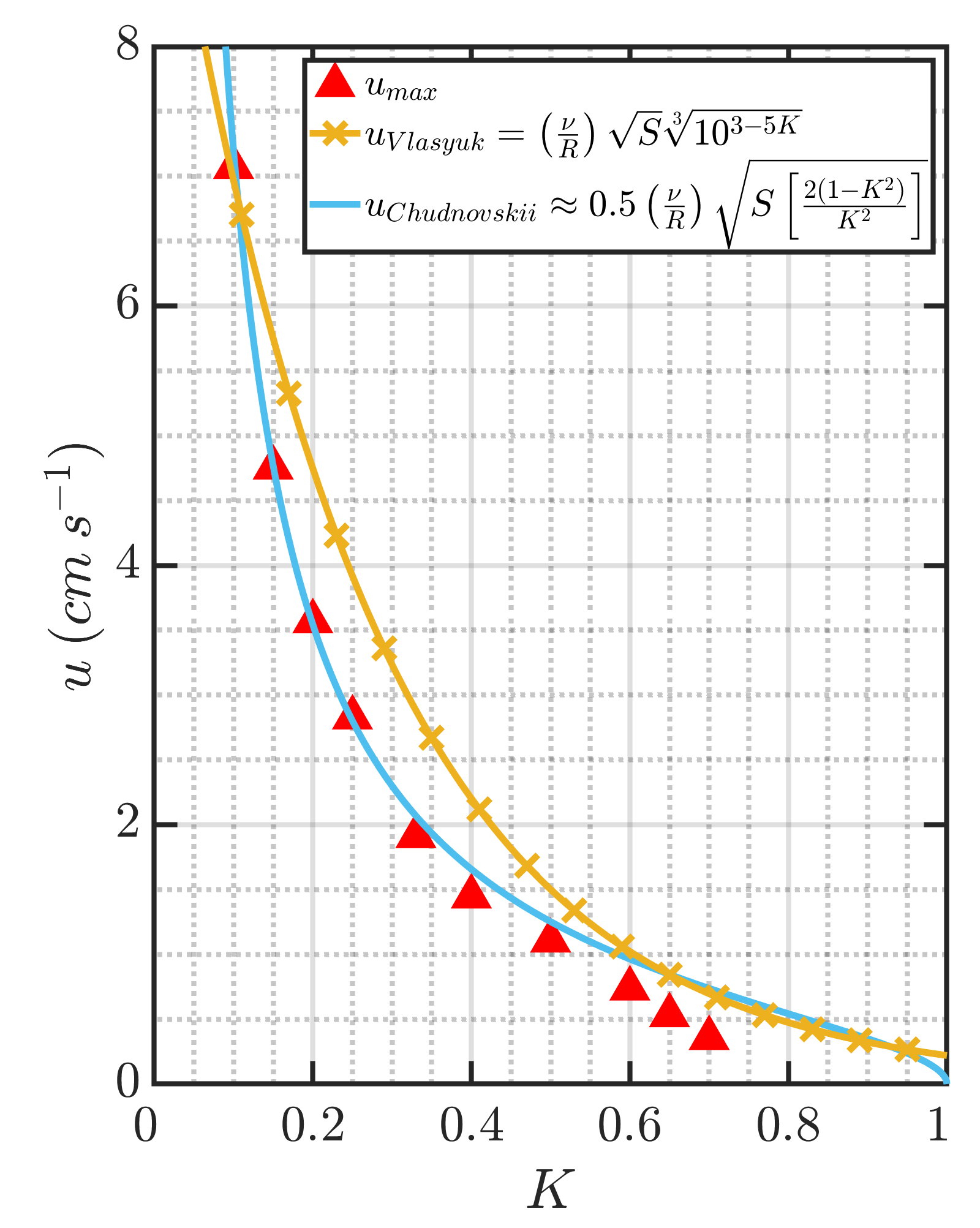}}
\centerline{(a) \hspace{175pt} (b)}
  \caption{The time-averaged r.m.s. velocity (a), and the maximum velocity (b) obtained from our 3D simulations are plotted as a function of $K$. These are compared with our present theoretical estimate and the estimates from the earlier literature \citep{vlasyuk1987effects,chudnovskii1989evaluating} respectively.}
\label{fig:u_theo_K}
\end{figure}

\begin{figure}  \centerline{\includegraphics[width=0.48\linewidth]{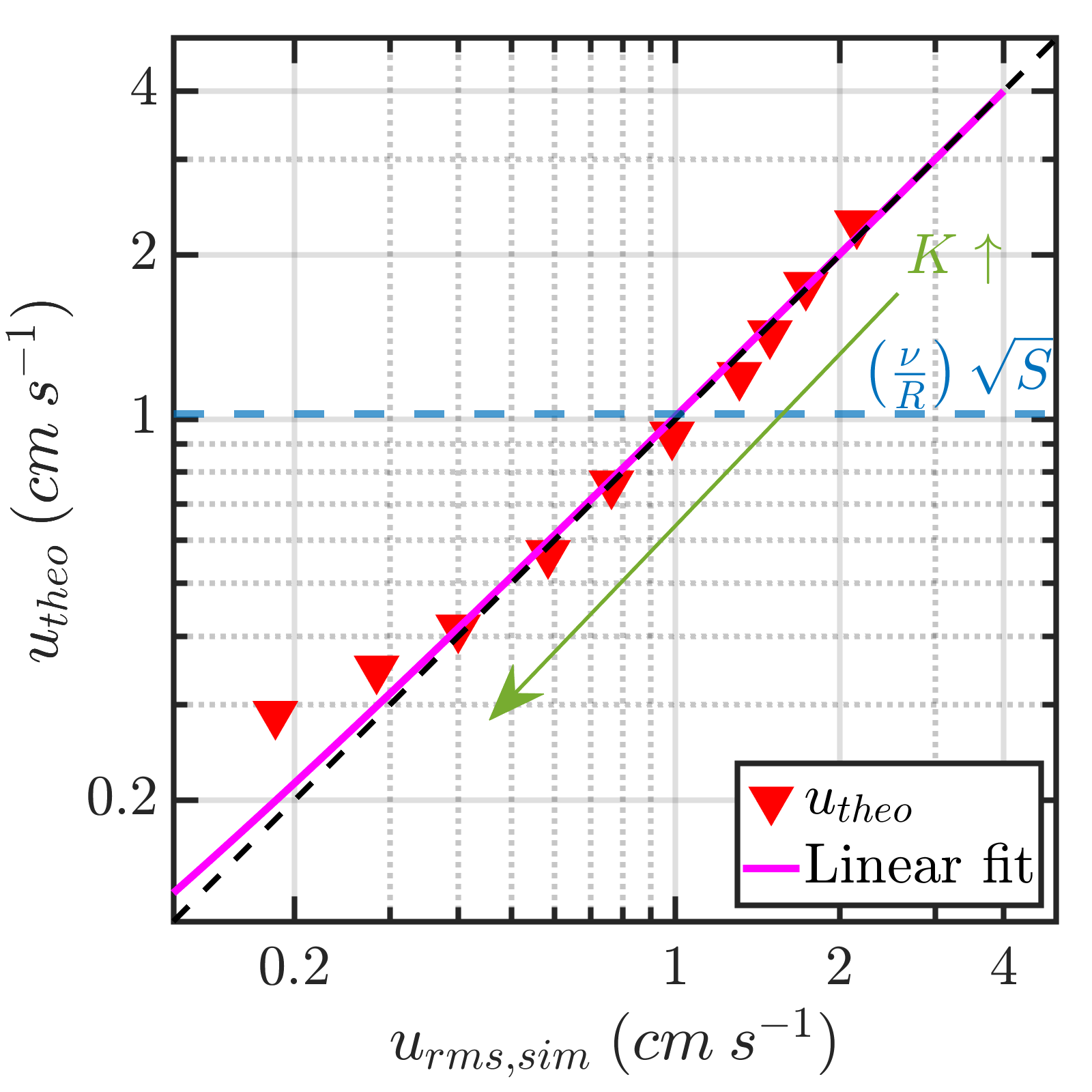} 
\includegraphics[width=0.48\linewidth]{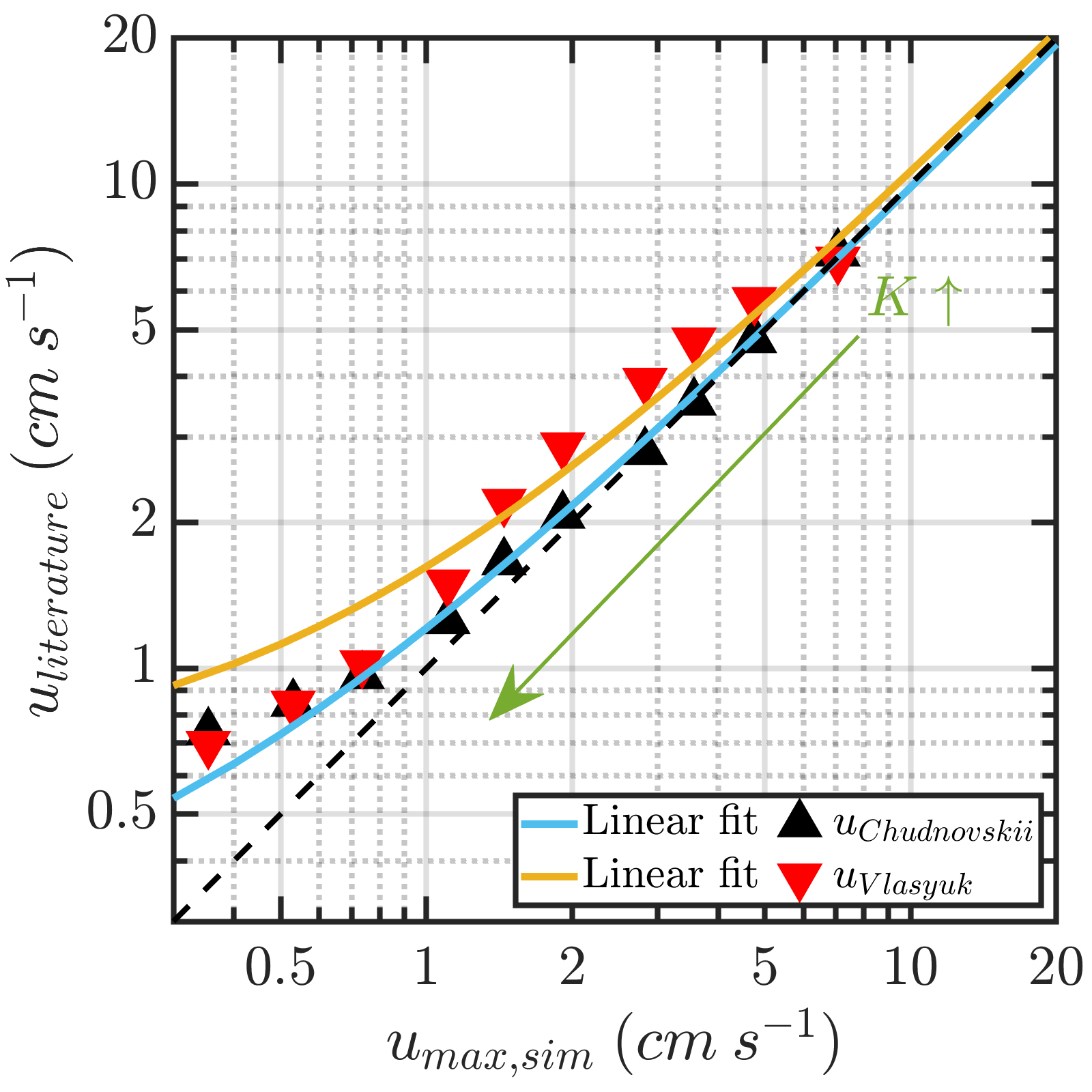}}
\centerline{(a) \hspace{175pt} (b)}
  \caption{(a) The $u_{theo}$ from our estimate is plotted against the $u_{rms}$ from our 3D simulations for varying $K$. They follow the linear trend, indicating the good predictive capability of our theoretical estimate. The dashed blue horizontal line shows the constant location for the $\nu\sqrt{S}/R$. (b) Velocity estimates from the literature \citep{vlasyuk1987effects,chudnovskii1989evaluating} are plotted against the $u_{max}$ from our simulations.}
\label{fig:u_theo_rms_K}
\end{figure}

Since $\boldsymbol{F}_{L,u} < 0$, the EVF is radially inward near the CC corners. (We expect the peak in $\boldsymbol{F}_{L,u}$ to be located just outside the CC corners.) This is the case for all CC radii up to $K=0.7$, though there is a difference in the EVF intensity and jet characteristics with the change in the CC radius thereafter, as discussed above. For $K\geq0.75$, we notice an unexpected flow near the CC corners. The flow is radially \emph{outward} near the current injection region. The location of the peak $\boldsymbol{F}_{L,u}$ in the domain is near the sidewall for these cases; however, it is negative. The peak in $\boldsymbol{F}_{L,u}$ near the sidewall is due to the local maximum in $B_{\theta}$. Such a magnetic field distribution is due to the uniform current application in the \emph{finite-size domain} (refer Biot-Savart law \autoref{BS_law}). This leads to a dominant radially inward flow at the mid-domain height and \emph{outward} flow near the CC region. This flow suppresses the initial (standard) EVF due to the diverging current lines. We find that the effects of uniform current and domain finiteness dominate at $K\geq 0.75$. The effect of the change in the nature of $\boldsymbol{F}_{L,u}$ distribution with the change in $K$ is also clearly evident in \cref{fig:unb_line}.

\subsubsection{An exact estimate of the average EVF velocity}   \label{stat_analysis_K}
We now compute the time-averaged root mean square (r.m.s.) and the maximum velocity in the volume
\begin{subeqnarray}
    u_{rms} & = & \dfrac{1}{\Delta t} \int_{t_1}^{t_2} \sqrt{\dfrac{1}{V_{\text{bulk}}}\int_{V_{\text{bulk}}} \left| \boldsymbol{u} \right|^2 \: \mathrm{d}V} \: \mathrm{d} t,     \label{average_u}\\[5pt]
    u_{max} & = & \dfrac{1}{\Delta t} \int_{t_1}^{t_2} \underset{\mathrm{x} \in V_{\text{bulk}}}{\text{max}} \left( \left| \boldsymbol{u} \right| \right) \mathrm{d} t,     \label{max_u}    
\end{subeqnarray}
where $\left| \boldsymbol{u} \right| = \sqrt{u_{r}^2 + u_{\theta}^2 + u_{z}^2}$ and $V_{\text{bulk}}$ is the volume of the bulk region, which is obtained by excluding the boundary region from the entire domain. From these, we can define the r.m.s. and the maximum Reynolds numbers, respectively, as
\refstepcounter{equation}
$$
    \Rey_{rms} = \dfrac{u_{rms}R}{\nu},  \quad
    \Rey_{max} = \dfrac{u_{max}R}{\nu},  
    \eqno{(\theequation{\mathit{a},\mathit{b}})}
    \label{Re_rms_max}
$$
based on the domain radius, $R$. Here we take $21\leq t \leq 50$ for $K\leq 0.75$, and $71\leq t \leq 100$ for $K=0.95$ since the flow develops very late in the latter case. Even though the EVF is mainly poloidal under steady-state conditions (the flow is axisymmetric and should have only $r,z$ components), we also consider the azimuthal velocity ($u_{\theta}$) in our calculations. As mentioned earlier, the flow becomes unstable at very low $K$ or high currents and thus starts to rotate. (This is very different from the \emph{swirl} flow, which can arise due to the interaction of the diverging/converging current with the \emph{axial} magnetic field). The continuity of the flow has to be satisfied in the numerical simulations. However, the kinetic energy of azimuthal rotation is at least one or two orders of magnitude lower than that of the poloidal flow \citep{herreman2019numerical}.

We focus on $K\in[0.1, 0.7]$ for the validation of our theoretical estimate, \cref{velocity_scale_2}. Choosing a proportionality constant, $\mathcal{C}_1$ for the best linear fit of $u_{theo}$ with the numerical simulations,
\begin{equation} 
    u_{theo} \approx \mathcal{C}_1 \sqrt{\dfrac{\mu_0 I^2}{4\pi^2 \rho R^2} \left[ \left( \dfrac{4}{3} \right) \dfrac{(1 - K)^2}{K} \right]},  \hspace{1cm} \text{with } \mathcal{C}_1 = 0.68.
    \label{velocity_scale_Final}
\end{equation}
Or, alternatively in terms of $S$,
\begin{equation} 
    u_{theo} \approx 0.68 \left( \dfrac{\nu}{R} \right) \sqrt{S \left[ \left( \dfrac{4}{3} \right) \dfrac{(1 - K)^2}{K} \right]}.
    \label{velocity_scale_Final_S}     \notag
\end{equation}
Note that, $\mathcal{C}_1 < 1$ is justified since the magentic field profiles \cref{B0_profiles_eqn} used to derive the estimate are large in comparison to the profiles computed from the Biot-Savart law \labelcref{BS_law}, which is evident in \citet[][refer figure 13(a) for $K=0.2$]{soni2024evaluating}. Also, $\mathcal{C}_1$ is $\mathcal{O}(1)$. Now, we define
\begin{equation}
    \Rey_{theo} = \dfrac{u_{theo}R}{\nu},    \label{Re_theo}
\end{equation}
based on the domain radius, $R$.

Similarly, we find a proportionality constant for Chudnovskii's estimate \citep{chudnovskii1989evaluating} for the maximum velocity. Thus,
\begin{equation} 
    u_{Chudnovskii} \approx \mathcal{C}_2 \sqrt{\dfrac{\mu_0 I^2}{2\pi^2 \rho R^2} \left[ \dfrac{1 - K^2}{K^2} \right]},  \hspace{1cm} \text{with } \mathcal{C}_2 = 0.5.
    \label{u_chudnovskii}
\end{equation}
Alternatively, it can also be written in terms of $S$,
\begin{equation} 
    u_{Chudnovskii} \approx 0.5 \left( \dfrac{\nu}{R} \right) \sqrt{S \left[\dfrac{2(1 - K^2)}{K^2} \right]}.
    \label{velocity_scales}     \notag
\end{equation}
The empirical estimate of Vlasyuk \citep{vlasyuk1987effects} is
\begin{equation}
    u_{Vlasyuk} = ({\nu}/{R}) \sqrt{S} \sqrt[3]{10^{3-5K}}.  \label{vlasyuk_formula}
\end{equation}
which we will use for comparison. Henceforth, $u_{theo}$ denotes the prediction of the time-averaged r.m.s. speed, and $u_{Chudnovskii}$ and $u_{Vlasyuk}$ denote the prediction of the maximum speed. Now, we use this fit \labelcref{velocity_scale_Final} to test the predictive capacity of the average EVF velocity. We measure the predictive capacity of these estimates with the help of the coefficient of determination, $\mathcal{R}^2$. A value close to 1 shows the best predictive capacity of the estimate. The $\mathcal{R}^2$ values are reported to three significant digits.

We now plot the time-averaged r.m.s. and maximum velocity obtained from our 3D simulations in \cref{fig:u_theo_K}(a) \& (b) respectively. The r.m.s. velocity clearly does not exhibit a linear dependence on $K$. This prompts the question: what functional dependence does $u_{rms}$ have on $K$? Our theoretical estimate \labelcref{velocity_scale_Final} suggests that $u\propto (1-K)/\sqrt{K}$. It predicts the r.m.s. velocity quite accurately with $\mathcal{R}^2 = 0.988$; thus helps in estimating the global average EVF velocity in the domain. In \cref{fig:u_theo_K}(b), we see that Chudnovskii's estimate \labelcref{u_chudnovskii} compares well with the maximum value, whereas the empirical estimate from Vlasyuk \labelcref{vlasyuk_formula} predicts an even higher maximum velocity. They have $\mathcal{R}^2$ values of $0.997$ and $0.962$ respectively. We see a decrease in the $\Rey$ (refer \autoref{table:cases_K}) with increasing CC radius (or decreasing $S_M$). The standard EVF parameter ($S$) would obviously predict the same velocity and hence the same $\Rey$ for different $K$ values. To further assess the accuracy of these estimates, we plot them against the $u_{rms}$ and $u_{max}$ obtained from our simulations, respectively in \autoref{fig:u_theo_rms_K}(a) \& (b). Their linear fit is also shown in these figures. Once again, we see that our theoretical estimate predicts the r.m.s. speed quite accurately. Also, the chosen proportionality constant, $\mathcal{C}_1 = 0.68$, closely aligns the linear fit to the $u_{theo} = u_{rms}$ line for almost all chosen $K\in[0.1,0.7]$ (see \autoref{fig:u_theo_rms_K}(a)). On the contrary, the maximum velocity is close to the literature estimates (\ref{u_chudnovskii} and \ref{vlasyuk_formula}) only for lower $K$ values (see \autoref{fig:u_theo_rms_K}(b)). Therefore, we can conclude that our theoretical estimate reliably predicts the r.m.s. velocity, and consequently, the r.m.s. Reynolds number for varying radius ratio.

\subsection{Effect of the applied current}       \label{I0_effect}

\begin{table}
  \begin{center}
\def~{\hphantom{3}}
  \begin{tabular}{lcccccccc}
      $K$  &  $I$ (\SI{}{\A}) & $N_d \times N_{\theta} \times N_z$  & $\Delta t$ (\unit{\s})   & $S (\times 10^4)$   &   $S_M (\times 10^4)$  & $\Rey_{theo}$  & $\Rey_{rms}$   &   $\Rey_{max}$
      \\[3pt]
       0.2  &   ~10   & $260 \times 120 \times 85~$ & 0.02~~  & ~~~2   & ~~~7  &   ~180   & ~116 & ~~313   \\
       0.2  &   ~20   & $260 \times 120 \times 85~$ & 0.02~~  & ~~~6   & ~~28  &   ~359   & ~267 & ~~706   \\
       0.2  &   ~30   & $260 \times 120 \times 85~$ & 0.02~~  & ~~15   & ~~63  &   ~539   & ~433 & ~1109   \\
       0.2  &   ~40   & $260 \times 120 \times 85~$ & 0.01~~  & ~~26   & ~111  &   ~718   & ~609 & ~1514   \\
       0.2  &   ~50   & $260 \times 120 \times 85~$ & 0.01~~ & ~~41   & ~174  &   ~898   & ~748 & ~1920   \\
       0.2  &   ~72   & $260 \times 120 \times 85~$ & 0.002~  & ~~85   & ~361  &   1292   & 1136 & ~2819   \\
       0.2  &   100   & $260 \times 120 \times 85~$ & 0.005~  & ~163   & ~697  &   1795   & 1632 & ~3934   \\
       0.2  &   ~~~129.10  & $260 \times 120 \times 85~$ & 0.004~  & ~272   & 1161  &   2317   & 2298 & ~5414   \\
       0.2  &   200 & $260 \times 120 \times 85~$ & 0.002~  & ~653   & 2787  &   3590   & 3726 & ~~8918   \\[5pt]
       0.33 &   555 & $290 \times 120 \times 95~$ & 0.0005  & 5031   & 8943  &   6431   & 6965 & 15167   \\
  \end{tabular}
  \caption{Details on the numerical simulations for various $I$ with the number of cells ($N_d \times N_{\theta} \times N_z$, in $d$, $\theta$, and $z$--directions respectively) and chosen time step to satisfy the \emph{CFL} criteria. The current density at the bottom boundary is $J_0\in [0.35, \: 7]$ \SI{}{\A/\cm^2} for $K=0.2$. For $K=0.33$ with $I=$ \SI{555}{\A}, it is \SI{17.68}{\A/\cm^2}. Reynolds number (based on $R$) for these cases is also reported along with the $S$ and $S_M$ values. The $S$, $S_M$, and $\Rey$ values are rounded off to the nearest integer.}
  \label{table:cases_I0}
  \end{center}
\end{table}

\begin{table}
  \begin{center}
\def~{\hphantom{3}}
  \begin{tabular}{lcccccccc}
      $K$  &    $I$    &  $|\boldsymbol{B_0}|_{max}$ & $(F_{L,u})_{max}$  & $|\langle \omega_{\theta} \rangle|_{max}$   & $|\langle \boldsymbol{u} \rangle|_{max}$   &   $\langle u_{z} \rangle_{max}$
      \\[3pt]
        &   (\SI{}{\A})   &  (G) & (\SI{}{\cm/\s^2})  & (\SI{}{1/\s})   & (\SI{}{\cm/\s})   &   (\SI{}{\cm/\s})
      \\[3pt]
       0.2  &   ~10   & ~1.70 & ~0.03  & ~~0.92   & 0.12  &   0.12   \\
       0.2  &   ~20   & ~3.40 & ~0.11  & ~~3.07   & 0.28  &   0.28   \\
       0.2  &   ~30   & ~5.11 & ~0.25  & ~~7.04   & 0.44  &   0.44   \\
       0.2  &   ~40   & ~6.81 & ~0.45  & ~12.15   & 0.61  &   0.61   \\
       0.2  &   ~50   & ~8.51 & ~0.70 & ~18.11   & 0.77  &   0.77   \\
       0.2  &   ~72   & 12.26 & ~1.45  & ~33.39   & 1.13  &   1.13   \\
       0.2  &   100   & 17.03 & ~2.79  & ~57.82   & 1.58  &   1.58   \\
       0.2  &   ~~~129.10  & 21.98 & ~4.65  & ~87.84   & 2.13  &   2.13   \\
       0.2  &   200 & 34.05 & 11.17 & 160.86  & 3.46   & 3.46   \\[5pt]
       0.33 &   555 & 57.53 & 19.37  & 268.76   & 5.08  &   5.08   \\
  \end{tabular}
  \caption{The maximum values used in normalizing the quantities. This is for the effect of $I$. The values are rounded off to the two significant digits.}
  \label{table:I_max_A1}
  \end{center}
\end{table}

\begin{figure}
\centering
\begin{tabular}{lccc}
\rotatebox{90}{\hspace{0.5cm} ${I=\SI{10}{\A}}$} &
\includegraphics[width=4cm]{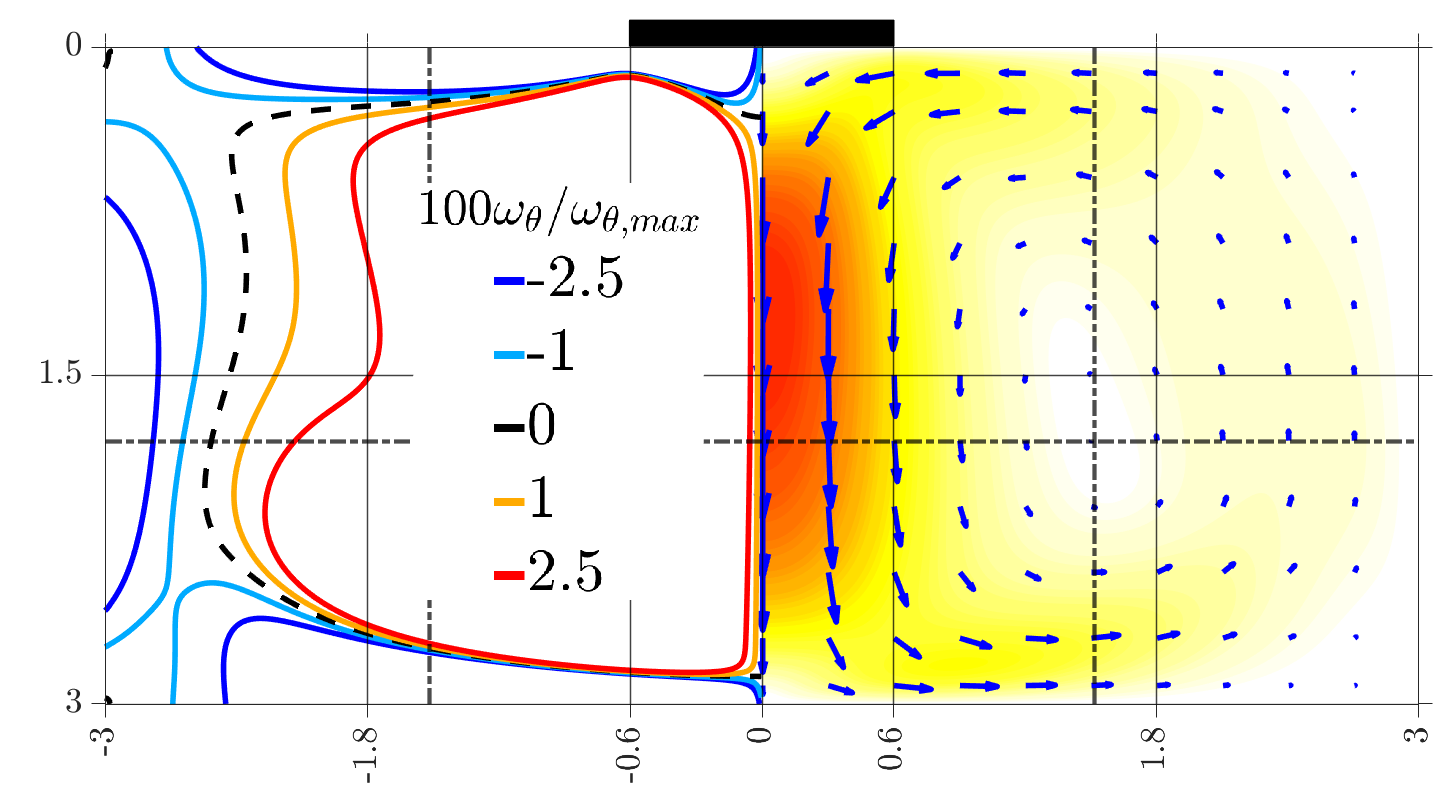} &
\includegraphics[width=4cm]{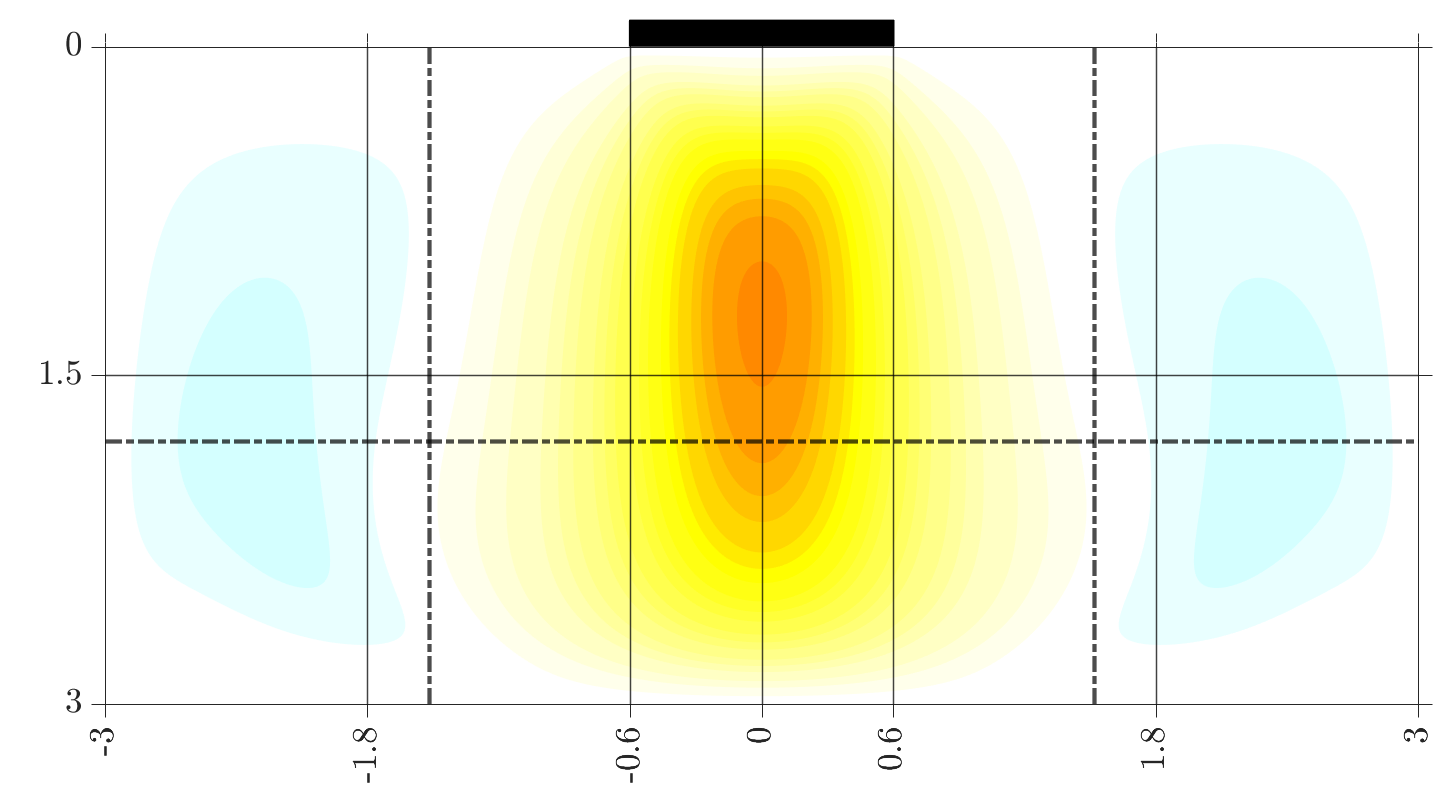} &
\includegraphics[width=4cm]{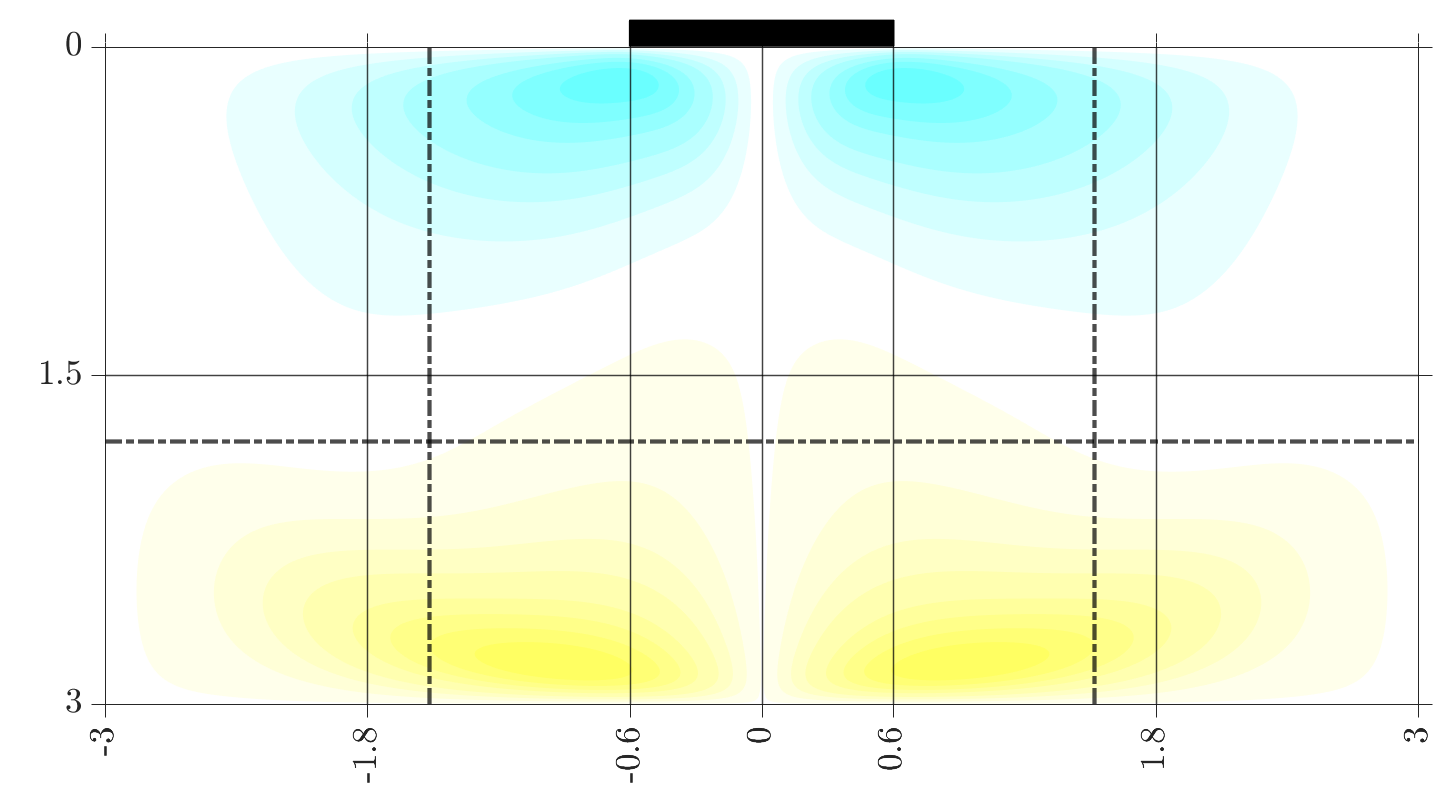} \\
\end{tabular}
\begin{tabular}{cc}
    \centering
\includegraphics[width=3.5cm]{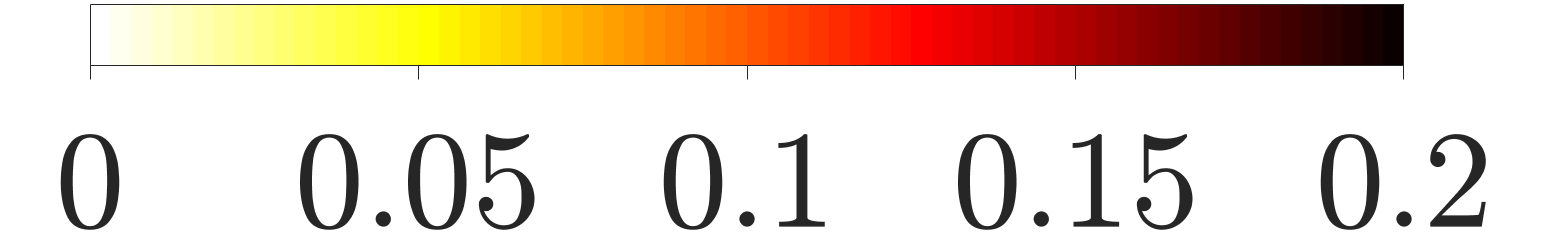} &
\includegraphics[width=3.5cm]{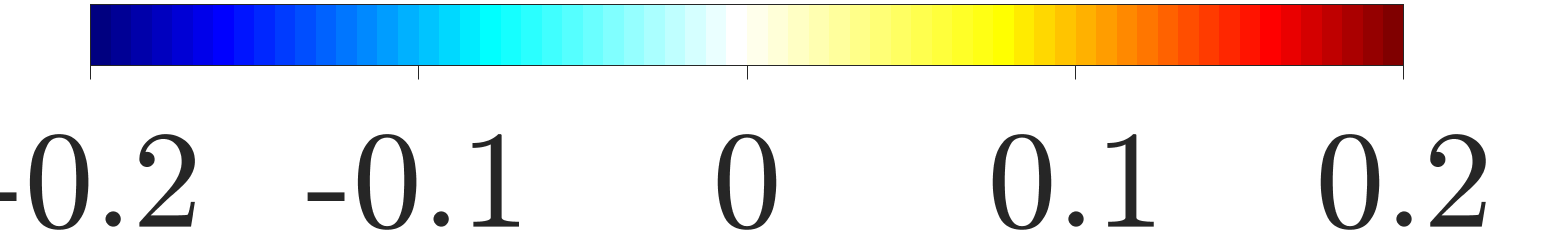} \\
    \end{tabular} \\[2pt]

\begin{tabular}{lccc}
\rotatebox{90}{\hspace{0.5cm} ${I=\SI{30}{\A}}$} &
\includegraphics[width=4cm]{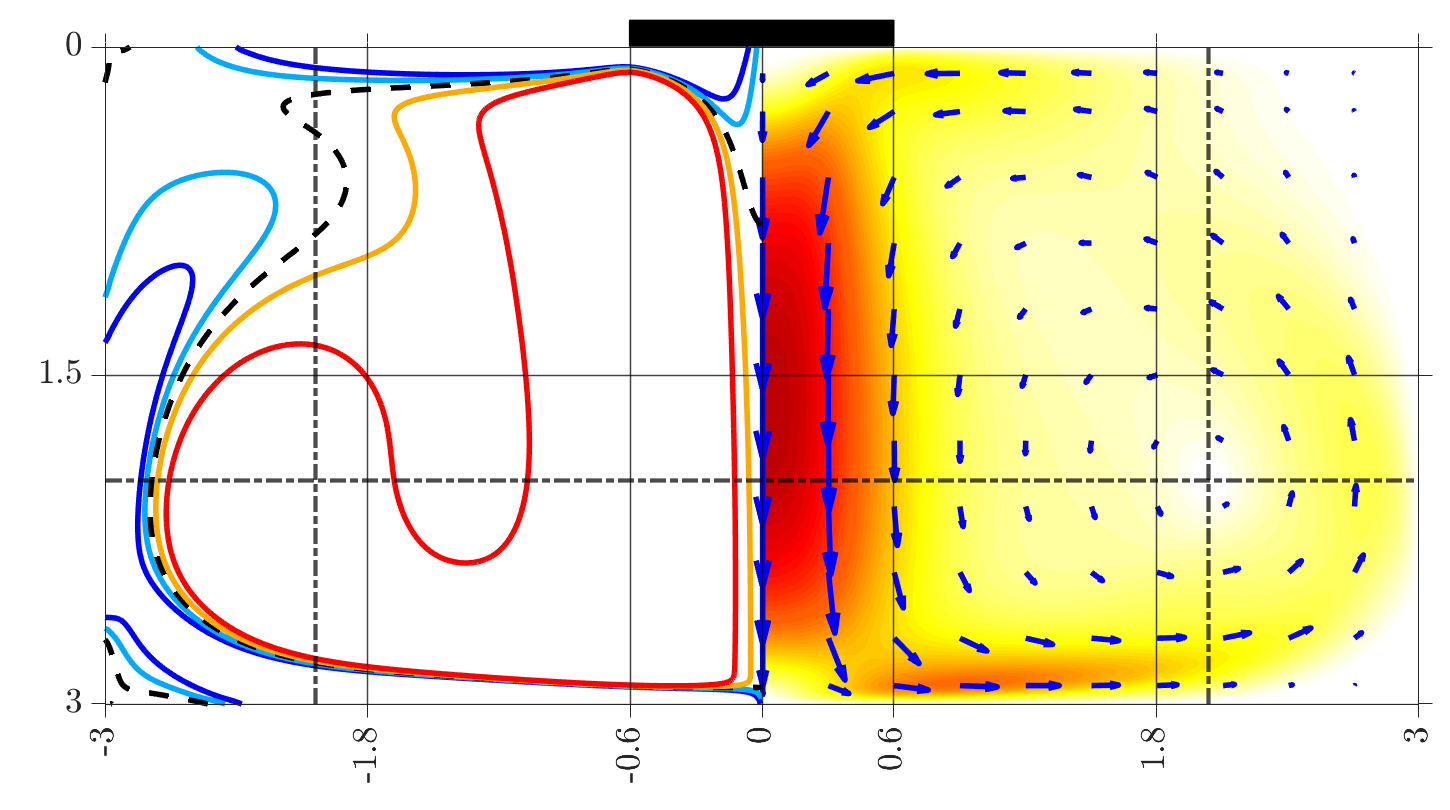} &
\includegraphics[width=4cm]{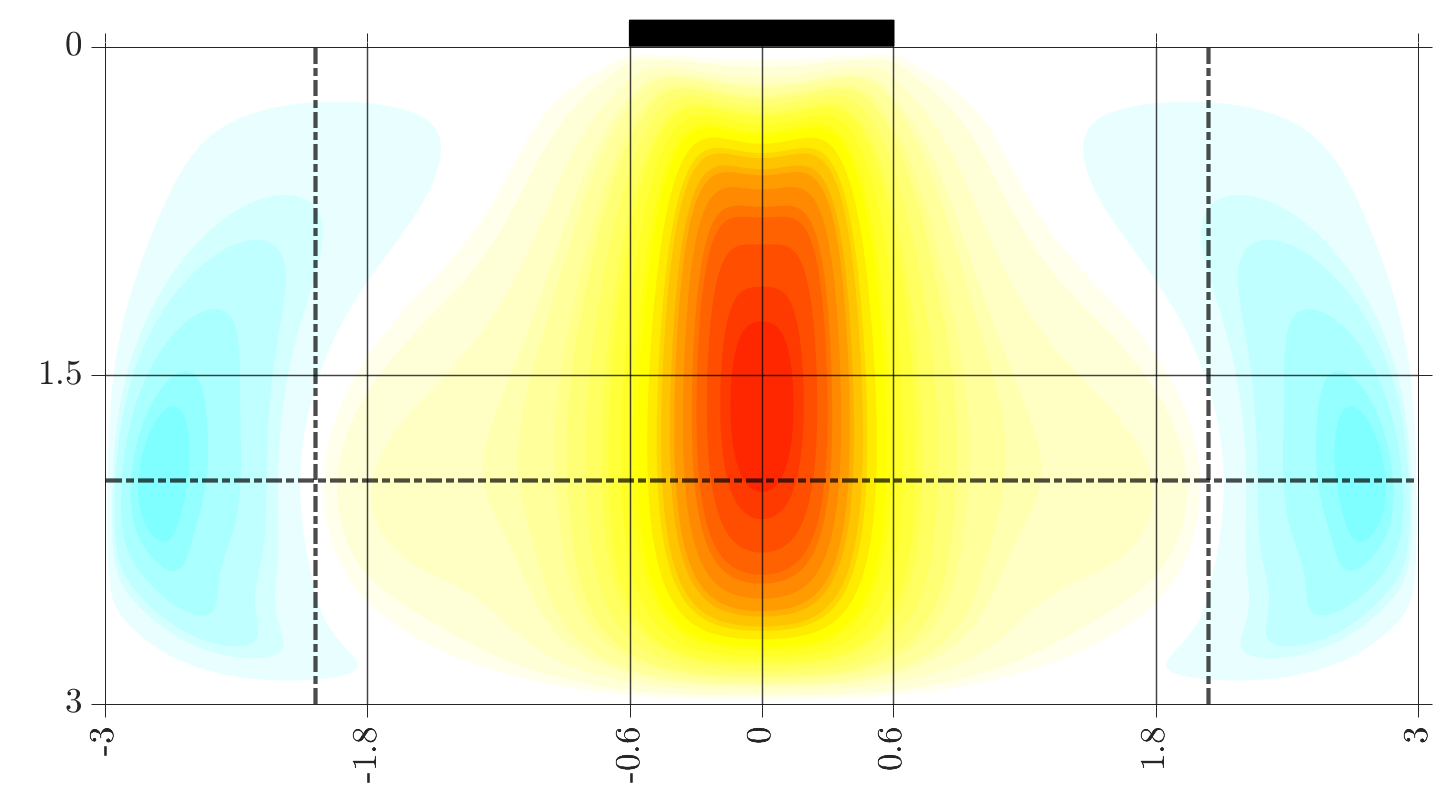} &
\includegraphics[width=4cm]{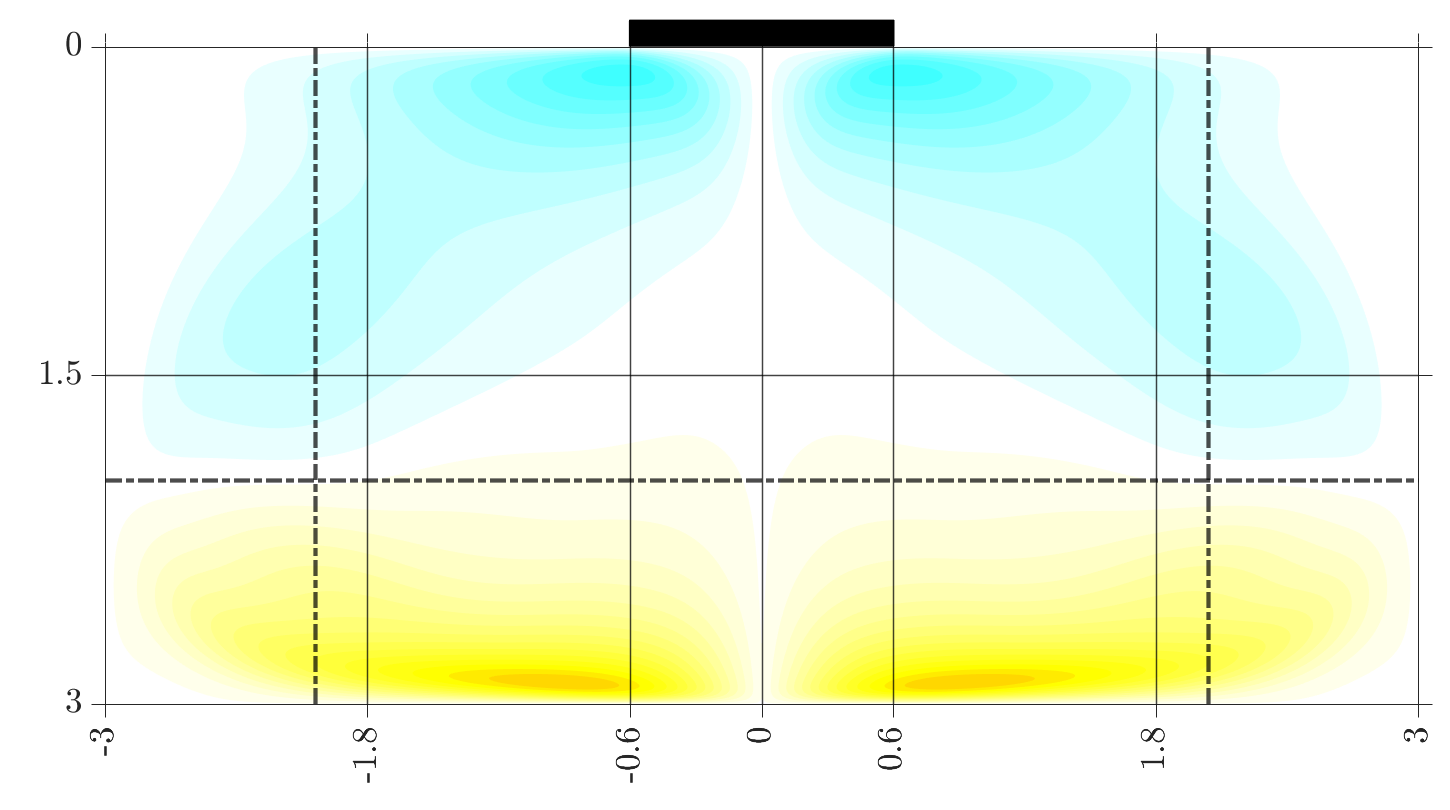} \\
\end{tabular}
\begin{tabular}{cc}
    \centering
\includegraphics[width=3.5cm]{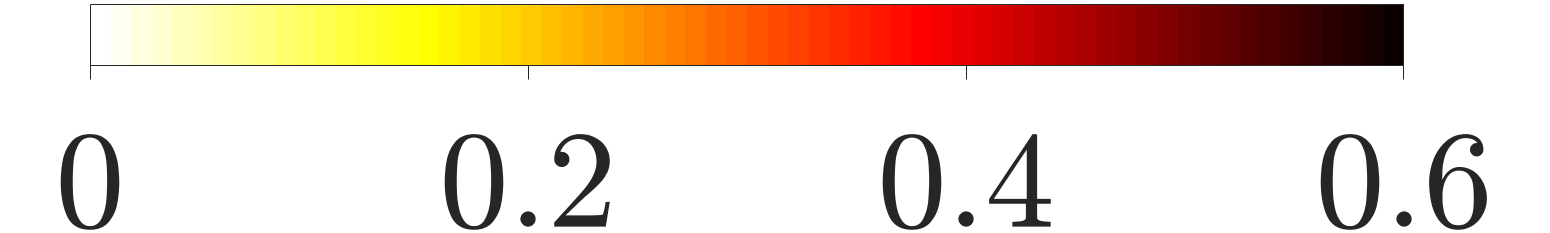} &
\includegraphics[width=3.5cm]{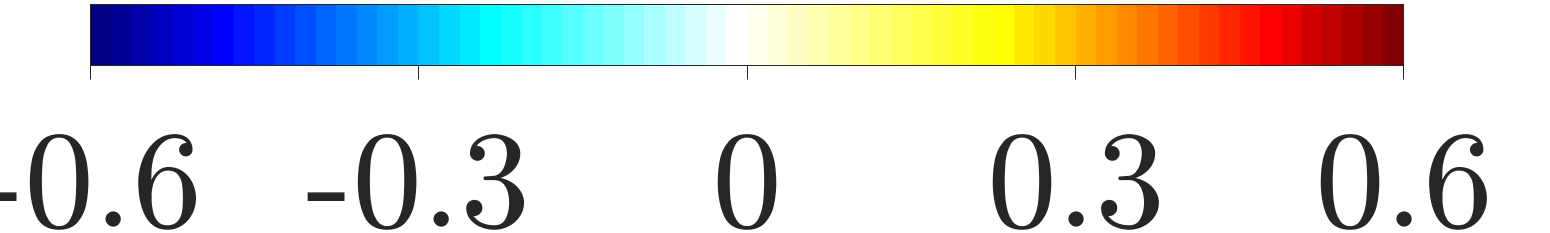} \\
    \end{tabular} \\[2pt]

\begin{tabular}{lccc}
\rotatebox{90}{\hspace{0.5cm} ${I=\SI{50}{\A}}$} &
\includegraphics[width=4cm]{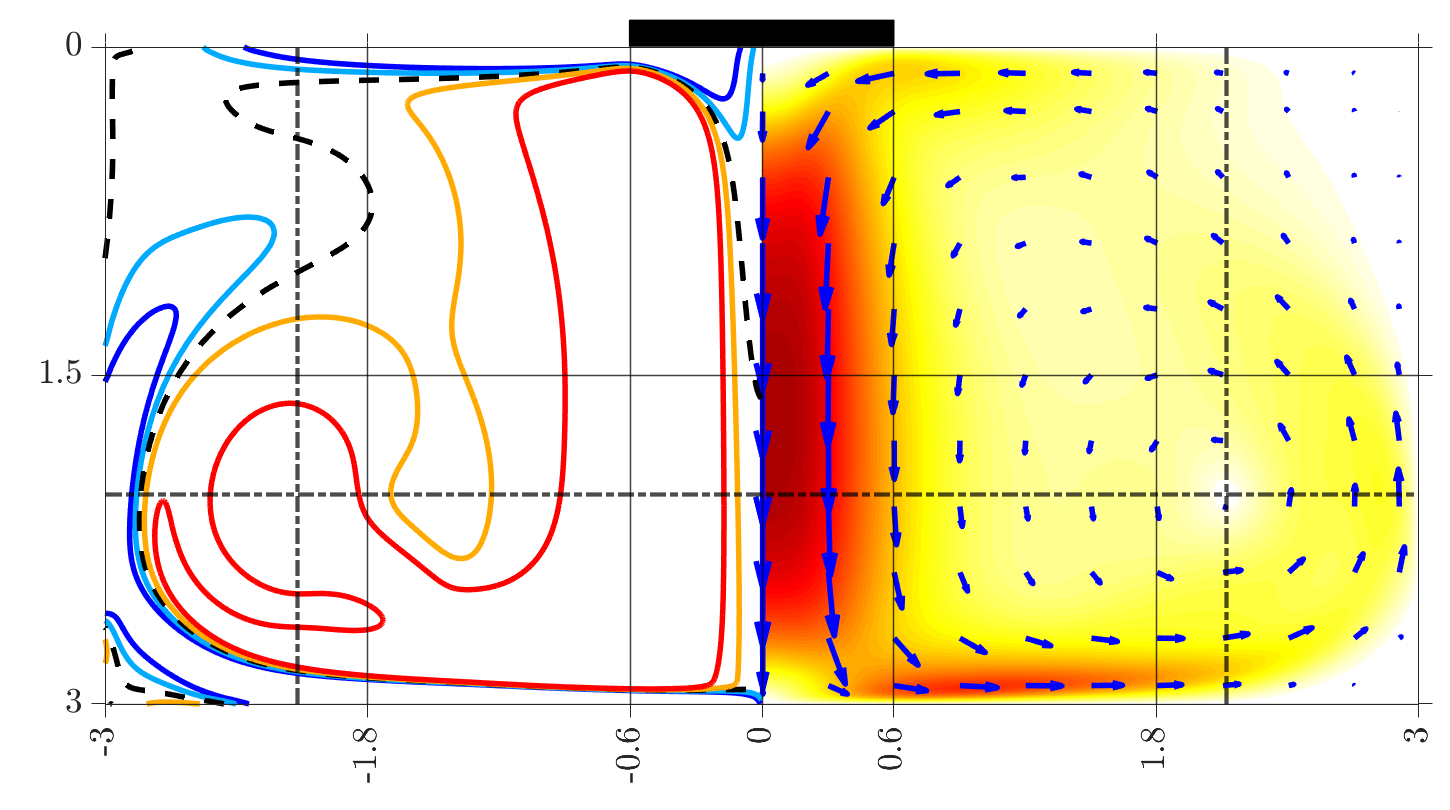} &
\includegraphics[width=4cm]{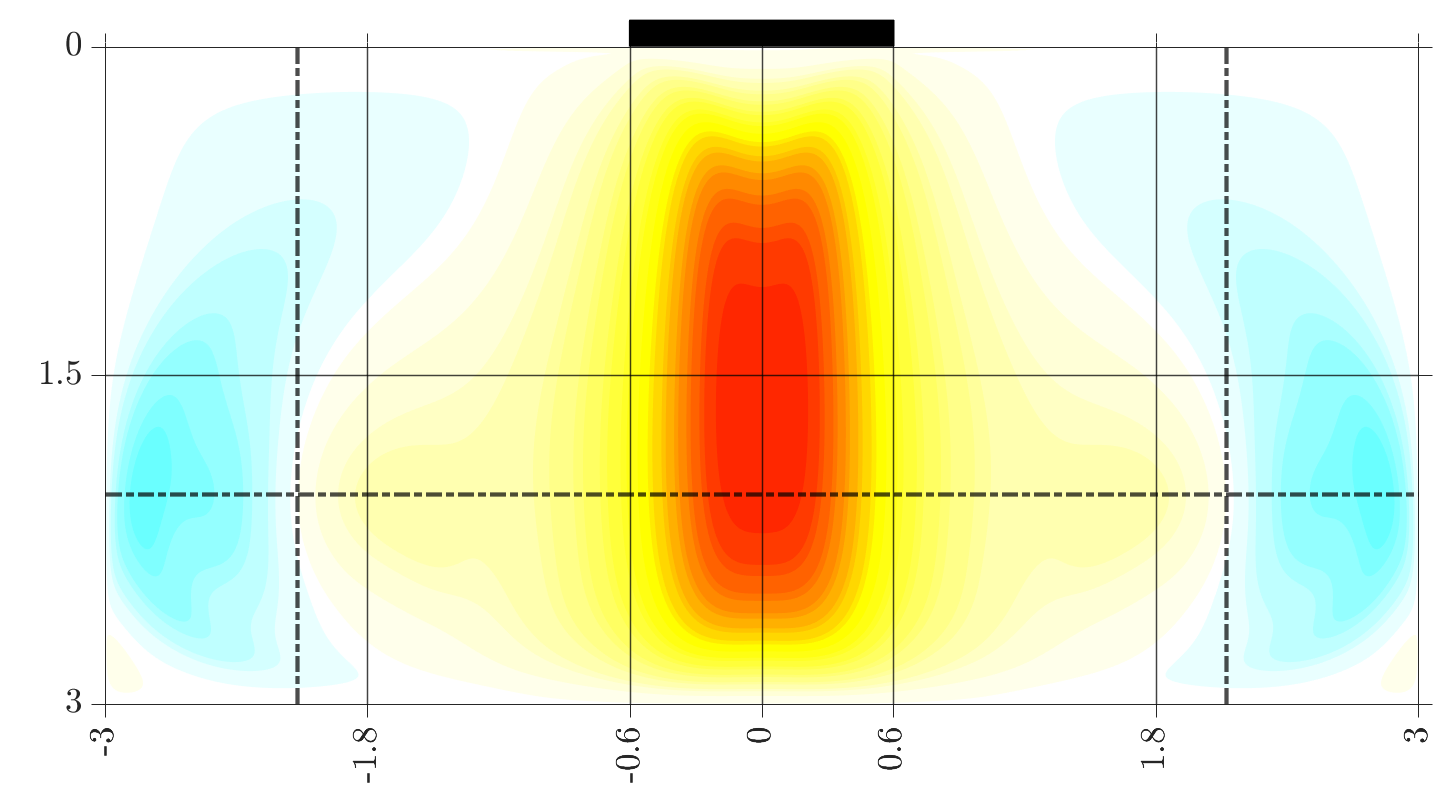} &
\includegraphics[width=4cm]{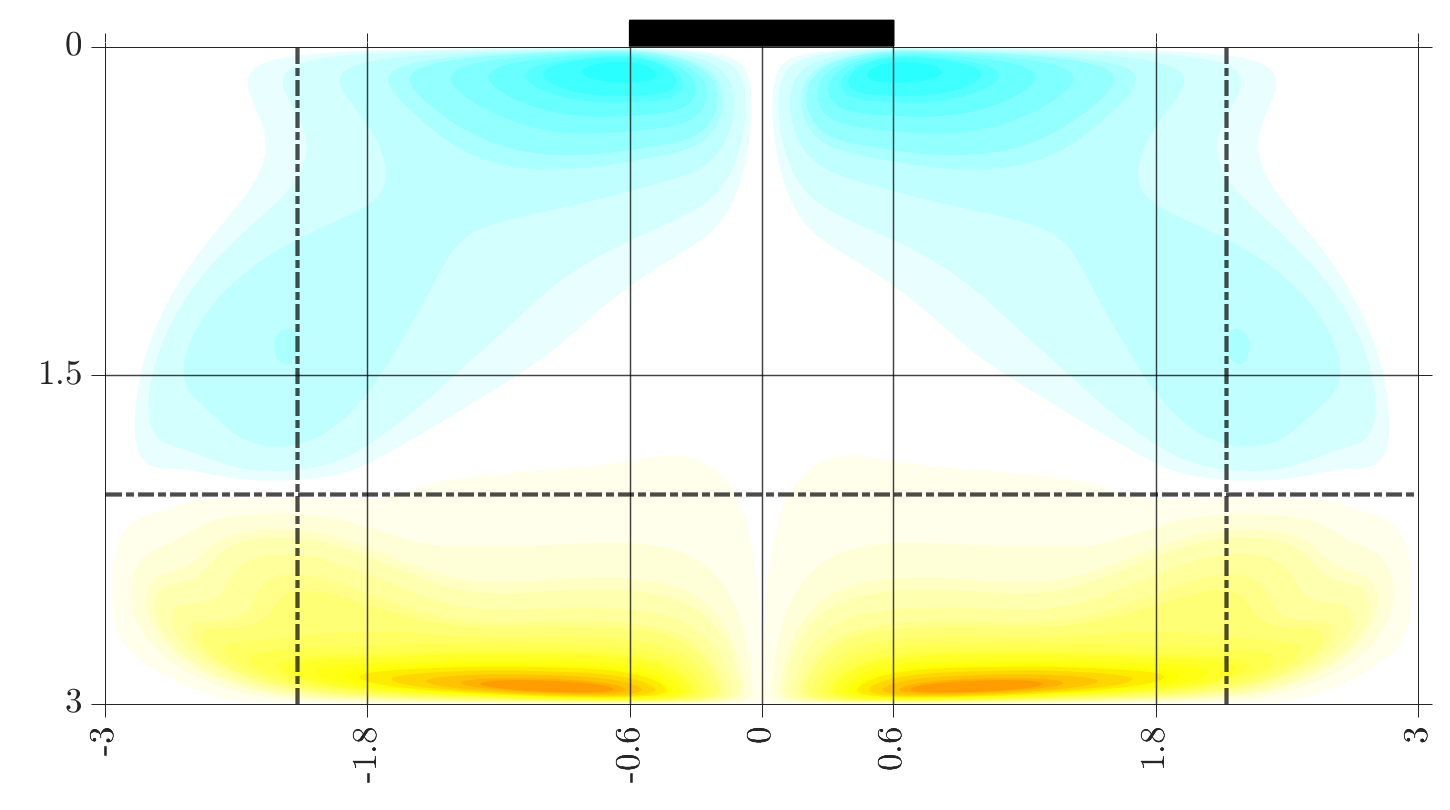} \\
\end{tabular}
\begin{tabular}{cc}
    \centering
\includegraphics[width=3.5cm]{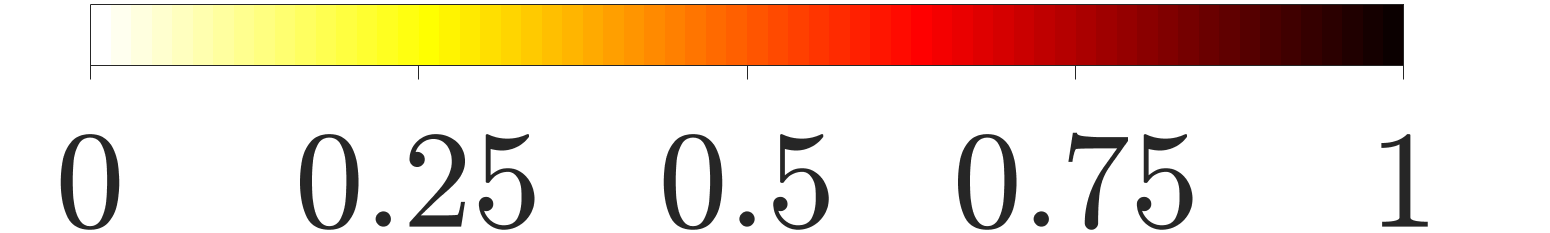} &
\includegraphics[width=3.5cm]{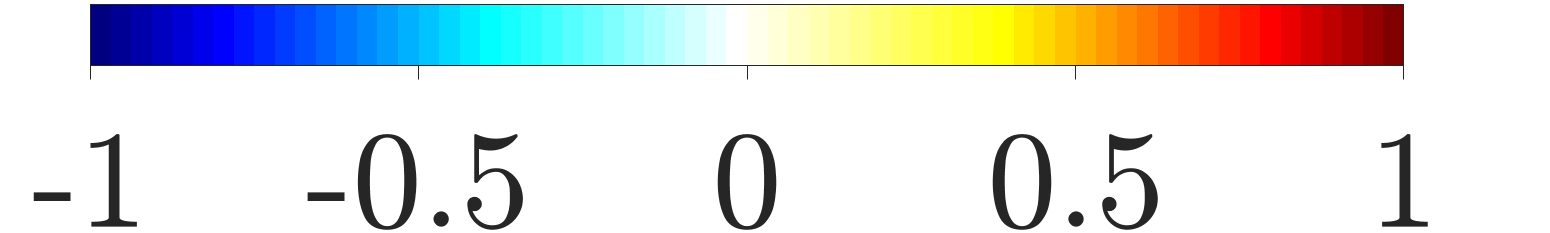} \\
    \end{tabular} \\[2pt]

\begin{tabular}{lccc}
\rotatebox{90}{\hspace{0.5cm} ${I=\SI{100}{\A}}$} &
\includegraphics[width=4cm]{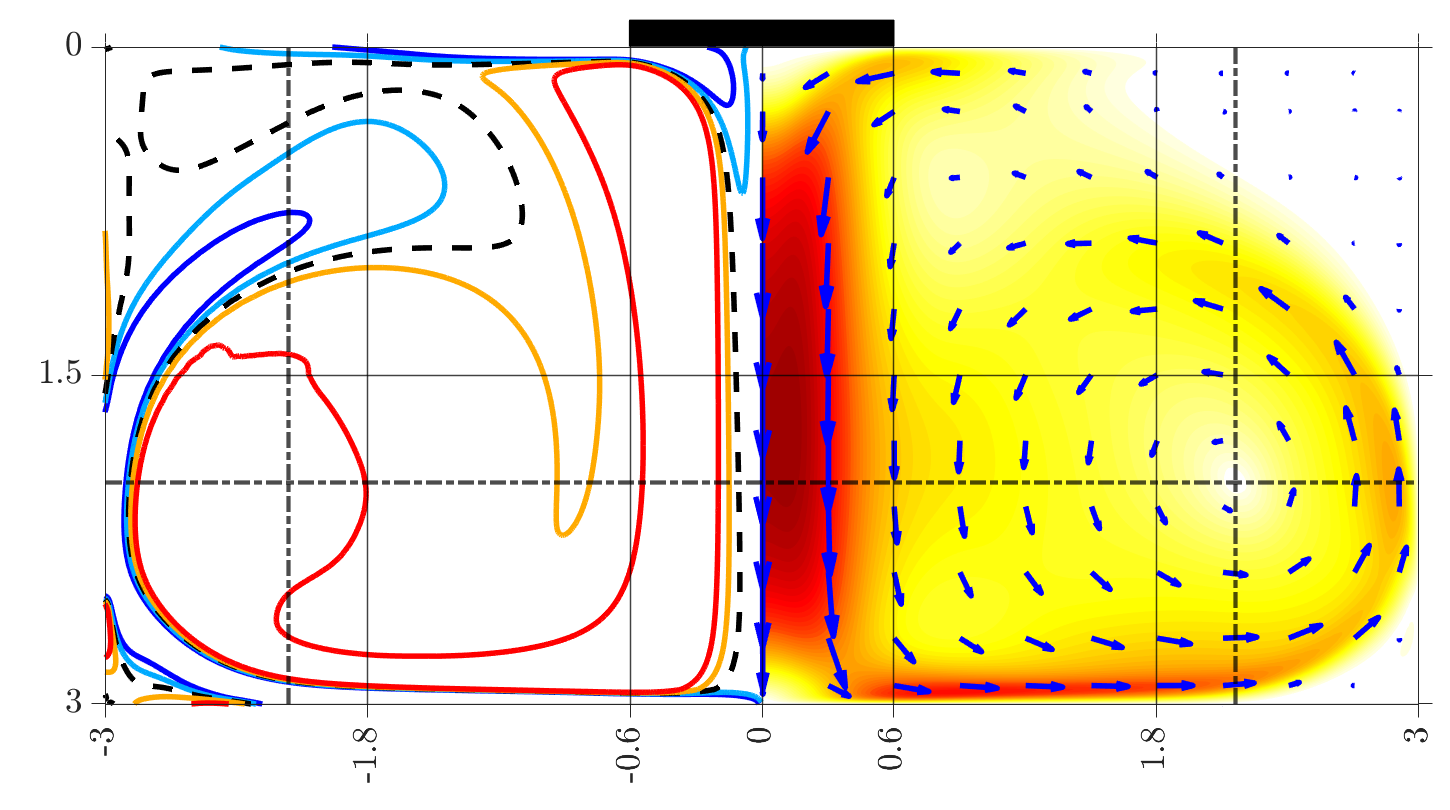} &
\includegraphics[width=4cm]{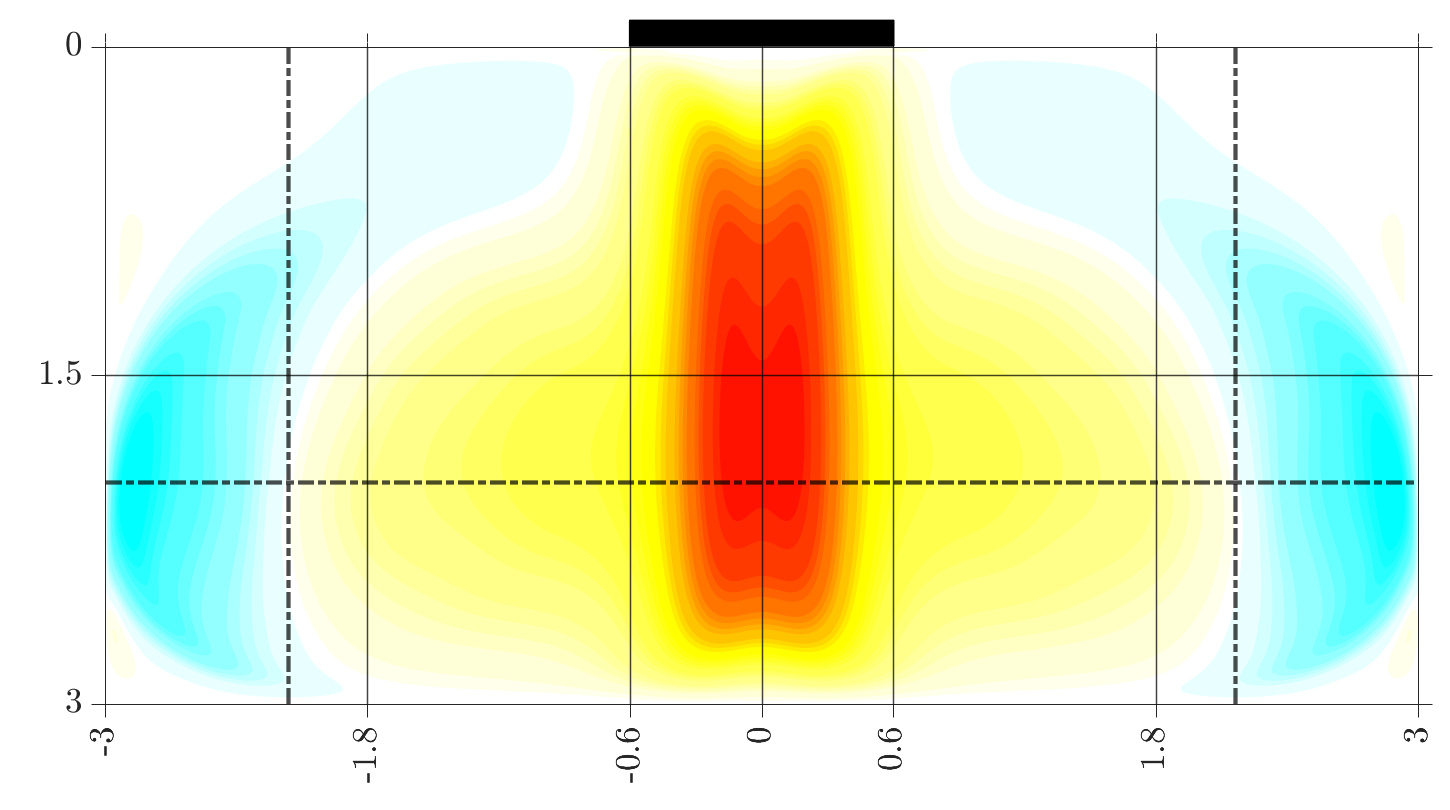} &
\includegraphics[width=4cm]{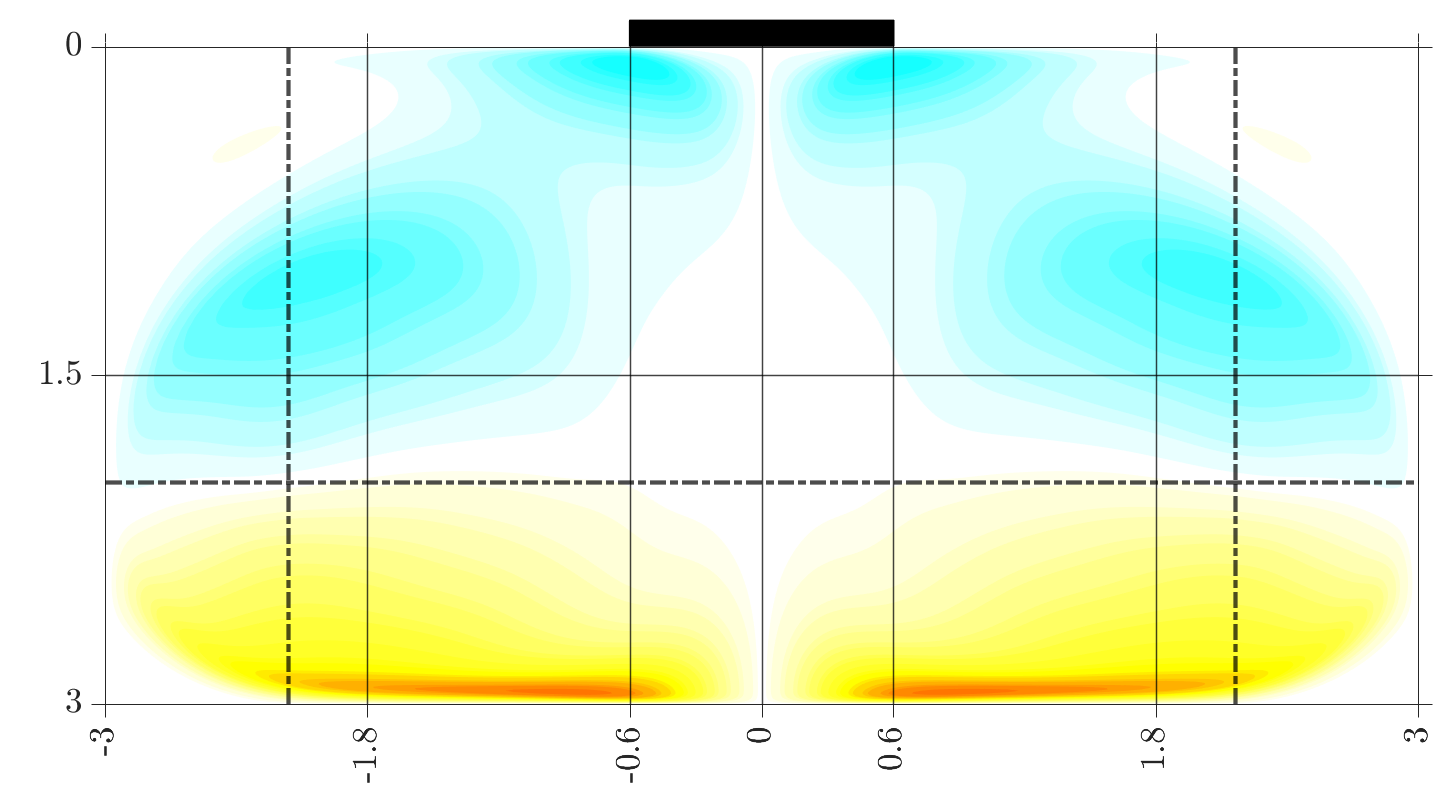} \\
\end{tabular}
\begin{tabular}{cc}
    \centering
\includegraphics[width=3.5cm]{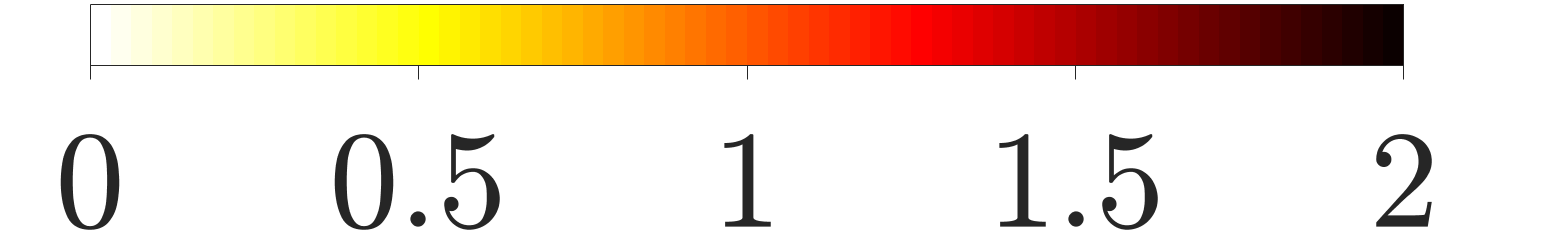} &
\includegraphics[width=3.5cm]{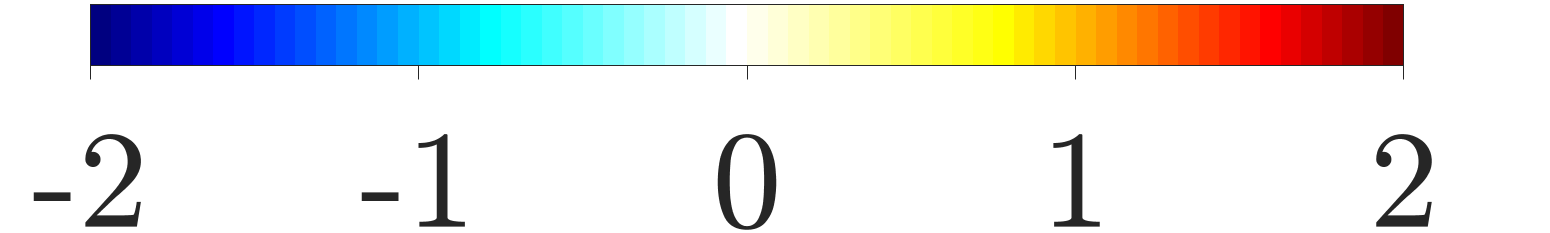} \\
    \end{tabular} \\[2pt]

\begin{tabular}{lccc}
\rotatebox{90}{\hspace{0.2cm} ${I=\SI{129.10}{\A}}$} &
\includegraphics[width=4cm]{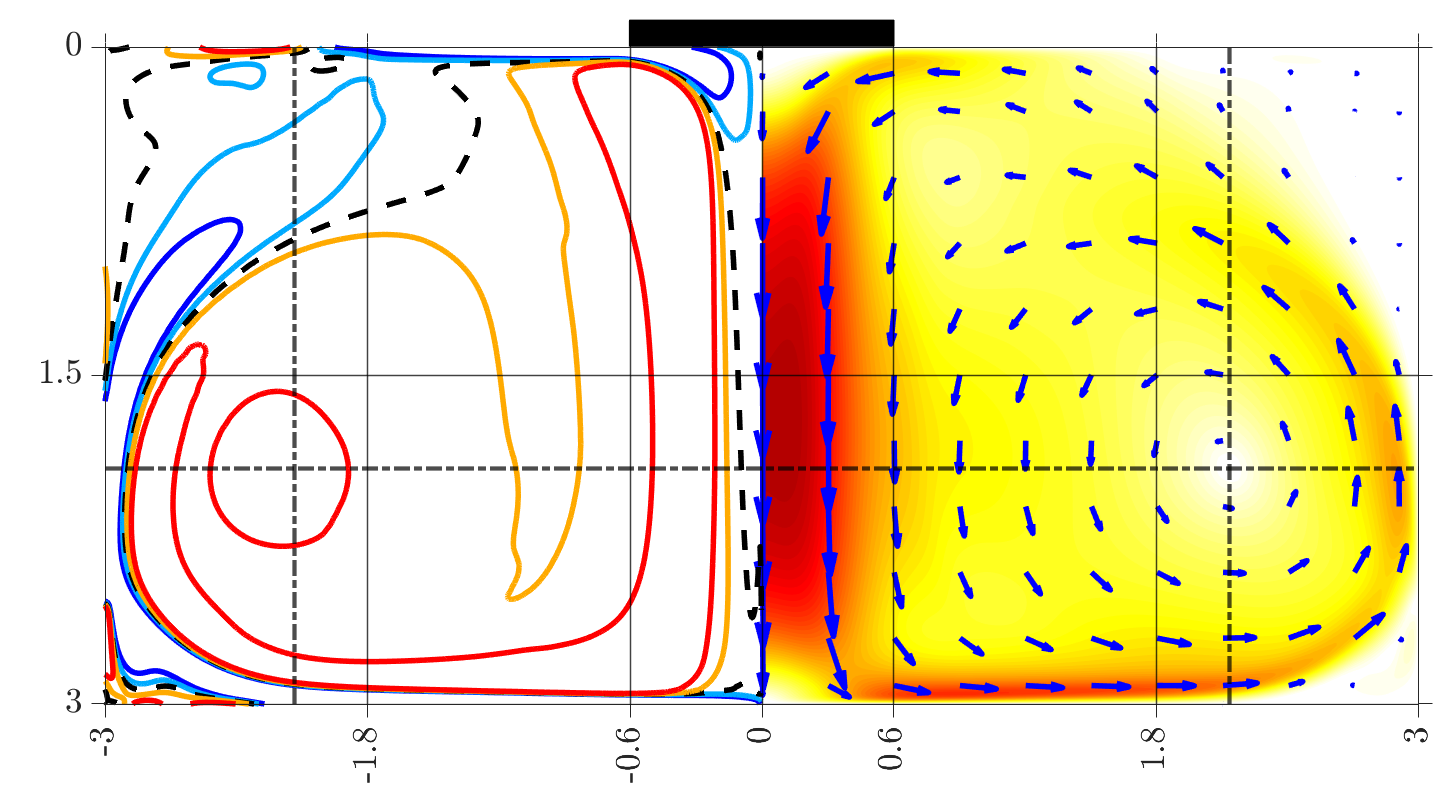} &
\includegraphics[width=4cm]{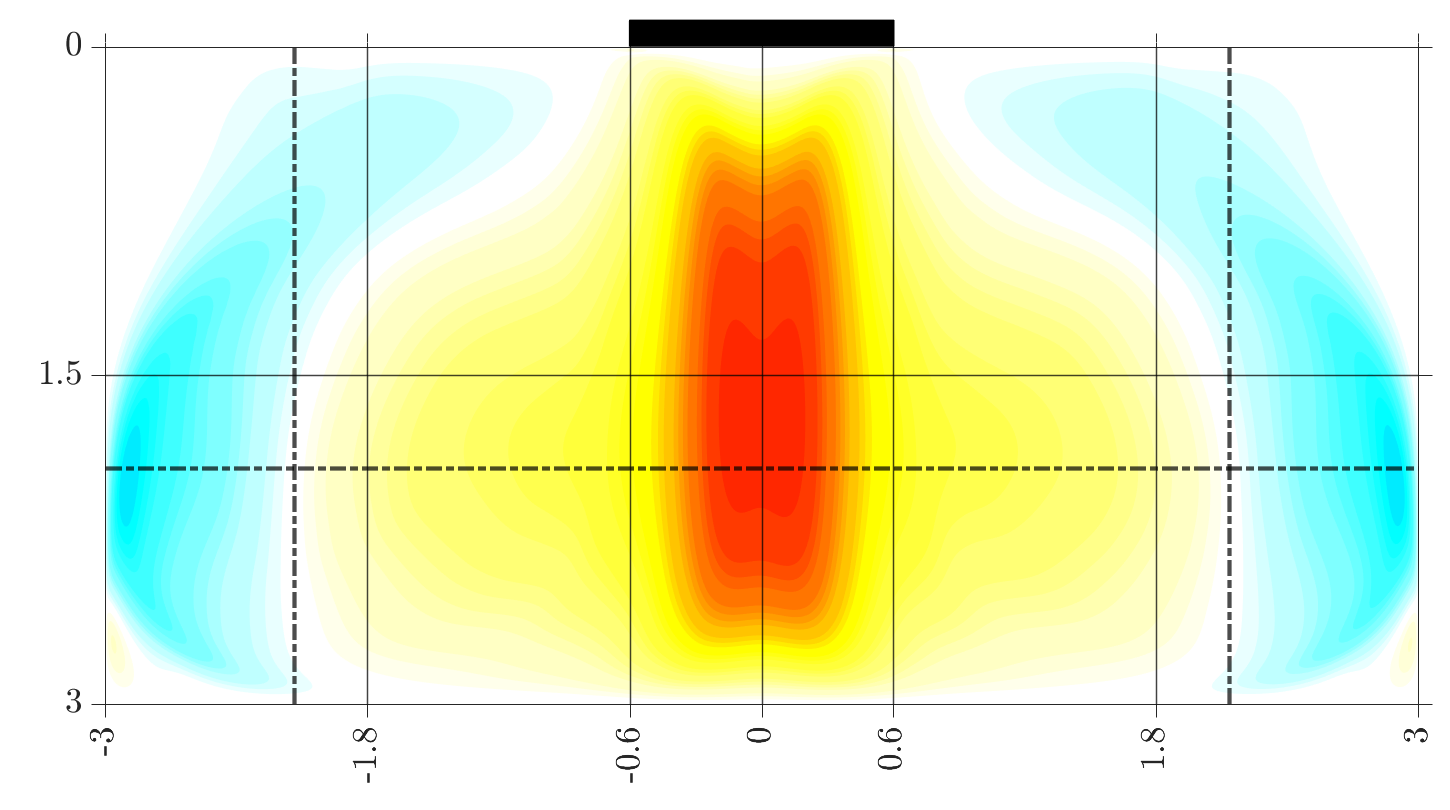} &
\includegraphics[width=4cm]{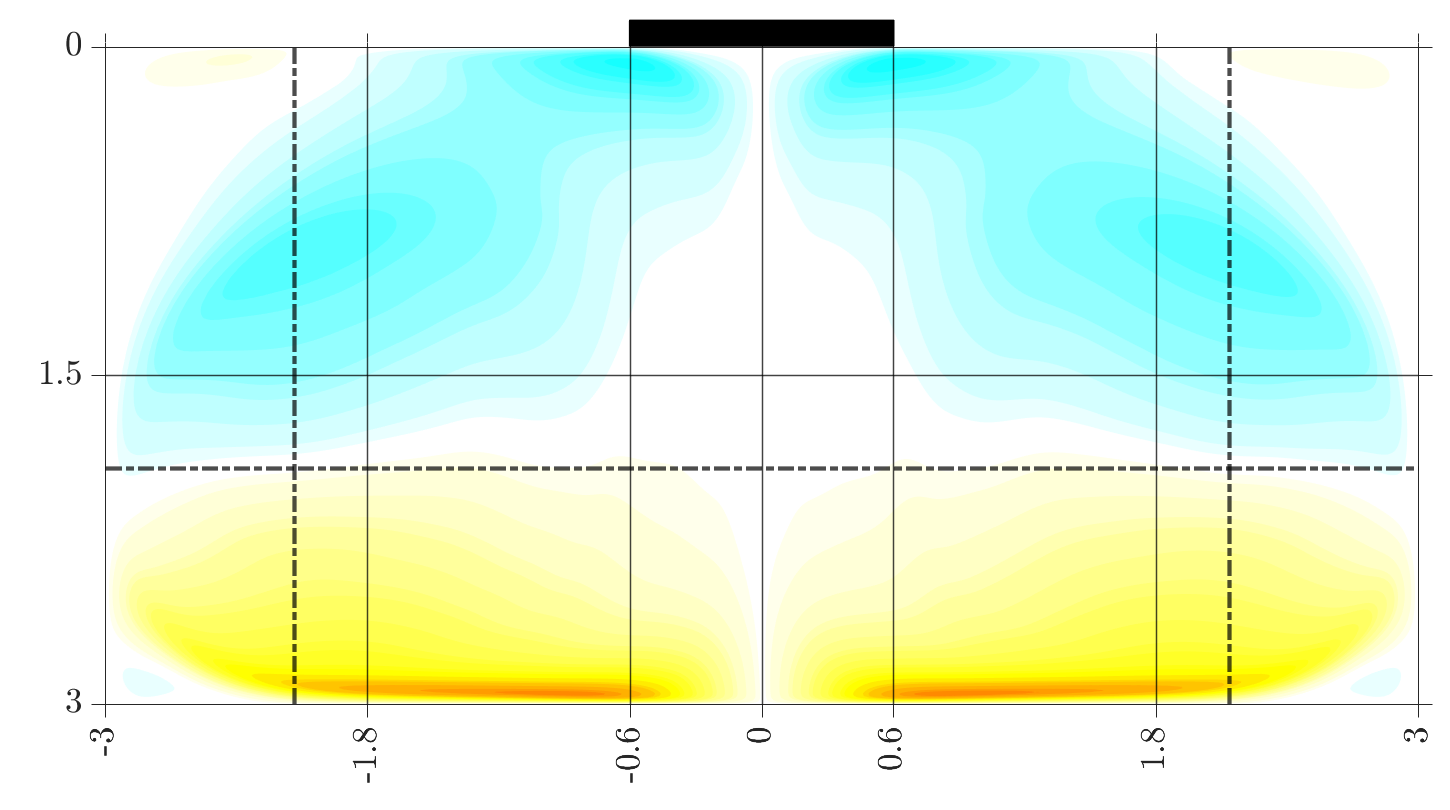} \\
\end{tabular}
\begin{tabular}{cc}
    \centering
\includegraphics[width=3.5cm]{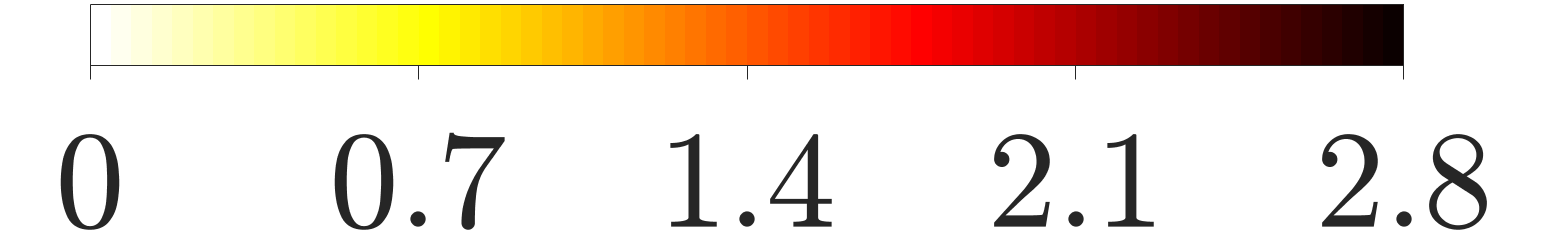} &
\includegraphics[width=3.5cm]{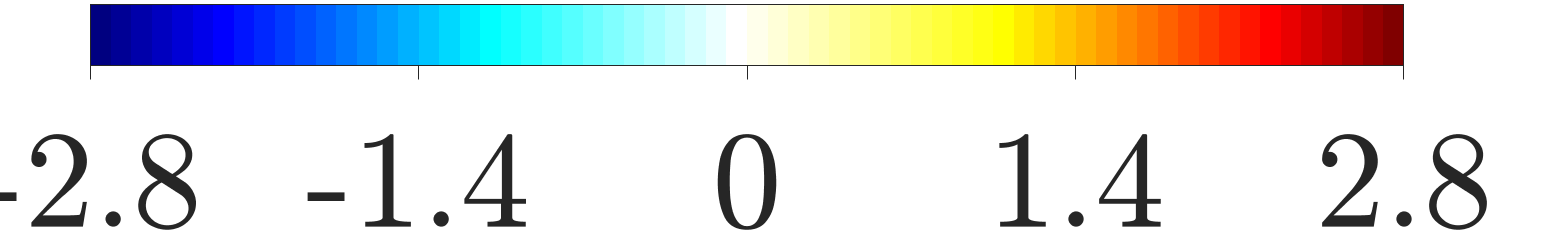} \\
    \end{tabular} \\[2pt]

\begin{tabular}{lccc}
\rotatebox{90}{\hspace{0.45cm} ${I=\SI{200}{\A}}$} &
\includegraphics[width=4cm]{Fig/contour_plots_v09/Vectors_XZ_slice_Umeanmag_50sec_K02_200A_v09.png} &
\includegraphics[width=4cm]{Fig/contour_plots_v09/XZ_slice_UmeanZ_50sec_K02_200A_v09.png} &
\includegraphics[width=4cm]{Fig/contour_plots_v09/XZ_slice_UmeanR_50sec_K02_200A_v09.png} \\
\end{tabular}
\begin{tabular}{cc}
    \centering
\includegraphics[width=3.5cm]{Fig/contour_plots_v09/Vectors_XZ_slice_Umeanmag_50sec_K02_200A_cb_v09.png} &
\includegraphics[width=3.5cm]{Fig/contour_plots_v09/XZ_slice_UmeanR_50sec_K02_200A_cb_v09.png} \\
    \end{tabular} \\[2pt]
\caption{(i) Time-averaged vorticity (left) and velocity magnitude (right), (ii) time-averaged axial velocity, and (iii) radial velocity. All the velocities are in \SI{}{\cm/\s}. These contours are shown in the xz-plane extracted from the 3D simulations of EVF for various applied currents, $I$. CC is shown in black. Please note that each case has a different colormap legend.}
\label{fig:contour_plots_I}
\end{figure}

In this section, we first discuss the EVF characteristics for $I\in[10,\:200]$\SI{}{\A} at a constant radius ratio, $K=0.2$. With an increase in the applied current value, we expect the flow intensity to increase. This is evident in the reported $\Rey$ for various $I$ in \autoref{table:cases_I0}. Note that both $S$ and $S_M$ also increase, as expected. (The case with $K=0.33$, $I=$ \SI{555}{\A} will be considered later in \S\labelcref{mod_param}.) The current path and location of the peak magnetic field remain unchanged regardless of variations in the applied current. This also applies to the distribution of the Lorentz force. At lower currents, the electro-vortex jet is too weak to reach the bottom boundary, see the \cref{fig:contour_plots_I}(i) \& (ii). In this scenario, the viscous effects are important. Consequently, the return flow near the sidewall is weak at lower currents. As the current magnitude increases, inertial effects become prominent, leading to the advection of positive vorticity near the bottom boundary and the formation of a toroidal vortex. Unlike the cases with varying $K$ considered in \S\labelcref{K_effect}, here the negative vorticity near the CC region is not advected much towards the axial region. This prevents the zero-vorticity line at the axis from being disrupted, ensuring that the electro-vortex jet does not bifurcate, as was the case in \autoref{fig:contour_plots_K}. (The flow and vorticity evolution movie can be found in the supplementary material.)

\begin{figure}  \centerline{\includegraphics[width=0.57\linewidth]{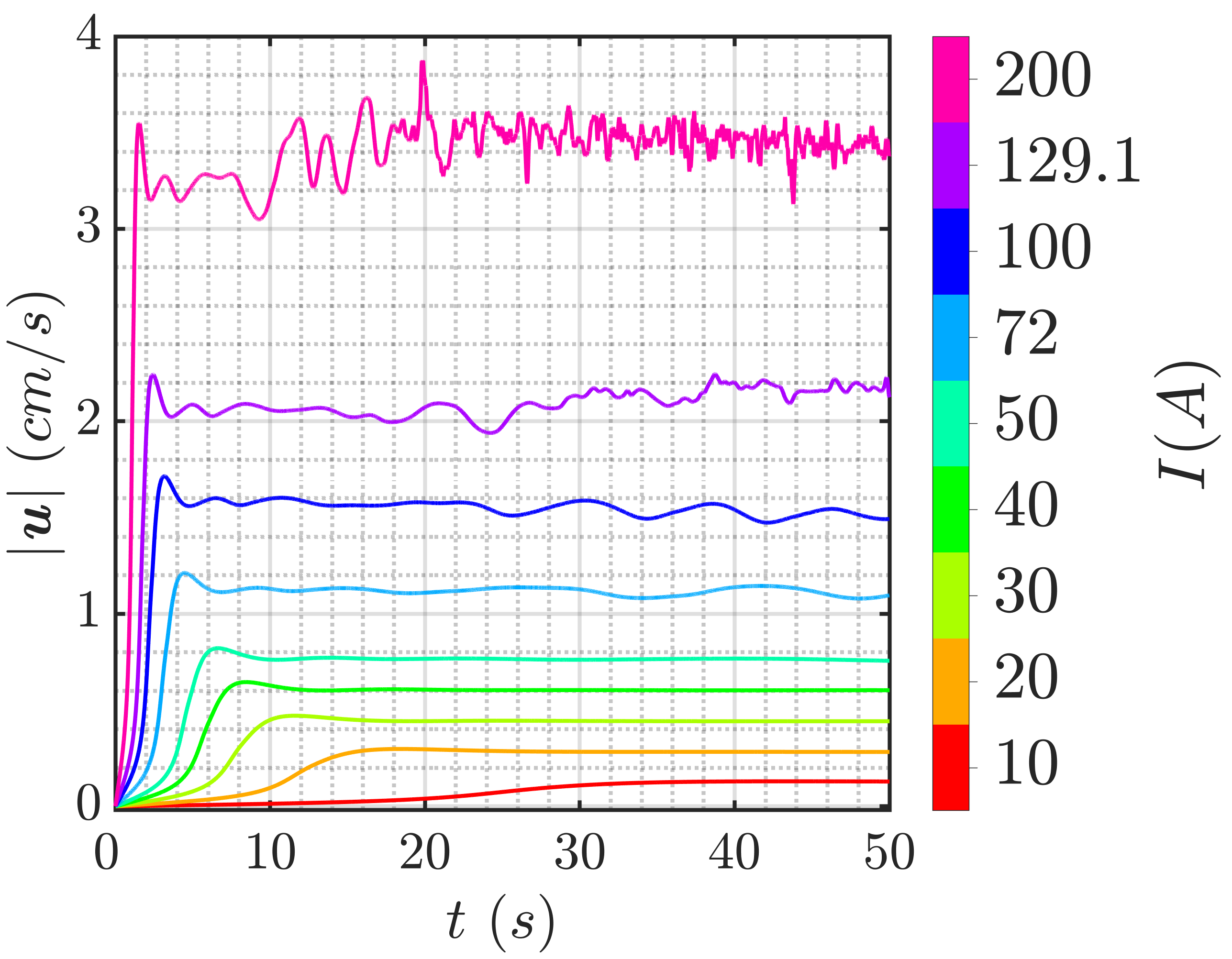} 
\includegraphics[width=0.42\linewidth]{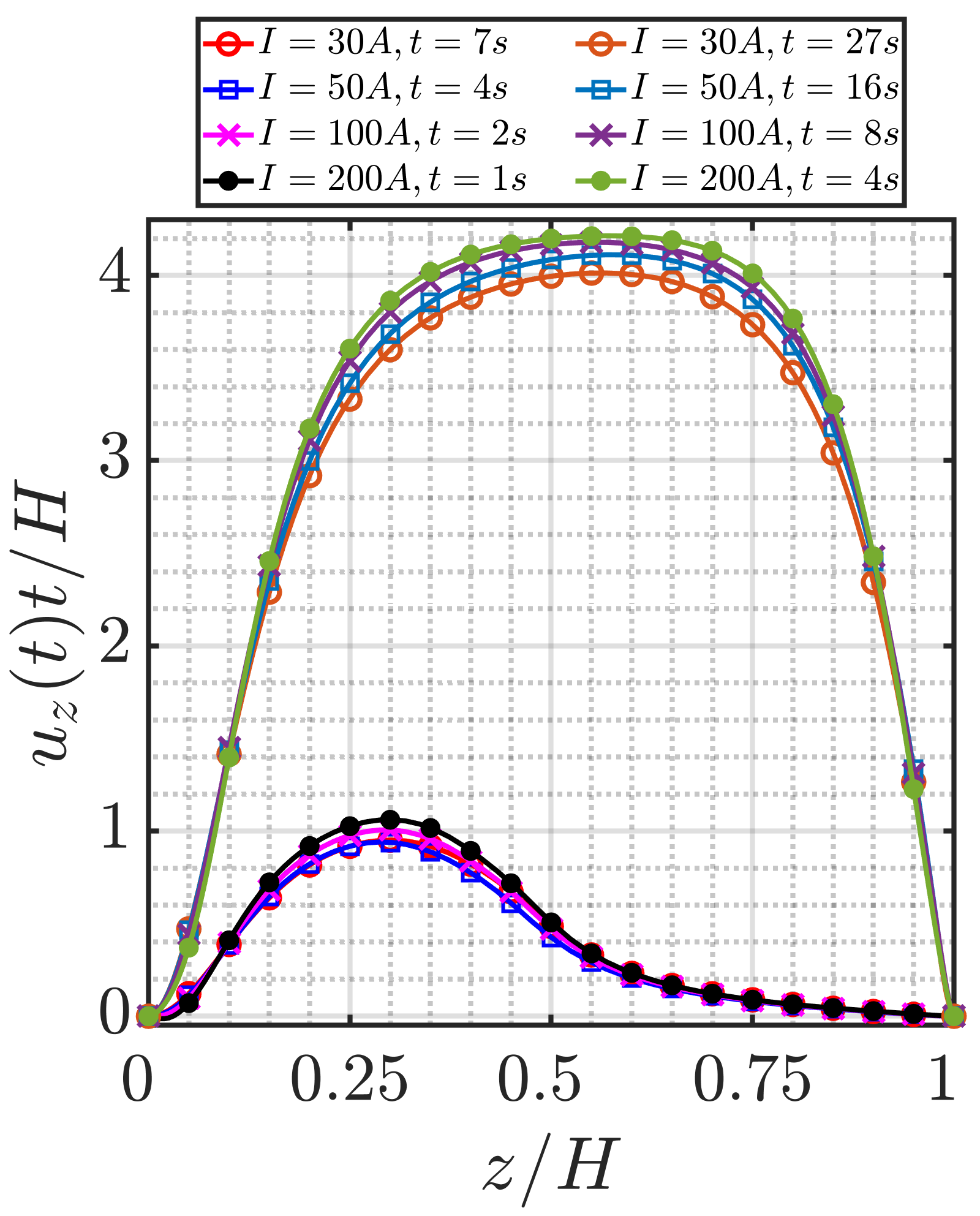}}
\centerline{(a) \hspace{175pt} (b)}
  \caption{(a) Time evolution of velocity magnitude at a point located at the centroid of the domain, and (b) testing the $u \propto I$ relationship for different $I$. Dimensionless axial velocity along the axis of the domain as a function of its height for two time instants: one before and the other after the jet hits the bottom boundary.}
\label{fig:u_coll_probe}
\end{figure}

By analyzing the velocity contours in \autoref{fig:contour_plots_I}, it appears that the flow at the lower current magnitudes is laminar and becomes turbulent at the highest current value (\SI{200}{\A}) under investigation. To cross-check, we plot a time series of velocity magnitude at a point positioned at the domain's centroid in \autoref{fig:u_coll_probe}(a). As anticipated, it reveals that the flow is steady and laminar at lower current values (up to \SI{50}{\A}), transitioning to turbulence with a further increase in the current strength. Recall that our estimate is valid for high $\Rey$ flows where the relation $u \propto I$ holds (see equation \autoref{velocity_scale_2}). To find the cut-off value below which this scaling is no longer applicable, we plot the dimensionless axial velocity for various currents at two time instants: one prior to and the other subsequent to the jet's contact with the bottom boundary, refer to \autoref{fig:u_coll_probe}(b). For instance, the curves for $I=$ \SI{100}{\A} at $t=$ \SI{2}{\s} and $I=$ \SI{200}{\A} at $t=$ \SI{1}{\s} should overlap if the scaling is valid. We observe that the velocity curves do largely overlap for all currents up to $I=$ \SI{30}{\A}. Thus, our estimate is valid as long as the balance between inertia and Lorentz forces holds, required for $u \propto I$.

\begin{figure}  \centerline{\includegraphics[width=0.48\linewidth]{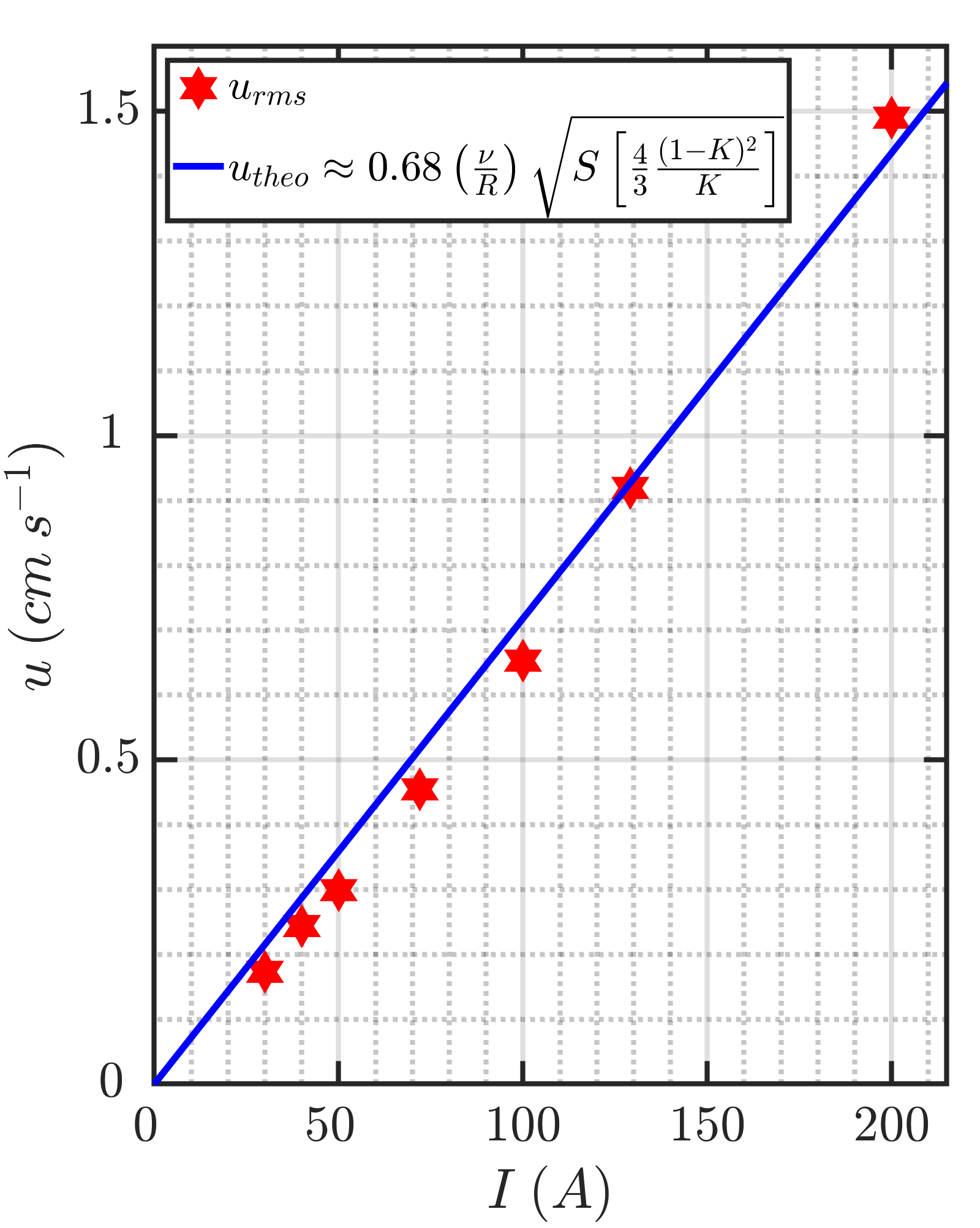} 
\includegraphics[width=0.48\linewidth]{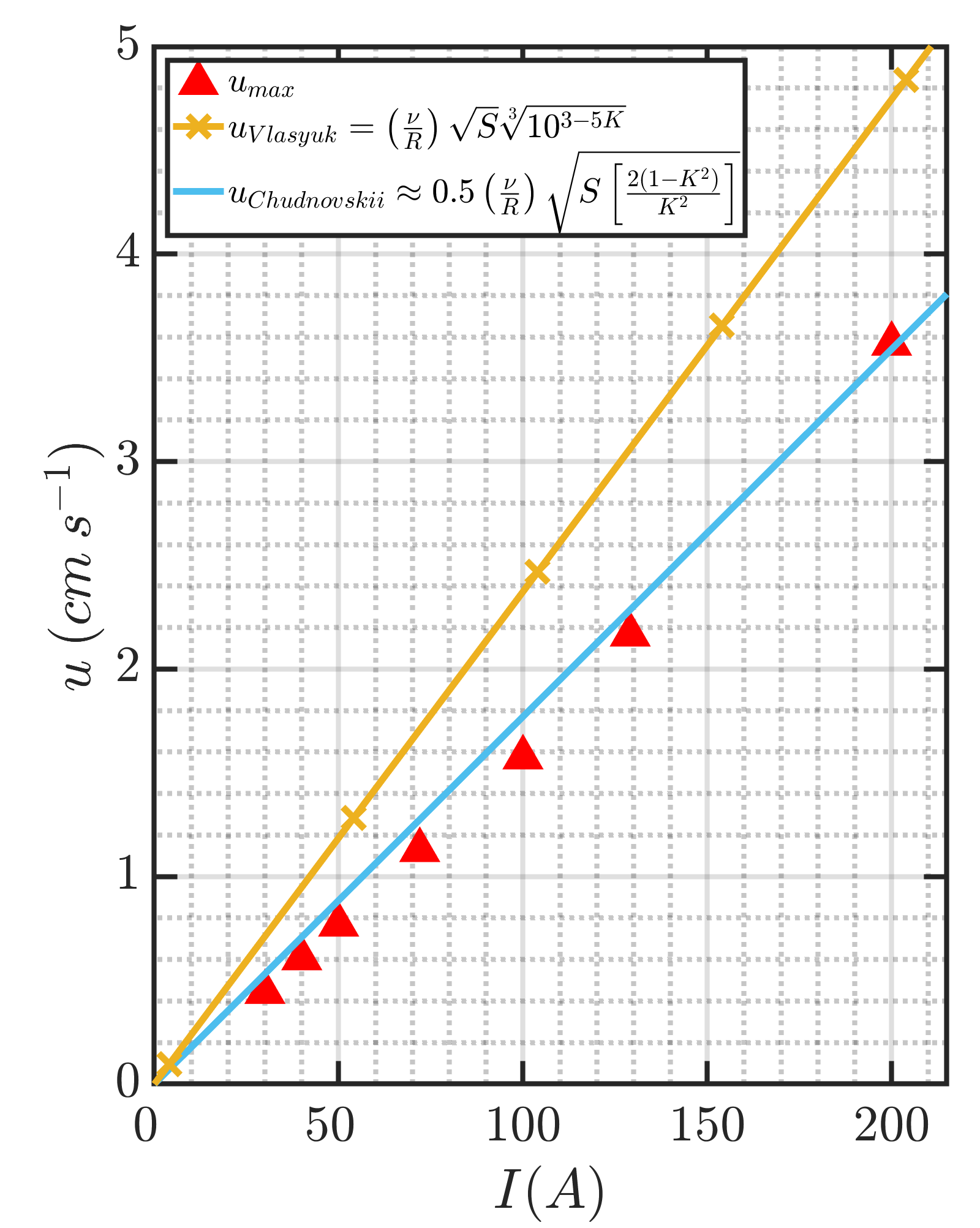}}
\centerline{(a) \hspace{175pt} (b)}
  \caption{(a) The time-averaged r.m.s. velocity is plotted as a function of $I$. The velocity varies linearly with $I$, indicating the scaling $u\propto I$ for high $\Rey$ flows. Our estimate closely captures this variation. (b) The maximum velocity as a function of $I$. This is compared with the estimates in the literature \cite{vlasyuk1987effects,chudnovskii1989evaluating}.}
\label{fig:u_theo_I}
\end{figure}

\begin{figure}  \centerline{\includegraphics[width=0.48\linewidth]{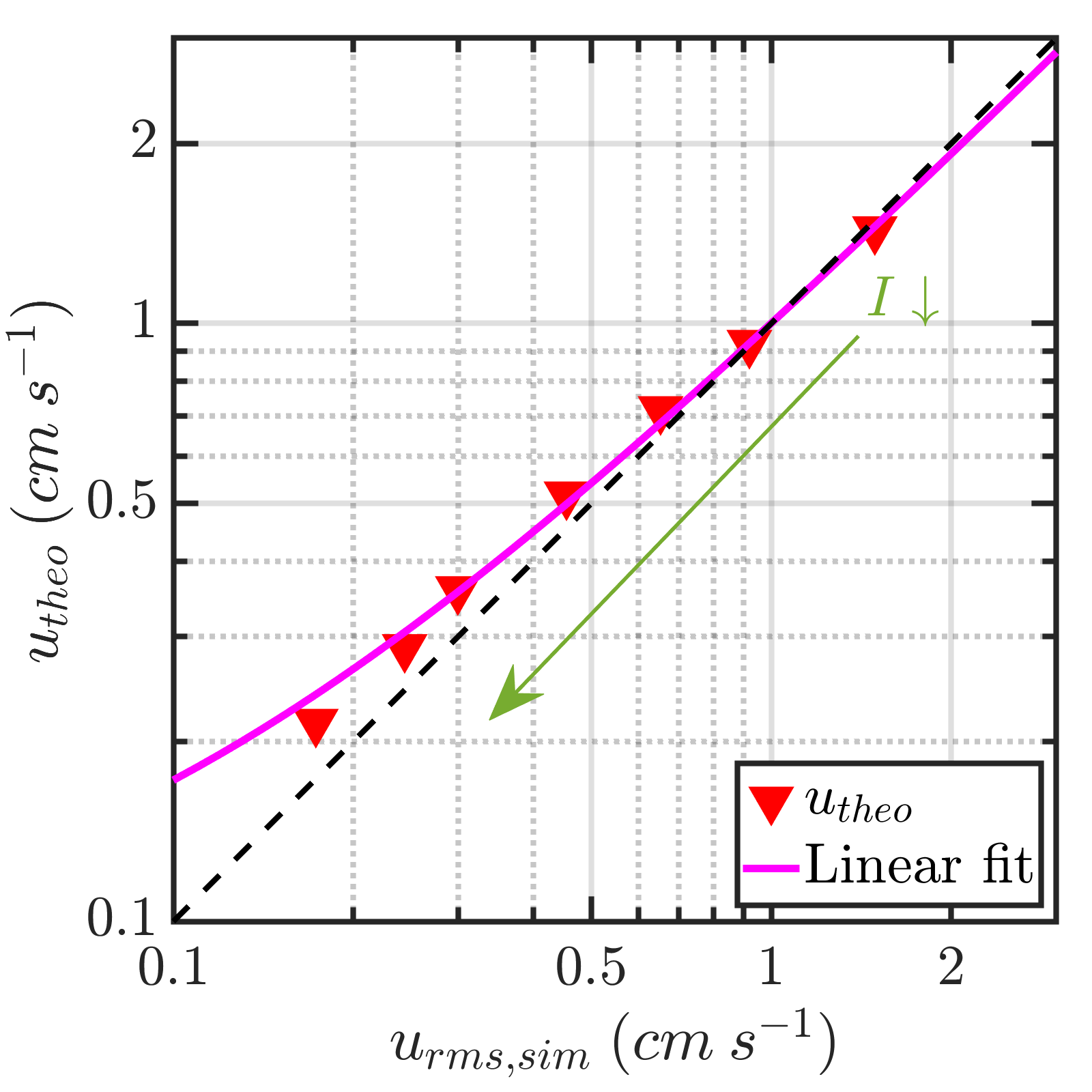} 
\includegraphics[width=0.48\linewidth]{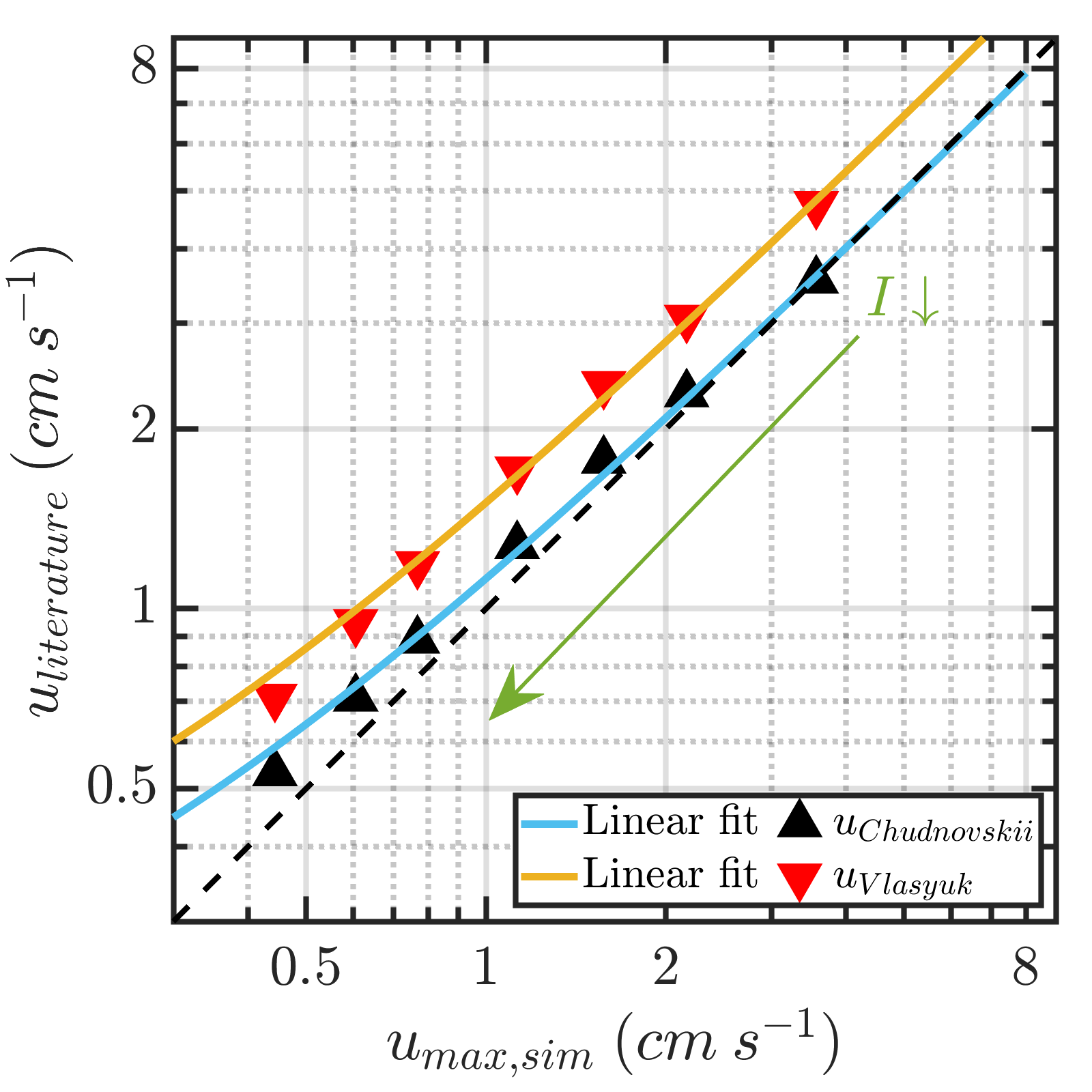}}
\centerline{(a) \hspace{175pt} (b)}
  \caption{(a) The $u_{theo}$ vs $u_{rms}$ for varying applied current values, and (b) estimates from the literature \cite{vlasyuk1987effects,chudnovskii1989evaluating} are plotted against the maximum velocity and their respective linear fit.}
\label{fig:u_theo_rms_I}
\end{figure}

Using the time interval $121<t<150$ for $I=$ \SI{10}{\A}, $71<t<100$ for $I=$ \SI{20}{\A} to \SI{40}{\A}, and $21<t<50$ for $I\geq$ \SI{50}{\A}, we calculate the r.m.s. and maximum velocity respectively using equations \cref{average_u,max_u}. These are reported in terms of $\Rey$ in \autoref{table:cases_I0}. We consider $I\in[30,200]$\SI{}{\A} for the validation of our estimate since $I\leq20$ \SI{}{\A} do not follow $u\propto I$ as discussed earlier. The r.m.s. velocity almost varies linearly with $I$ (see \autoref{fig:u_theo_I}(a)) indicating the applicability of the scaling $u\propto I$. Our theoretical estimate \labelcref{velocity_scale_Final} closely follows these data points with $\mathcal{R}^2 = 0.997$. On the other hand, Chudnovskii's estimate \labelcref{u_chudnovskii} closely predicts the maximum velocity (see \autoref{fig:u_theo_I}(b)) whereas Vlasyuk's estimate \labelcref{vlasyuk_formula} is close to the maximum velocity at lower currents and overpredicts it for larger currents. To further assess the accuracy of these estimates for varying $I$, we plot them against the $u_{rms}$ and $u_{max}$, respectively in \autoref{fig:u_theo_rms_I}(a) \& (b). The linear relationship between $u_{theo}$ and $u_{rms}$ is quite close for higher currents (see \autoref{fig:u_theo_rms_I}(a)), as expected. However, it slightly deviates at lower currents, due to low $\Rey$. This observation underscores the validity of our theoretical estimate for both laminar and turbulent flows. An analogous plot for the maximum velocity with the estimates of the literature \labelcref{u_chudnovskii} and \labelcref{vlasyuk_formula} is shown in \autoref{fig:u_theo_rms_I}(b) where the linear fit follows similar observations.

\section{Modified EVF parameter: combined effects of changing \texorpdfstring{$K \:\& \: I$}{K \& I}}          \label{mod_param}
\begin{figure}  \centerline{\includegraphics[width=0.6\linewidth]{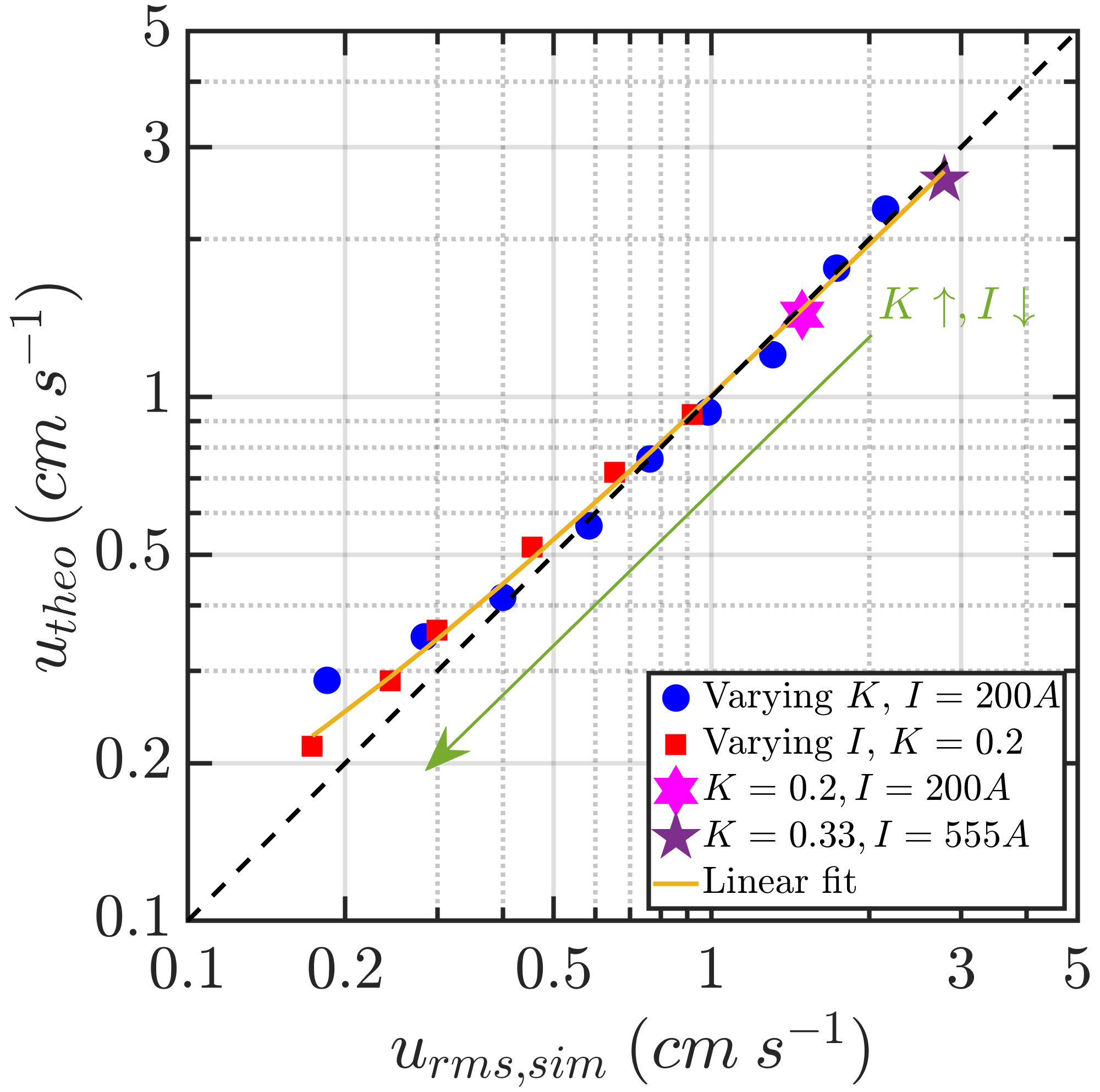}}
  \caption{Testing the validity of our theoretical estimate. The $u_{theo}$ is plotted against the r.m.s. velocity for the combined data set reported in \autoref{table:cases_K,table:cases_I0}. The linear trend indicates the strong validity of our estimate with $\mathcal{R}^2 = 0.991$.}
\label{fig:combined_u}
\end{figure}

Let us combine all the simulation results, reported in \cref{table:cases_K,table:cases_I0}, to validate our theoretical estimate for various $K\in[0.1,0.7]$ and $I\in[30,555]$\SI{}{\A} in a global manner. The $u_{theo}$ is nearly aligned with the $u_{rms}$ obtained from simulations, as shown in \autoref{fig:combined_u} with $\mathcal{R}^2 = 0.991$. Furthermore, we observed that the chosen proportionality constant, $\mathcal{C}_1 = 0.68$, facilitates close alignment of the data sets with the $u_{theo} = u_{rms}$ line. This demonstrates the robust validity of our theoretical estimate \labelcref{velocity_scale_Final}. This also allows us to extend our estimate to obtain a \emph{modified} EVF parameter
\begin{equation}
    S_M  = S \left[ \left( \dfrac{4}{3} \right) \dfrac{(1 - K)^2}{K} \right]   \notag
\end{equation}
that incorporates the effect of radius ratio, $K$. The term outside the $[\cdot]$ in this equation is the standard EVF parameter, $S = \mu_0 I^2/4\pi^2 \rho \nu^2$, used in the literature. Note that this is applicable for high $\Rey$ flows where the scaling $\Rey_{rms} \propto \sqrt{S_{M}}$ holds true. Here, $\Rey_{rms}$ is based on the time-averaged r.m.s. velocity, $u_{rms}$, and the domain radius, $R$. We also propose a new scaling relationship $u \propto I(1-K)/\sqrt{K}$ for this scenario; as a revision to the existing scaling $u\propto I$.

\begin{table}
  \begin{center}
\def~{\hphantom{3}}
  \begin{tabular}{lccccccccc}
 & Metal  & $K$  &    $I$    &  $S$ & $S_M$  & $\Rey_{max}$   & $\Rey_{rms}$   &   $u_{max}$  &   $u_{rms}$
      \\[3pt]
\citet{mohammad2025current} & Ga & 0.0636 & ~80 & \num{5.13e05} & ~\num{9.44e06} & 7950 & 2089 & 4.13 & 1.09 \\



\citet{herreman2019numerical} & Mg & 0.2~~~ & 250 & \num{2.07e06} & \num{8.85e06} & 4988 & 2023 & 7.78 & 3.16 \\

 & Sb & 0.2~~~ & 250 & \num{7.38e06} & \num{31.49e06} & 9411 & 3816 & 3.90 & 1.58 \\
  \end{tabular}
  \caption{Estimates for the r.m.s. and the maximum values for a couple of studies in the literature using \labelcref{velocity_scale_Final,u_chudnovskii} respectively. The current is in \SI{}{\A} and velocities are in \SI{}{\cm/\s}.}
  \label{table:u_estimate}
  \end{center}
\end{table}

\begin{figure}  \centerline{\includegraphics[width=1\linewidth]{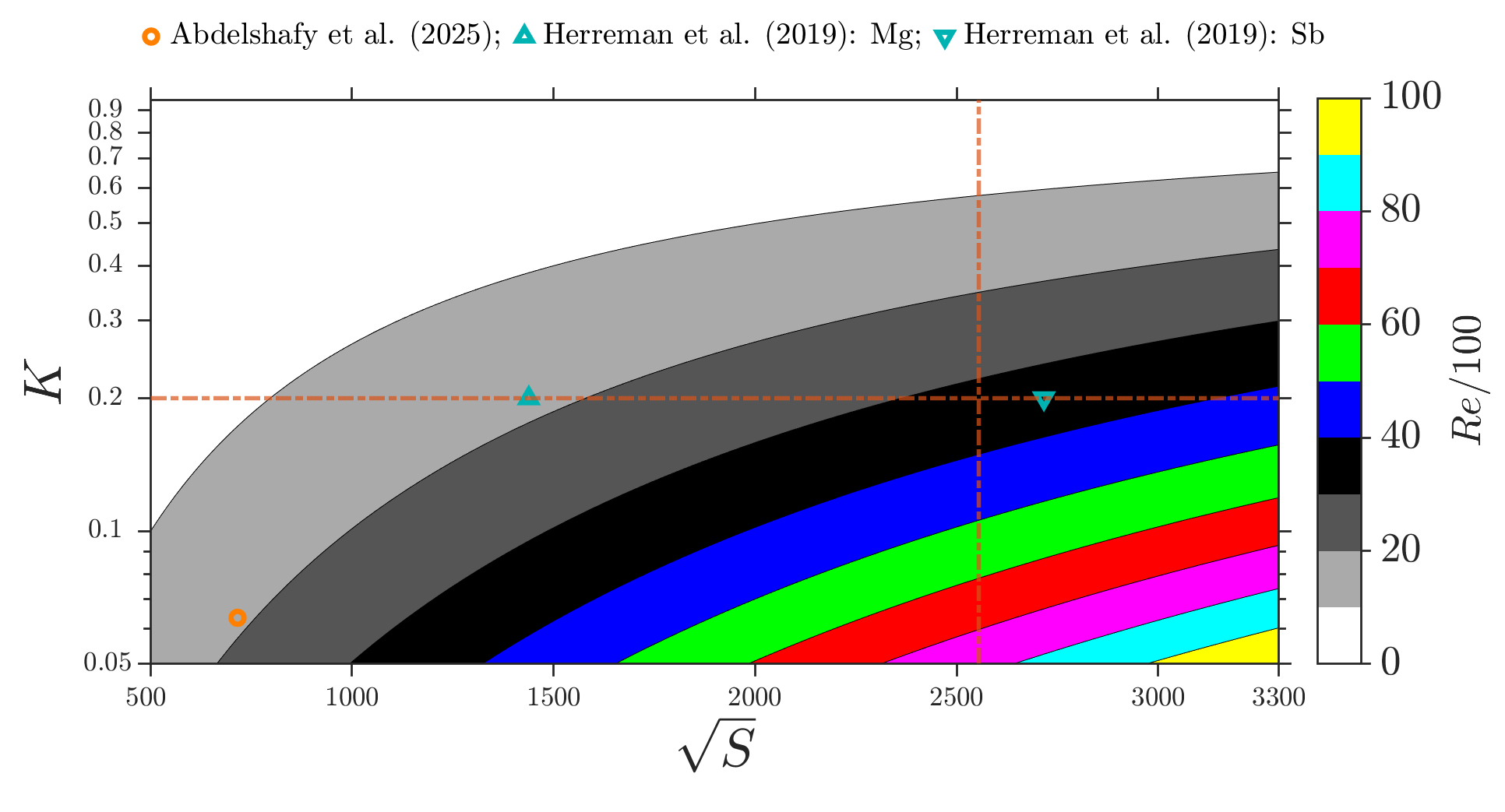}}
  \caption{The Reynolds number as a function of $K$ and $\sqrt{S}$. The orange dashed-dot lines denote the chosen parameters to validate our theoretical estimate with numerical simulations.}
\label{fig:K_I_map}
\end{figure}

An outcome of this study is the variation of $\Rey$ with $K$ and $S$ that can be helpful in the design of an LMB current collector. For this purpose, let us plot $\Rey$ with $K$ and $\sqrt{S}$ in \autoref{fig:K_I_map} using \eqref{velocity_scale_Final,Re_theo}. Once again, it can be noted that $\Rey$ is strongly dependent on $K$ for the same $S$. Using the parameters ($K$, $I$, and material properties) in the literature \citep{mohammad2025current,herreman2019numerical}, we estimate the r.m.s. and the maximum velocity from \eqref{velocity_scale_Final,u_chudnovskii} respectively, listed in \autoref{table:u_estimate}. The maximum EVF velocity reported by \citet{herreman2019numerical} for Mg electrode is \SI{8}{\cm/\s} which compares well with our calculation. Note that the recent work by \citet{mohammad2025current}, being an experimental study with Earth's external magnetic field present, does not contain any measured EVF velocity for us to compare. Nevertheless, their typical velocity of \SI{0.5}{\cm/\s} is close to our r.m.s. estimate. In both investigations, the empirical estimates of \citet{vlasyuk1987effects} were used to characterize the strength of EVF. As evident from \autoref{fig:u_theo_I}(b), this is not recommended for large currents. Moreover, the r.m.s. velocity, rather than maximum, is a better representation of a typical EVF velocity. Thus, our theoretical estimate combined with the CFD results will be helpful in the selection of the CC radius for a given current (or current density) by accounting for the strength of the electro-vortex flow.

\section{Conclusions}    \label{sec:conclusion}
In this article, we investigate the role of the current collector radius ($r_0 = K R$) on the electro-vortex flow (EVF) characteristics in a cylindrical liquid metal domain using both analytical and numerical approaches. Using the vorticity transport equation, we derive a new theoretical estimate of the \emph{average} EVF velocity, given as
\begin{equation} 
    u_{theo} \approx 0.68 \sqrt{\dfrac{\mu_0 I^2}{4\pi^2 \rho R^2} \left[ \left( \dfrac{4}{3} \right) \dfrac{(1 - K)^2}{K} \right]}.
    \notag
\end{equation}
Our estimate proposes a modified scaling law, $u \propto I (1-K)/\sqrt{K}$ for high $\Rey$ regime; a revision to the existing law, $u \propto I$. This estimate incorporates the effect of diverging current lines through the radius ratio ($K$); an important parameter that governs the EVF. Extending our estimate leads to the \emph{modified} EVF parameter, $S_M$, an alternative to $S$, which clearly incorporates the nature of functional dependency on $K$, in addition to current strength and fluid properties. This leads to $\Rey \propto \sqrt{S_M}$. We also revisit and combine the formulations of the maximum EVF velocity \eqref{eq16}, applicable for large currents, in \citet{chudnovskii1989evaluating} and \citet{davidson2017MHD}, and provide a physical justification and numerical validation.

We validated our theoretical estimate with extensive numerical simulations using our custom code in \textsc{OpenFOAM}. It demonstrates excellent agreement with the numerical results within the range $K\in[0.1,0.7]$ and $I\in[30,555]$\SI{}{\A}. We also evaluated the predictive capacity of the estimates of \citet{vlasyuk1987effects} and \citet{chudnovskii1989evaluating} for the \emph{maximum} strength of EVF at high $\Rey$. We found that Chudnovskii's estimate \cref{u_chudnovskii} better captures the nonlinear trend for $u_{max}(K)$ as compared to that by Vlasyuk's estimate \cref{vlasyuk_formula}, widely used in the literature.


Using vorticity dynamics and the Prandtl-Batchelor theorem \citep{batchelor1956steady}, we explain key characteristics of the electro-vortex flow, including the driving mechanism, formation and evolution of the electro-vortex jet. Notably, our numerical results reveal distinct flow behaviours for $K\geq 0.75$ in confined domains -- features not previously reported in the literature. In such cases, the canonical EVF is significantly altered by flow effects arising from the finite size of the domain. These findings also point to promising directions for future work, such as investigating the impact of varying the domain height ($R \neq H$). Even though our study is focused on standard electro-vortex flow (EVF), the analytical methods can be extended for practical flows where other multiphysics phenomena -- such as swirl -- could be present. This may be useful for the design of current collectors in liquid metal batteries. In perticular EVF in Na-Zn batteries is reported to be important for the battery design \citet{duczek2024fluid}.

\vspace{0.5cm}
\textbf{Acknowledgements.} The authors would like to thank Ravi Kant for the fruitful discussions and anonymous reviewers whose constructive feedback helped refine the content of the article. SS wishes to thank Dr. Tom Weier, Dr. Gunter Gerbeth, Prof. Caroline Nore, and Dr. Ashish Mishra for the insightful discussions and comments.  AR wishes to thank IRCC, IITB and SERB for the financial support.

\vspace{0.5cm}
\textbf{Declaration of Interests.} The authors report no conflict of interest.

\bibliographystyle{jfm}
\bibliography{jfm_references}

\end{document}